\documentclass[twocolumn,nolinenumbers,twocolappendix]{aastex631}
\usepackage{graphicx}
\usepackage{MnSymbol}
\usepackage{placeins}
\usepackage{hyperref}
\usepackage{cleveref}
\usepackage{gensymb}
\usepackage{tablefootnote}
\graphicspath{{Figures/}}
\usepackage{nameref}
\Crefname{figure}{Fig.}{Figs.}
\usepackage{tikz}
\usetikzlibrary{shapes.geometric}
\begin{document}

\newcommand{\affilCEA}{ Université Paris-Saclay, Université Paris Cité, CNRS/AIM, CEA-Saclay/DAp, F-91191 Gif-sur-Yvette, France }

\title{SNR~1987A: {\it Spitzer} data from days 6000 to 8000 revisited}

\correspondingauthor{P.\ Bouchet}
\email{Patrice.Bouchet@cea.fr}

\author[0000-0002-6018-3393]{Patrice Bouchet}
\affiliation{\affilCEA}

\author[0009-0007-5200-1362]{René Gastaud}
\affiliation{Université Paris-Saclay, Université Paris Cité, CEA-Saclay/DEDIP, F-91191 Gif-sur-Yvette, France}

\author[0000-0001-6492-7719]{Alain Coulais}
\affiliation{LIRA, Observatoire de Paris, Université PSL, Sorbonne Université, \\
Université Paris Cité, CY Cergy Paris Université, CNRS, Paris, France}
\affiliation{\affilCEA}
 
\author[0000-0001-8403-8548]{Richard G. Arendt} 
\affiliation{Center for Space Sciences and Technology, University of Maryland, Baltimore County, Baltimore, MD 21250, USA}
\affiliation{Code 665, NASA/GSFC, 8800 Greenbelt Road, Greenbelt, MD 20771, USA}
\affiliation{Center for Research and Exploration in Space Science and Technology, NASA/GSFC, Greenbelt, MD 20771, USA}

\begin{abstract}

An excess emission has been observed by {\it Spitzer} in the 3 – 5~\micron\ range of the SNR~1987A spectrum. It is generally argued that this excess could be due to the presence of warm amorphous carbon dust in the equatorial ring (ER) around the supernova, but the proposed models all have problems. This prompted us to present an alternative view on the interpretation of the Spectral Energy Distribution (SED) of SNR~1987A from the near-IR wavelengths to the radio frequencies (from 3~\micron\ up to 1.4 GHz), between 6000 and 8000 days after outburst. We argue that the origin of that excess could be attributed instead to a free-free emission. We show that under very specific conditions (the free-free is self-absorbed at a cut-off frequency imposed by the mass of the emitting region), it could be produced by collisional heating of the gas. We then discuss the time evolution of the various components of the SED. We establish a linear relationship between the growth of the warm carbon dust mass and that of the silicates dust during the analyzed period. Finally, we build the {\it Spitzer} light curves and we show that our models reproduce the observations pretty well, although our study clearly favors the free-free case. In conclusion, we argue that the free-free model provides a formally very good description of the data, however the model does require some very specific parameter choices, and results in an unusually low temperature for the ionized gas. 

\end{abstract}

\keywords{ supernovae: individual: SN~1987A (124); Core-collapse supernovae (304); Supernova remnants (1667);
Circumstellar dust (236); Infrared astronomy (786); Light curves (918)}

\section{Introduction}
\label{sct:intro}

Astronomers have followed the full evolution of the iconic supernova SN~1987A, the closest optical supernova (SN) in 400 years, across the entire electromagnetic spectrum, owing to its close proximity within the nearby Large Magellanic Cloud (LMC) 
\citep[$d = 49.6$~kpc,][]{Pietrzynski2019}, as it completes its transformation into a SNR (see \citealp{McCray1993,Woosley1997,McCray2016} for reviews).  It is therefore natural that SNR~1987A was among the first targets selected for observation during the first observations carried out by the {\em James Webb Space Telescope} (JWST) (see for instance \citealp{Arendt2023,Jones2023,Larsson2023,Bouchet2024,Fransson2024,Matsuura2024}).  

The first possible spatially-resolved detection of near- and mid-IR emission for any SN was reported by \citet{Bouchet2004}, using the Gemini South 8-m telescope at the position of SN~1987A on day 6067 since the explosion. Following that, {\it Spitzer} observations of SNR~1987A spanned more than a decade.
Since day $\sim$4,000 after outburst, the mid-IR emission has been dominated by dust, particularly residing in the inner equatorial ring (ER) rather than from the ejecta. Indeed, the ground-based images of \cite{Bouchet2004} showed this directly, and was confirmed by the lower resolution {\it Spitzer} images that required deconvolution (or modelling) \citep{Arendt2016, Arendt2020}. Decomposition of the marginally-resolved emission also confirms mid-IR domination by dust in the ER.

The remnant of the supernova SN 1987A is being born. Numerous articles report on recent observations in the whole electromagnetic spectrum. Those include, for instance, \cite{Larsson2016,Larsson2019b} for the visible range (and references therein), \cite{Indebetouw2014} and \cite{Matsuura2017, Matsuura2019, Matsuura2022}
(and references therein) for the far-IR and (sub)-mm range, and \cite{Manchester2005}, \cite{Staveley2007}, \cite{Cendes2018}, and \cite{Zanardo2018} (and references therein) for radio observations. \cite{Matsuura2015, Matsuura2019, Matsuura2022} propose a model of the spectral energy distribution of SNR~1987A which includes in particular a cold dust component in the ejecta of $\sim0.5 \pm0.1 $~$M_\sun$ at $\sim20$~K discovered by {\em Herschel} \citep{Matsuura2011, Matsuura2015} and confirmed by ALMA \citep{Indebetouw2014}. The timing for its condensation and the location of this cold dust formation in the remnant of SN 1987A has been studied by \cite{Wesson2015}.

On the other hand,~\cite{Bouchet2006}, \cite{Dwek2010}, \cite{Arendt2016}, and \cite{Arendt2020} analyze 
{\it Spitzer} observations, and interpret the 3 - 20~\micron\ spectrum as the 
sum of 1.2 x 10$^{-6}$~$M_\sun$ of silicates dust at $\sim190$~K at day 7554 together with a warm amorphous carbon component at $\sim525$~K to account for the excess emission seen at the shorter wavelengths. Several models have been proposed based on the presence of warm dust, with various different compositions, but they all present difficulties \cite[for instance]{Kangas2023}. Moreover, as noted by \cite{Dwek2010} the potential presence of carbon with the silicates dust in the ER is surprising, defying the common paradigm that either carbonaceous or silicates dust can form in the wind of the progenitor, depending on the C/O abundance ratio in the outflow. 

The observations therefore suggest either that CO formation did not exhaust all the carbon in the outflow or that the co-existence of the two dust components is a manifestation of a binary origin of the ER, in which case the progenitor of SN~1987A could have shared a common envelope with a less massive carbon-rich star either at the time of the explosion or at least in the not-too-distant past as was first proposed by \cite{Hille1989} and \cite{Podsiadlowski1990}. Indeed, as shown by ~\cite{Podsiadlowski2003}, \cite{Morris2007, Morris2009}, and \cite{Podsiadlowski2017}, such a merger can result in a BSG and perhaps a complex circumstellar system, including  the two outer rings observed by the {\em Hubble Space Telescope} (HST) \citep{Morris2009}. On the other hand, \cite{Woosley1997} have argued that single star models might yet prove to be the correct explanation for the progenitor Sk -69$\degree$202, provided new rotational or convection physics can simultaneously result in a blue star and explain the ring structure. Later on, \cite{Woosley2002} pointed out that even if it is possible (within the current uncertainties of stellar evolution) to make assumptions that cause models of single stars to return to the blue, these assumptions are not natural and to get the merger \citep[postulated by][]{Podsiadlowski2017} to happen just 20,000 years before the supernova, would make SN~1987A an uncommon even. 

\cite{Menon2017} propose 84 pre-supernova models (16 -- 23~$M_\sun$) of binary mergers for blue supergiant progenitors. Within the parameter space studied, the majority of their pre-supernova models are compact, hot BSGs with effective temperature $>$12,000~K and radii of 30 -- 70~$R_\sun$ of which six match nearly all the observational properties of Sk -69$\degree$202. This work cannot be considered therefore as 100\% conclusive. Later on, ~\cite{Menon2019} argued that the explosions of their binary merger models exhibit an overall better fit to the light curve of SN 1987A than previous single star models: their merger model that best matches the observational constraints of the progenitor of SN~1987A and the light curve is a model with a radius of 37~$R_\sun$, an ejecta mass of 20.6~$M_\sun$, an explosion energy of $1.7 \times  10^{51}$ erg, a nickel mass of 0.073~$M_\sun$, and a nickel mixing velocity of 3000 km~s$^{-1}$, which is in good agreement with the observations. These authors note, however, that several single star models also match the observations, and produce reasonably good fits with the light-curve shape.

Finally, ~\cite{Utrobin2019} showed that none of the single-star progenitor models proposed for SN 1987A to date satisfies all constraints set by observations (the blue color of the progenitor, the rings surrounding it, the progenitor's chemical anomalies, the characteristics of the supernova explosion, and general consistency with the theory of massive stars) as reviewed by \cite{Podsiadlowski1992b}. The binary origin of SN~1987A is thus currently the most accepted model by the community.

However, it is a fact that all the models built under the hypothesis RSG - BSG, be the progenitor a single or a binary star, suffer from serious caveats as noted by various authors. \cite{Woosley2002}, for instance, point out that if 5\% of all massive stars end their lives as blue supergiants because of merger with a companion ~\citep{Podsiadlowski1992}, one would expect most (other) supernovae in the LMC to occur in red supergiants and further that a few percent of all supernovae, even those occurring in regions of solar metallicity, would be like SN 1987A. So far observations do not test this prediction. \cite{Parthasarathy2006}, on another hand, note that if Sk -69$\degree$202 was a merger product, rapid rotation of a few hundred kms$^{-1}$ is expected, while the spectra of Sk -69$\degree$202 had shown no evidence for rapid rotation. 

\cite{Parthasarathy2006} then suggest that it is possible that Sk -69$\degree$202 never was an RSG, since \cite{Pastorello2005}  showed that the progenitor of SN 1998A also was a BSG. In spite of the strong observational selection against the discovery of subluminous events, those authors cite the existence of several other Type II SNe with (fragmentary) light curves possibly similar in shape to SN 1987A and SN 1998A, which suggests that explosions of BSGs are not too uncommon. This would be inconsistent with explanations that require fine tuning (as do all of the SN~1987A progenitor models based on the RSG - BSG scenario). Instead, there may be some regular channel for producing SN 1987A–like events without invoking a binary origin of the SN~1987A precursor, as claimed by \cite{Parthasarathy2006}.

In conclusion, the presence of lukewarm carbon together with silicates in the ER is unlikely. This prompted us to look for another plausible explanation to account for the excess emission at short infrared wavelengths. 

The collision between the blast wave of SN~1987A and the inner ER predicted to occur sometime in the interval 1995 -- 2007 \citep{Gaensler1997,Borkowski1997} was still underway during the epoch considered here, although the shock wave was already interacting with the material outside the ring. Since $\sim$5,000 days post-explosion, bright multi-wavelength emission has been produced by the shock interaction between the ER and the forward shock \citep{McCray2016}.
After day $\sim8500$, this emission has been observed to be fading in IR, optical, and soft X-rays \citep{Fransson2015,Larsson2019b,Arendt2020,Alp2021,Maitra2022}, suggesting that shocks are disrupting the dense ER and that the blast wave has passed through it already at the time of the observations reported in the present work \citep{Fransson2015}. Overall, at that time we were witnessing the interaction of the SN blast wave with its surrounding medium, creating an environment that was rapidly evolving at all wavelengths.
In summary, since its explosion, SN~1987A has evolved from an SN, dominated by the emission from the radioactive decay of $^{56}$Co, $^{57}$Co, and $^{44}$Ti in the ejecta, to an SNR whose emission is dominated by the interaction of the SN blast wave with its surrounding medium \citep{Larsson2011}. 

The circumstellar medium consists of an ER flanked by two outer rings \citep{Burrows1995}, possibly part of an hourglass structure \citep{Chevalier1995,Sugerman2005}. The free expansion of the dense inner ejecta has simultaneously continued within the ER, revealing a highly-asymmetric distribution in progressively greater detail. At UV-optical (UVO) wavelengths, ``hot spots" have appeared inside the ER \citep{Pun1997}, and their brightness varies on timescales of a few months \citep{Lawrence2000}. These spots are inward protrusions of the dense ($\sim$10$^4$~cm$^{-3}$) ER, embedded in the lower density ($\sim$10$^2$~cm$^{-3}$) HII region interior to the ring \citep{Groning2008b}.

New hot spots have continued to appear as the entire inner rim of the ER has become lit up by the interaction with the blast wave.
HST images with equivalence to $R$-band (WFPC2/F675W, ACS/F625W, and WFC3/F675W; see \citealp{Larsson2021}) obtained between 1994 and 2009 revealed a necklace of such hot spots, nearly filling a lighted ring.
Monitoring at X-ray wavelengths with the {\em XMM-Newton}, {\em Chandra}, and at radio frequencies, showed that while soft X-rays followed the optical and IR evolution, reaching maxima at $8,000 - 10,000$ days, hard X-rays and radio have shown a steady increase in flux \citep[e.g., Fig.\ 4 in][]{Alp2021}. 

Comparison of the Gemini $11.7~\micron$ image with {\em Chandra} X-ray images, {\em Hubble} UVO images, and Australia Telescope Compact Array (ATCA) radio synchrotron images shows generally good correlation across all wavelengths \citep{Bouchet2006}.
If the dust resides in the diffuse X-ray emitting gas then it is collisionally heated. The IR emission can then be used to derive the plasma temperature and density, which were found to be in good agreement with those inferred from the X-rays \citep{Dwek2010}.
Alternatively, the dust could reside in the dense UVO knots and be heated by the radiative shocks that are propagating through the knots. 

\cite{Arendt2020} give a comprehensive summary of the whole Spitzer observations data set: they reveal in particular that decomposition of the marginally resolved emission confirms that the emission at $3.6$ and $4.5$~\micron\ is dominated by the ring, not the ejecta, and that the infrared morphological changes that occur during the reported observations resemble those seen in both the soft X-ray emission and the optical emission. They also note that the integrated ER light curves at 3.6 and 4.5 \micron\ are more similar to the optical light curves than the soft X-ray light curve, though differences would be expected if dust is responsible for this emission and its destruction is rapid.  

\cite{Arendt2020} suggest a model to fit the light curves ({\it Spitzer}, soft X-rays, and optical) with a Gaussian function convolved with an exponential decay. Unknowns of their model are the nominal date for the peak of the interaction, the scale for the rise in the emission (which starts when the shock sweeps into the ER), and the scale for its fading. Although the fits obtained by these authors are acceptable, the fact that the values of these parameters turn out to be surprisingly similar at the different wavelengths prompts us to suggest a different approach in order to fit the near-IR (3.6, 4.5 and 5.8~$\micron$) {\it Spitzer} light curves.

The work presented here could thus be considered as a complement of \cite{Arendt2020}'s. It is more specifically a new version of \cite{Matsuura2019}'s analysis of the Spectral Energy Distribution (SED) of SNR~1987A, that focuses on the time variability of each of its components, with a special emphasis on the origin of the near-IR excess emission first observed with {\it Spitzer} at its shortest wavelengths (i.e. 3.6 -- 5.8 \micron). 

We describe the decomposition of the SED into several components in \S~\ref{sct:decomposition}, and we discuss the data used in our analysis in \S~\ref{sct:observations}. We present our fitting procedure and we show the results of our fits for the warm carbon dust model in \S~\ref{sct:results}. We then discuss the free-free emission in \S~\ref{sct:ff} : we first explore its possible origin (\S~\ref{sct:origin}), we then propose a self-absorbed model for this emission (\S~\ref{sct:selfabsorbed}), and we give the results of our fits for that model (\S~\ref{sct:results_ff}). We show the evolution with time of the various parameters which define the components of the SED for the two models in \S~\ref{sct:evolution}, and we build the resulting theoretical light curves and discuss their agreement with the {\it Spitzer} data in \S~\ref{sct:lc_theoretical}. Finally, we give our conclusions in \S~\ref{sct:conclusion}.  

\section{Decomposition of the SED}
\label{sct:decomposition}
We decompose the IR-Radio SED of SNR~1987A into 5 components associated with the inner ejecta or with the ER (which dominates the 3.6 and 4.5 $\mu$m emission):
\begin{enumerate}
\item{Assuming emission from the ER, \cite{Dwek2010}, \cite{Arendt2016}, and \cite{Arendt2020} interpret the {\it Spitzer} observations with the sum of 2 dust components with:
\begin{equation} \label{eq:fnuspitzer}
F^{{\rm Spitzer}(i)}_{\nu} = \sum{\frac{4\kappa_{i}(\nu)}{4\pi d^{2}}\ M_{i}(t)\ \pi B_{\nu}[T_{i}(t)]}
\end{equation}
where $i$ = 1 for amorphous carbon and $i$ = 2 for silicates; $\kappa_i(\nu)$ are the mass absorption coefficients of the dust taken from \cite{Draine1984} and \cite{Rouleau1991}, for the silicates and amorphous carbon
respectively; $d$ is the distance to the SN; and $B_\nu$ is the Planck function.}

\item 
In a partially ionized gas, the free electrons accelerate in the vicinity of ions and produce free-free radiation (free electrons also recombine to form neutral atoms, emitting line radiation while cascading towards the ground level).

The general expression giving the free-free continuum flux that arises from the ionized fraction of a hydrogen envelope (as do the
recombination lines) is given by: 
\begin{equation} \label{eq:fnu-ff-general}
F_\nu^{\rm ff} = C~Z^2~n_e n_{\rm i}~ T_{\rm ff}^{-1/2} ~\exp({-h\nu}/{kT_{\rm ff}})\ g
\end{equation}
with $g$ being the Gaunt factor. 
[Equation~(\ref{eq:fnu-ff-general}), can be found for instance, in the lectures given by Frank van den Bosch (Yale Physics Department)\footnote{\url{https://www.astro.yale.edu/vdbosch/astro320-summary27.pdf}}, or  J.S. Kaastra (Leiden University)\footnote{\url{https://ned.ipac.caltech.edu/level5/Sept08/Kaastra/Kaastra3.html}}.] 

For SNR~1987A, we will see in \S\ref{sct:free-free} that the temperature that we find is less than 5000~K, and the number density is less than $10^{10}$~cm$^{-3}$, so the gaunt factor is unity.
\cite{Wooden1989} modifies this expression by introducing the ionization coefficient $\xi$, and this was later reformulated by \cite{Wesson2015}, to give (in units of W~m$^{-2}$~{Hz}$^{-1}$):
\begin{equation} 
F_\nu^{\rm ff} = 5.733 \times 10^{-15} (1 - \xi)^2 \exp({-h\nu}/{kT_{\rm ff}})  T^{-1/2}_{\rm ff} t^{-3}\ .
\end{equation}
This expression was formulated for the free-free emission associated with the hydrogen envelope in the ejecta at early times (up to day 777 after outburst), and not within the ER. Still, the analytic calculations performed by \cite{Wooden1989} only assume a uniform density, a constant ionization fraction, and a maximum velocity of the envelope of 3000~km s$^{-1}$. These assumptions could be legitimately applied to the ER. However, the time dependence of the ionization rate is derived from the radioactive decays that power the light curves at the epoch analyzed by these authors, which doesn’t apply to our analysis. We will therefore use for our fitting procedure the equation:
\begin{equation} \label{eq:fnu-ff}
F_\nu^{\rm ff} = A~(1 - \xi)^2 ~\exp({-h\nu}/{kT_{\rm ff}}) ~T^{-1/2}_{\rm ff} 
\end{equation}
where $A = {5.733 \times 10^{-15}}/{(7200 -5500)^3}$ (assuming that the blast wave reaches the ER at day 5500), and $\xi = \xi(t)$.
The time dependence in our Equation~(\ref{eq:fnu-ff}) is thus moved from an explicit $t^{-3}$ to an unspecified functional form for $\xi(t)$. 
In other words, $(1-\xi)^2$ serves as a proxy for the total mass. Changing $T_{\rm ff}$ shifts the spectrum in wavelength [via $\exp({-h\nu}/{kT_{\rm ff}} $)] but $(1-\xi)^2$ is needed to adjust the overall normalization which is otherwise fixed by an assumed gas mass hidden in the coefficient $A$. \cite{Wooden1989} and \cite{Wesson2015} assume 10~M$_{\sun}$ which is far too much for the ER, but they were looking at the supernova at much earlier times, and not at the remnant. 
    
\cite{Laki2011} use a free-free emission component in their decomposition of the overall SED for the
two periods 2004 -- 2005 and 2007 -- 2010. These authors give an excellent description of the SED fitted by a cold dust and
a synchrotron components, to which is added a free-free contribution limited by the near-IR data (they quote that it may include contributions from hot dust and/or atomic line emission). To represent the free-free radiation, they use for their fitting procedure:

\begin{equation} \label{eq:fnu-ff-alpha}
F_\nu^{\rm ff} \propto \nu^{\alpha_{\rm ff}}
\end{equation}
with $\alpha_{\rm ff}$ constant ($\approx$~0.1) over the period 2004 -- 2010 days. Their formulation is therefore distinct from ours, which is due to the fact that the present work aims at deriving from our fitting procedure a time dependence of the evolution of the ionization coefficient.

\item \cite{Matsuura2019} include an additional, warmer dust component to account for an observed excess emission at 30 and 70 \micron\ by SOFIA near day 10000. This component is likely from the ring. The SOFIA data were obtained at days 10731—10733, well after the period studied in the present work. We stress that it is very important to not mix data too far apart in time, because the emission is evolving (Figure~\ref{fig:lc_IRAC} shows that it changes by factors greater than the uncertainties in only $\sim$100 days). Whether or not that component is significant, it is not relevant for the analysis of the period considered in our work.

\item Cold dust in the ejecta accounts for the emission at Herschel and ALMA wavelengths. Following \cite{Laki2011} the far-IR to sub-mm emission can be represented by a modified Planck curve of thermal dust emission:
\begin{equation} \label{eq:fnu-bnu-alpha}
F_\nu^{\rm IR} \propto B_\nu(T)\ \nu^\alpha_{\rm IR}.
\end{equation}

\cite{Indebetouw2014} use an amorphous carbon dust at 26~$\pm$~3~K (Eq. 1) and a mass of $0.23\pm~0.05~M_{\sun}$. \cite{Cigan2019} found a 18-23 K dust temperature for a mixture of carbon and silicates dust.

\item Synchrotron radiation from the ER accounts for emission at the ATCA frequencies \citep{Zanardo2014, Cendes2018, Cigan2019}. 
The mm-dm radio emission is matched by a power law with a coefficient typical of synchrotron radiation [note that $\alpha$ here is distinct from $\alpha$ in Eq.~(\ref{eq:fnu-bnu-alpha})]:
\begin{equation} \label{eq:fnu-alpha}
F_\nu^{\rm radio} \propto  \lambda^\alpha_{\rm radio}.
\end{equation}

\end{enumerate}

The overall SED of SNR 1987A between 3~$\mu$m and 1~cm can then be represented either by:

\begin{equation} \label{eq:fnu1total}
F_\nu^{\rm total} = F_\nu^{\rm Spitzer(1)} + F_\nu^{\rm Spitzer(2)} + F_\nu^{\rm IR} + F_\nu^{\rm radio}
\end{equation}
or
\begin{equation} \label{eq:fnu2total}
F_\nu^{\rm total} = F_\nu^{\rm ff} + F_\nu^{\rm Spitzer(2)} + F_\nu^{\rm IR} + F_\nu^{\rm radio}
\end{equation}
depending on which model we choose for the interpretation of the excess emission at the shortest IR wavelengths.

Note that we do not include the line emission of any components, as the total flux of any of the  observed IR lines is negligible compared to the continuum when measured in the broad bands
used by {\it Spitzer} \citep[e.g.][]{Arendt2016} and at longer wavelengths. 

\section{Observations}
\label{sct:observations}
\subsection{Data Used}
We are using data from the {\it Spitzer} Space Telescope \citep{Werner2004,Gehrz2007} with its 2 instruments IRAC \citep{Fazio2004} and MIPS \citep{Rieke2004}. Unfortunately on May 15, 2009 (around day 8000) {\it Spitzer} ran out of its liquid helium coolants and entered in its warm phase. Therefore, the last {\it Spitzer} measurements longward of 5.8~\micron\ were obtained at day 7983, and only observations at $3.6$ and $4.5~\micron$ are available after that date. We use also data from the AKARI mission \citep{Murakami2007} with its 2 instruments IRC \citep{Onaka2007} and FIS \citep{Kawada2007}, from the {\it Herschel} Space Telescope  \citep{Pilbratt2010} and its two instruments PACS \citep{Poglitsch2010} and SPIRE \citep{Griffin2010},  from the Large APEX BOlometer CAmera 
\citep[LABOCA,][]{Siringo2009} and the Submillimeter APEX BOlometer CAmera  
\citep[SABOCA,][]{Siringo2010}, and from the Australia Telescope Compact Array 
\citep[ATCA,][]{Staveley1992}. Table~\ref{tab:telused} displays the characteristics of all the instruments that have been used in our study (wavelengths, references and epochs of observations).

All of the SNR~1987A {\it Spitzer} data that we use are from \cite{Arendt2020}, who discuss them extensively. They are reproduced in  Figures~\ref{fig:lc_IRAC} and \ref{fig:imageshort}.

\subsection{The Light Curves}
\label{sct:lightcurves}

Figure~\ref{fig:lc_IRAC} displays the light curves obtained by {\it Spitzer} until day 10,378 (July 23, 2015) reported by \cite{Arendt2016}. Note that more data have been obtained throughout the entire mission, until day 11,886 \citep[Sept. 10, 2019,][]{Arendt2020}.
%Figure~\ref{fig:imageshort} shows unambiguously that the emission observed at $3.6$ and $4.5\micron$ originates indeed from the ring.
Figure~\ref{fig:imageshort} shows that the emission contributing to the light curves at $3.6$ and $4.5\ \micron$ by {\it Spitzer} is dominated by the ring, which is supported 
by JWST NIRCam observations at a later date \citep[day 12975][]{Arendt2023,Matsuura2024}.
The near-infrared light curves illustrate the history of SN 1987A during the analyzed period (after the interaction between the shock wave and the inner equatorial ring at day $\sim$5500). 

\cite{Woltjer1972} and then \cite{Reynolds2008} describe the hydrodynamic evolution of SNRs and identify 4 phases that reflect the different physical mechanisms that occur.  
However, we emphasize that the description of these evolutionary phases (free expansion, Sedov phase expansion without dust destruction, and then a partial dust destruction) should be considered as the idealized expansion of a spherically symmetric SNR into a uniform medium (or one that may have a radial density profile). Solutions usual focus on radius, velocity, temperature, etc. as a function of time. In this case, only a small portion of the spherical blast wave is impeded by the ER,
and the density enhancement of the ER is also very confined in the radial direction. As pointed out by \cite{Jones1998}, the distinct phases can last more or less time, and may occur simultaneously in different regions of the remnant. The situation seems less like idealized SNR expansion, and more like numerical modeling efforts for shocks overtaking and crushing small clouds.
 
The period analyzed in the present work, $\sim$6000 -- $\sim$8000~days after outburst, (i.e. March 16, 2002 to January 18, 2009) doesn’t correspond to a free expansion. Indeed, the interaction with the dense ring had already significantly decelerated the ejecta by this point, with shocks speeds of a few hundred km s$^{-1}$ in the dense knots and on the order of 2000 km s$^{-1}$ in the less dense gas between the knots. See, for example, \cite{Groning2008b}, \cite{Tegkelidis2024}, \cite{Frank2016}, and \cite{Ravi2024}. The period considered here is actually limited by the existence of the longer wavelength {\it Spitzer} data. It turns out that the period of cryogenic Spitzer data that is analyzed here happens to correspond to the rising light curve. It corresponds also to the monotonic increase at all frequencies of the ATCA light curves reported by \cite{Zanardo2017} (which then after slows down), together with decreases in the rate of expansion of the radius and the asymmetry of the emitting region after day $\sim$~7500.

Spitzer data used in the analyzed period include 9 days. However, the available data are not always sufficient to properly adjust all the components of the SED on each of the dates mentioned. It is therefore necessary to make some assumptions. This is particularly the case for the cold dust in the ejecta and in a lesser extent for the silicates dust in the ring. To be able to formulate these necessary hypotheses, we must select some days for which the SED is best defined. Day 7200 was chosen because it is the mean date between AKARI \citep{Seok2008} and ATCA observations (\citealp{Manchester2005, Staveley2007, Zanardo2010, Cendes2018, Zanardo2018}). We interpolate the {\it Spitzer} data for that date. We have also selected day 7400 to make use of the \cite{Laki2011} measurement with APEX and the LABOCA instrument at 345 GHz (870~\micron) which allows us to properly scale the cold dust component at dates prior to its discovery. However, it is clear that this cold dust component is very poorly constrained during the whole period analyzed in the present work, including for the two selected dates (days 7200 and 7400). Since the cold dust component is in the inner ejecta, we assume that it is not changing dramatically (as the ER). We therefore make the assumption that the flux measured by {\it Herschel} between days 8467 and 8564 \citep{Matsuura2011} can be applied at earlier dates.
Finally, we use ALMA and SOFIA data only for fitting the SED of SNR~1987A at day $\sim$10,000 (see Appendix), since those data  were obtained at even later epochs [ALMA: days 9174 -- 9351, \cite{Indebetouw2014}; SOFIA: days 10731 -- 10733, \cite{Matsuura2019}]. 

We note that \cite{Laki2012a} report flux measurements of $19.6\pm5.6$~mJy at 870~\micron\ during the period July - September 2011 (days 8907 - 8969), which are quite close to the ones performed at day 7401 giving a value of 21 $\pm$4~mJy. That is, the flux at that specific wavelength has not varied significantly between day $\sim$~7400 and day $\sim$~9000 (and thus, {\it a fortiori} until day $\sim$~8000). Moreover, Figure~3 of \cite{Laki2012b} clearly shows that although the synchrotron emission varies quite significantly between year 2004 and 2011, the cold dust component is relatively stable during that period. These two observations confirm the validity of our hypothesis that the cold dust component in the ejecta can be considered constant between days 6000 and 8000.

In order to better constrain the parameters of the cold dust component, the 1.3 mm measurement made with the ESO-Swedish Telescope at La Silla (SEST) could have been of interest: the flux seems to have been very slightly increasing from 7.6 $\pm$ 2.5 mJy at days 888 -- 1290 \citep{Biermann1992} to 11 $\pm$ 4 mJy at days 1650 -- 2250 \citep{Bouchet1996}. \cite{Laki2012a} use the value 7.6 $\pm$ 2.5 mJy at day $\sim$ 1290, and their decomposition of the SED shows that this data point comprises both cold dust emission at 33~K at a $\sim$ 2 mJy level and the synchrotron radiation for $\sim$ 5.6 $\pm$ 2.5 mJy. We do not use however any SEST measurements because not only were they obtained at a time very far from the one analyzed in the present work, but also because the flux density that was measured at that time was most probably originating from the ejecta.

It should be kept in mind, however, that the accepted flux covering the 80 – 870~\micron\ range for days 7200 and 7400 (hence defining the cold dust component) was measured with {\it Herschel} at days $\sim$8500. We maintain that this caveat doesn't hamper the overall results of the present work, since the cold dust component is lying in the ejecta.

\section{Results}
\label{sct:results}
\subsection{The Fitting Procedure}

In all our fitting procedures we convolve the $\kappa_i$ coefficients
and the observed fluxes with the overall instrumental transmission through the band pass of each filter of the instruments. The measurement errors ($\sigma_i$) and width of the filters are drawn in the displays of the SEDs.  For illustrating the goodness of our fits, the residuals $\chi^2$ values are indicated:

\begin{equation} \label{eq:chi2}
\chi^2 = \sum\frac{(x_i - m_i)^2}{\sigma_i^2}, 
\end{equation} 
Where $m_i$ and $x_i$ are the fitted and measured data points, respectively.
This is the ``bestnorm" definition by \cite{Markwardt2009} for weighting the residuals so that each deviate has a gaussian sigma of 1.0. We also define a $\Delta$ value, whose definition is very close to the standard reduced $\chi^2$, often noted {$\chi^2_\nu$}, $\nu$ being the degree of freedom. In $\Delta$ we replace $\nu$ by the number of filters, which is also the number of data points.

\begin{equation} 
\Delta = \sqrt\frac {\chi^2}{n}, 
\label{eq:Delta}
\end{equation}
$n$ being the number of data points. 

We think that $\Delta$ is a better indicator of the goodness of our fits, since it is dimensionless and does not depend therefore upon the unit. However, it depends upon the error. The authors of the measurements that we are using give errors which are surely correct for each instrument, but we run into problems when we combine different instruments. Thus, a different choice for the value of the errors will give a different $\chi^2$, but will not reflect the goodness of the fit. Furthermore, fixing the degree of freedom will modify the $\chi^2$ even if everything else is constant.

A fitting procedure \citep[in our case, MPFIT,][]{Markwardt2009} requires a relative error for each measurement. This relative error is not the error on the measurement given by the observations, but a weight applied to each data point. It is larger, in fact, because we deal with different instruments, and the model is not perfect (in other words, the deviation of the data computed by the model to the measured data can be due to the model). For all the fits shown in the present analysis, we found that the best results are obtained with a relative error fixed to 10\% ($\Delta Flux/Flux < 0.10$) for the infrared data (up to $\lambda < 1$~mm) and 1\% ($\Delta Flux/Flux < 0.01$) for the ATCA data ($\lambda > 1$~mm). Interpolations between different dates for {\it Spitzer} data (see Tables~\ref{tab:data7200} and~\ref{tab:data7400}) were obtained after a linear smoothing of the data for each filter.

Finally, SNR~1987A has been monitored with ATCA until day 8014 \citep{Zanardo2010}. We have thus interpolated Table~1 of these authors (after having fit by a degree 3 polynomial to each of the 4 frequencies light curves) in order to obtain the values corresponding to the 9 dates of the {\it Spitzer} data. 

Our fitting procedure yields two parameters characterizing each dust component or the free-free emission: their temperatures and their masses or the ionization coefficient in the case of the free-free, together with a $\chi^2$ value for each fit which allows to select the best one. Both parameters defining a component are clearly correlated. We perform a Monte-Carlo simulation (20,000 draws) to show this correlation in the case of the day 7400 (as an example, but relationships of the same kind apply for each individual fit shown in the present paper).
These correlations between the two parameters that define each component of the SED are shown in Figure~\ref{fig:correlation_wc} for the warm carbon case. This figure illustrates the true uncertainties in the parameters of each component.

\subsection{The Case of Warm Carbon Dust}

The data used for days 7200 and 7400 are given in Tables~\ref{tab:data7200} and \ref{tab:data7400}. The results of the fits to the SED for these days in the case of the warm carbon dust are shown in Figure~\ref{fig:sed_wc}. The parameters resulting from the fits which define the components of the SED at days 7200 and 7400 for this model, are given in Table~\Ref{tab:parameters_wc}. 

Guided by these results for Days 7200 and 7400, we now apply the model to other days of {\it Spitzer} observations, on which the SED was less completely sampled. To overcome the difficulties in characterizing the cold dust  component due to the lack of data for the 9 {\it Spitzer} data points, we will assume that for the whole period 6000 -- 8000 days the value of the temperature and of the mass of this component are the mean of the values derived for days 7200 and 7400: i.e. M$_{(Cold~Dust)}$ = 0.24~$\pm$0.03~M$_\sun$ and T$_{(Cold~Dust)}$ = 23~$\pm$0.6~K for the warm carbon dust model. We note that \cite{Laki2012b} give a good compromise for the temperature at T = 27~K, while \cite{Matsuura2011} report T$\sim$20~K, \cite{Indebetouw2014} found T = 26~$\pm$3~K, and \cite{Cigan2019} derive T = 18 – 23~K. Regarding the mass, all these authors stress that it is strongly dependent on the dust composition and the grain size, the value mostly accepted being in the range 0.2 – 0.4~M$_\sun$. Our adopted values agree then with the ones previously reported.

We then fit the {\it Spitzer} data obtained between days 6000 and 8000 with the warm carbon hypothesis. Our results are shown in Figure~\Ref{fig:individual-9d-wc-fits}.

\cite{Kangas2023} used adaptive optics imaging and integral field spectroscopy from the ESO Very Large Telescope, together with images from the Hubble Space Telescope, to study the near-infrared flux and morphology in order to lay the groundwork for JWST observations of the system. They provide measurements of the continuum flux in the NIR at relevant epochs, and they argue that there are several different components expected to contribute to the continuum at these wavelengths. In particular, they use the same 0.70 spectral index as \cite{Jones2023} and \cite{Larsson2023}, similar to the values found at days 7200 and 7400 for the warm carbon dust model. The main difference between those authors and us is that in our model the radio synchrotron doesn't contribute to the near-IR flux, while it appears to be a very small fraction in theirs. 

\section{The Free-Free Emission}
\label{sct:ff}
\subsection{The Possible Origin of the Free-Free Emission}
\label{sct:origin}

If free-free emission is invoked to explain the excess observed at short IR wavelengths, we must first explore its possible origin. 

\subsubsection{Radio active decay} 
\label{subsct:ffradioactive}

In order to explain a near-IR excess observed at day 1316 together with a positive detection achieved at 1.3~mm 
\citep{Bouchet1991,Bouchet1996, Wooden1989}, \cite{Laki2011} and \cite{Wesson2015}
suggested that free-free (or Bremsstrahlung) radiation was probably contributing to the energy spectrum. In those previous epochs, the heating source was the radioactive decay of short-lived radioisotopes including $^{56}$Co and $^{57}$Co.
The electrons giving rise to the free-free emission observed at that
time were located mainly in the hydrogen envelope, and the observed free-free emission was produced by the radioactive decays of Co isotopes in the ejecta. This is obviously not the case for the current work since we know that the near-IR excess originates from the ER (see Figure~\ref{fig:imageshort}). Moreover, after day $\sim$1800 the only isotope that still produces radioactive heating, hence $\gamma$-rays, is $^{44}$Ti with a decay timescale of 87 yrs \citep{Fransson2002}. Although one could always think that due to the interaction of the ejecta with the ring, a little of $^{44}$Ti could be there, we must admit that to make this assumption would be at least hazardous. 

\subsubsection{X-rays} 
\label{subsct:ffXrays}

Free electrons either locally produced in the ER via ionization by shocks or Cosmic Rays (CRs), or present in the ejected metal-rich core which penetrate the ER, could contribute to the free-free continuum flux and electron scattering optical depth, provided there is a heating source. If we exclude a radioactive decay of $^{44}$Ti in the ER, that source could be provided from the X-rays emitted by the interaction between the supernova ejecta and the circumstellar material, and from the ambient interstellar UV field. Clearly, as we will show later in \S\Ref{sct:results_ff}, our fits yield such a low temperature for the ionized gas that they most probably would exclude such a possibility.

\subsubsection{Cosmic Rays} 
\label{subsct:ffCRs}

Alternatively, non-thermal low energy cosmic rays are known to be the most efficient ionizers and could well be also at play \citep{Tu2024}. Indeed, we have seen in \S\ref{subsct:ffradioactive} that CRs generated by the radioactive decays that powered the supernova at early times have largely dissipated by now, and ionization by CRs from the general LMC ISM would be constant, and would have existed before the blast wave hits the ER.

\cite{Tu2024} study the chemistry jointly induced by shocks and CRs in the remnant W28, and use physical parameters obtained from observations: a preshock density $n_{\rm H} \sim~2 \times 10^5~{\rm cm}^{-3}$, 
a CR ionization rate $\zeta = 2.5 \times 10^{15}$ s$^{-1}$, and a shock velocity $V_{\rm s} = 15-20$ km~s$^{-1}$. We note that there is no inconsistency between these parameter values and the ones that could be attributed to the remnant of SN~1987A. 
\cite{Tu2024} note also that low-energy CR protons will not contribute to $\gamma$-ray
emission, but they are the dominant ionization sources in dense regions which are difficult to penetrate with ionizing UV emission. CR protons ionize molecular
hydrogen gas mainly through:

\begin{equation}
{\rm H}_2 \longrightarrow {\rm H}_2^+ + e^- 
\end{equation}
followed by
\begin{equation}
{\rm H}_2^+ + {\rm H}_2 \longrightarrow {\rm H}_3^+ + {\rm H}.
\end{equation}

However, there is no evidence for molecular hydrogen in the ER as reported by \cite{Fransson2016}, who show that there is H$_2$ only in the ejecta. We note that \cite{Matsuura2024} report the detection of faint H$_2$ crescents. Not only do the authors claim that these crescents are located between the ejecta and the ER, but also the observations were made at days 12974 and 12975, therefore very distant in time from observations discussed here,  while the authors argue that they are new features. The discussion about the interaction between H$_2$  and CR is therefore not relevant in the context of free-free emission from the ring.

\subsubsection{Collisional Heating} 
\label{subsct:ffcollisions}

The dust in the ER is thermally coupled to the gas, and more likely collisionally heated by the ambient gas, rather than radiatively heated by UVO photons \citep{Dwek2015}. Since the free-free emission originates from the ring, the matter is hydrogen dominated. Hydrogen cannot be ionized at the low temperature found from our fitting procedure (see later in \S\Ref{sct:ff}), and any ionization would then be fossil ionization. However, in a supernova environment, shocks can generate and accelerate particles, and collisional ionization in shocks could be at play. As formulated by \cite{Tipler2008} in their Equation (8-14) for equipartition:
\begin{equation} \label{eq:ekT}
\langle E\rangle = \frac{3}{2} kT.
\end{equation}

\cite{Spitzer1978} derives a relation between temperature behind the shock and the shock speed 
[Eq.~12-24]. 
\begin{equation} \label{eq:spitzer}
  T =   \frac{3 \mu}{16 k} V^2_{\rm s} = \frac{0.061 \mu}{k} \frac{E}{\rho r^3_{\rm s}}.
\end{equation}

Applying the mean mass per particle of $0.61~m_{\rm H}$ and assuming the equipartition discussed by the author, the left side of Eq.~(\ref{eq:spitzer}) can be revised to:

\begin{equation} \label{eq:vT}
V_{\rm s}~[{\rm km~s}^{-1}] = \sqrt{T[{\rm K}]/13.8}.
\end{equation}

Note that these computations are just a rough guideline since real shocks in supernovae and supernova remnants are often messy, dynamically changing (not in equilibrium), and moving through a medium that may have enhanced metal abundances.  Still, Eq. (\ref{eq:vT}) provides a convenient relation between the shock speed and post-shock temperature. 

 \cite{Shull1979} argue that the shock speed could be much faster if there has been significant cooling behind the shock. Indeed, Figure~3 and Table~3 of these authors show that, given some time, the high temperature behind faster shocks can drop down to a few $10^3$~K level. Their Table~3 indicates that such cooling takes hundreds of years. However, we can argue that their models are all likely propagating into a lower density than the one prevailing in the ER. Then, assuming a higher pre-shock density in SN~1987A, we would expect the overall evolution to run faster. Indeed, \cite{Raymond1979} present models with high pre-shock densities where the cooling time down to $10^3$~K is of the order of decades (see models Z, AA, FF and JJ in his Table~1). 

Recent studies based on the analysis of the emission lines in the ER show that the unshocked ring has a density of 1.5 - 5~$\times 10^3$ cm$^{-3}$ {\bf [up to 1 -- 4~$\times 10^4$ in the densest clumps \citep{Groningsson2008a}]} and a temperature $T = 6.5\times 10^3$ -- $2.4\times 10^4$~K (\citealp{Groning2008b,Tegkelidis2024}). 
These authors report a large range of ionization stages with densities of $4\times10^6$ -- $10^7$~cm$^{-3}$ and a temperature $T \sim 1.5\times 10^6$~K behind the shock, and velocities $\sim$300~kms$^{-1}$ (in agreement with Eq.~\ref{eq:vT}). To account for the observed range of speeds, they invoke a significant cooling behind the shock. \cite{Groning2008b} report that optical emission lines of the low ionization species originate from the photoionization zone behind the radiative shock with temperatures of $\sim$10$^4$~K. They argue that immediately behind a $\sim$300~kms$^{-1}$ shock the temperature rises to $T \sim 1.5\times 10^6$~K (this is where part of the soft X-rays and the coronal lines originate). Then, the gas undergoes a thermal instability and cools to $\sim$10$^4$~K in a very thin region. Finally, the gas temperature is stabilized by photoionization heating and a partially ionized zone takes over, while the temperature decreases drastically to very low values. These authors argue moreover that most of the emission comes from a region of thickness of only $\sim 10^{12}$~cm, and that low ionization lines all come from this zone although the emission extends to lower temperatures compared to the \ion{H}{1} and \ion{He}{1} lines. In the [\ion{N}{1}] region they observed temperatures as low as 4 -- $5\times 10^3$~K.

\cite{Tegkelidis2024} propose a similar scenario: the higher densities are situated closer to the center of the ER, and the postshocked gas collapses thermally and cools down to $\sim$10$^4$~K (at these temperatures the cooling is balanced by photoionization heating). They observe that  most hot spots, spatially resolved, exhibit elongation that suggest a short cooling time. They note that since it is significantly shorter than the characteristic time for the blast wave to traverse the obstacle, the emission arises only from a thin layer behind the shock.

We will see in \S~\ref{sct:results_ff} that our fits yield a mean temperature of the ionized gas $T \sim 1400$~K. Such a low temperature implies a very high density contrast to slow the forward shock that much in the ER, and/or strong post-shock cooling as discussed above. Although this value is still slightly lower than the lowest temperature reported in the very thin layer of the [\ion{N}{1}] zone \citep{Groning2008b,Tegkelidis2024}, we assume that the mechanism  proposed above could give credit to this temperature.
  
Although we could justify such a low free-free temperature, barely below the limit set by \cite{Groningsson2008a} and \cite{Tegkelidis2024}, we must still ensure that our model does not conflict with the observations. The tightest constraint on the longer wavelength free-free emission from the ER is probably $S(870~\micron) < 21$~mJy at day 7401 \citep{Laki2011}. We will see that our free-free spectra derived for that day and all the others during the period studied are well below that value (see Figures~\Ref{fig:sed_ffcut13}, \Ref{fig:sed_ffcut40}, \ref{fig:sed_9d_ffcut13} and  \ref{fig:sed_9d_ffcut40}). 
It is then not possible to find out at what wavelength the emission becomes (if it does) optically thick.

\subsection{Self-absorbed free-free emission?}
\label{sct:selfabsorbed}

The ionization fraction is defined as:
\begin{equation} \label{eq:xi}
   (1 - \xi) = {n_e}/{n_{\rm i}}
\end{equation}
where $\xi$ is the fraction of neutrals according to the definition given by \cite{Osterbrock1989}, $n_e$ is the number density of the free electrons and $n_{\rm i}$ is the number density of ions. 

The hypothesis that the ionized gas is optically thin in the whole spectral range analyzed here  doesn't seem realistic. We can thus estimate the mass of the ionized gas by considering the wavelength at which the free-free emission becomes optically thick. \cite{Draine2010} gives the value of the free-free optical depth $\tau_\nu$. 
\begin{equation} \label{eq:taunu}
\tau_\nu \approx 1.091~T_4^{-1.323}  \nu_9^{-2.118} \frac {n_{\rm i}}{n_{\rm p}} \frac {EM} {10^{25}\ {\rm cm}^{-5}}
\end{equation}
where $EM = \int{n_e n_{\rm p} ds}$ is the emission measure, $\nu_9$ the frequency in GHz, and $n_{\rm p}$ the number density of protons.
By definition, $\tau_\nu = 1$ at the cut-off frequency $\nu_{9,\rm cut}$ of the free-free emission. Then, Eq. (\Ref{eq:taunu}) becomes:
\begin{equation} \label{eq:taunu1}
1 \approx 1.091~T_4^{-1.323}  \nu_{9,\rm cut}^{-2.118}  \frac {n_{\rm i}}{n_{\rm p}}  \frac {EM} {10^{25}\ {\rm cm}^{-5}}.
\end{equation}
We assume ${n_{\rm i}}/{n_{\rm p}} = 1$. 

Assuming that the cut-off frequency is 13~GHz (2.3 $\times$ 10$^4$ \micron)
\begin{equation}
EM = 1.091^{-1} \times T_4^{1.323} \times \nu_9^{2.118} ~10^{25}~{\rm cm}^{-5}
\end{equation}
\begin{equation}
EM = 0.9166 \times 0.14^{1.323} \times 13^{2.118} \simeq 1.6~10^{26}~{\rm cm}^{-5}
\end{equation}
\cite{Groning2008b} and \cite{Tegkelidis2024} 
showed that behind the shock, the averaged density is:
\begin{equation}
n_e \simeq7~10^6~{\rm cm}^{-3}
\end{equation}
The corresponding path length is then:
\begin{equation}
\label{eq:length}
l = EM/n_e^2 \simeq 3.2~10^{12}~{\rm cm}.
\end{equation}
This value is slightly larger than the estimated thickness of only $\sim$10$^{12}$~cm of the zone from where most of the emission comes \citep{Groning2008b}. On the other hand, \cite{Tegkelidis2024} estimate the size of the hot spots to be $\sim$4 -- 6 $\times$ 10$^{16}$~cm, which gives credit to our calculated path length. 

Taking the ER radius to be $R = 6.2 \times 10^{17}$~cm (\citealp{Bouchet2006,Tegkelidis2024}), the volume would then be:
\begin{equation} \label{eq:volume}
V = 2{\pi}R{\pi}\left(\frac{l}{2}\right)^2 \simeq 3.1 \times 10^{43}~{\rm cm}^3
\end{equation}

which yields a mass:
\begin{equation} \label{eq:mass}
M = {\mu}m_{\rm H}n_e V \simeq 10^{-50}\ V ~{M_\sun}
\end{equation}
where $\mu$ = 1.28 is the mean atomic weight of the gas.
We obtain:
\begin{equation}
M = 3 \times 10^{-7}~M_\sun 
\end{equation}
Retrieving from \cite{Osterbrock1989} a recombination coefficient of:
\begin{equation}
\alpha = 1 \times 10^{-12}~{\rm cm}^2~{\rm s}^{-1}
\end{equation}
gives the recombination timescale:
\begin{equation} \label{eq:trec}
t_{\rm rec} = 1/{(\alpha} n_e) = 1.4 \times 10^5s \approx 39~{\rm h}. 
\end{equation}
Indeed, the duration $t_{\rm d}$ of the emission is $t_{\rm d} \gtrsim $~2000 days, or $4.8 \times 10^4$ hours (assuming that the interaction between the shock wave and the inner equatorial ring started at day $\sim$5500).
Therefore, the total mass will be: 
\begin{equation} \label{eq:masstotal}
M_{\rm Total} = M \times t_{\rm d}/t_{\rm rec} = \frac{4.8 \times 10^4}{39}\ M
\end{equation}
which yields: 
\begin{equation}
M_{\rm Total} = 3.6 \times 10^{-4}~M_\sun
\end{equation}
\cite{Tegkelidis2024} calculate the mass of the emitting substructures of the hot spots, based on the light curves, and find a value of :
\begin{equation}  
M \sim 0.24 (n_e/10^{6}~{\rm cm}^{-3})^{-1}\  10^{-2}M_\sun = 3.4 \times 10^{-4}M_\sun
\end{equation}
(with $n_e = 7 \times 10^{6}~{\rm cm}^{-3}$). Both values agree thus quite well.

On the other hand, these authors report that using the filling factor and the estimated sizes of the hot spots that they compute, they derive a mass of 
$0.8 (n_e/10^{6}~{\rm cm}^{-3})^{-1}\ 10^{-2}M_\sun$ = $5.6 \times 10^{-2}M_\sun$ instead. If we set the free-free cut-off frequency at 40~GHz ($\sim$7500~\micron), the same calculations as above yield:
\begin{equation}
EM = 1.7 \times 10^{27}~{\rm cm}^{-5}
\end{equation}
\begin{equation}
l = 3.5 \times 10^{13}~{\rm cm} 
\end{equation}
\begin{equation}
 M = 3.7 \times 10^{-5}~M_\sun
\end{equation}
and finally:
\begin{equation}
M_{\rm Total} = 4.4 \times 10^{-2}~M_\sun
\end{equation}
Which agrees quite well with \cite{Tegkelidis2024}'s result in that case. 
We note that these authors point out that there are additional spots not included in their calculation. 

Note also that \cite{Mattila2010} find a total mass of the ionized gas of $\sim5.8 \times 10^{-2}~M_\sun$ within the inner ring,  corresponding to the higher density structure of the ER, while \cite{Lundqvist1996} estimate the mass of the ER ionized by the initial flash, to be $\approx 4.5 \times 10^{-2}~M_\sun$. All these results are thus consistent.

To estimate the dust-to-gas mass ratio, we use the mean silicates dust mass (around day 7400) of $\approx~6\times 10^{-6} M_\sun$ (see Figure~\ref{fig:evolution_silicates}), which leads to:
\begin{equation}
M_{\rm Dust}/M_{\rm Gas} \sim 1.2 \times 10^{-3}
\end{equation}
This is lower than the \cite{Dwek1987}'s estimate of $\sim 7.5 \times 10^{-3}$ in the Cas~A and Cygnus Loop supernovae remnants and \cite{Clark2023}'s value of $\sim 4 \times 10^{-3}$ for the LMC. That may indicate that dust is slightly depleted in the ER.  

We emphasize that these are mass estimates of the denser regions within each spot and not the total mass of the hot spots or the ER. For instance, \cite{Tegkelidis2024} anticipate an additional mass contribution ($\sim1.2$ -- $3.6\times 10^{-2}$~M$_\sun$) from the less dense components of the hot spots. 

Our calculations show that the final mass is not only defined by the cut-off frequency, but also that the relationship between both is strictly one-to-one. It is clear that we can find a cut-off frequency yielding any observed total mass, provided the computed path length remains smaller than the sizes of the hot spots as reported by \cite{Tegkelidis2024}. That means that our model can be adjusted to be coherent with the observations. We still argue, then, that dust collisions could be a source of free-free emitting gas heating, even though the temperature that is derived in that case implies a very slow shock speed. Indeed, the majority of the dust might instead reside in hotter (X-ray emitting) post-shock regions at much lower densities. On the other hand, we note that \cite{Hollenbach1989} have built models using very dense medium (up to $n_e = 10^6~{\rm cm}^3$) for shock speeds as low as 30~km~s$^{-1}$, and hydrogen column densities in the $10^{15} - 10^{22}~{\rm cm}^{-2}$ range, therefore in the same regime as the one proposed here for SNR~1987A.  These models are intended for supernova remnants in interstellar molecular clouds. Still, all of them present a fast cooling down to temperatures less than hundreds of degrees. 

These encouraging results prompt us to  make the assumption that our free-free model could be valid under the condition that the emission is self-absorbed (in that regime, one probes the Rayleigh-Jeans part of the Planck curve corresponding to the electron temperature). We stress, however, that since we can't observationally verify the cutoff because it is swamped by the synchrotron emission, this value of $\nu_{9,\rm cut}$ is not truly arbitrary but it is set by the total mass of the free-free ionized gas.

Although we have shown in this section that, under certain conditions, 
free-free emission from relatively cool ($T\sim1400 \pm 50$ K) gas could account 
for the observed emission, it may seem surprising 
that this component is not secondary to free-free emission from hotter gas 
that dominates the optical/IR line emission. A possible resolution of this may
be found in noting the temperature and density dependence of free-free emission,
as given in Eq. (\ref{eq:fnu-ff-general}). 
First, if rough pressure equilibrium is maintained as the gas cools, then the densities
$n_e$ and $n_i$ are inversely proportional to the temperature. Next, the 
exponential term in Eq. (\ref{eq:fnu-ff-general}) converges to 1.0 at higher $T_{\rm eff}$,
although for the range 1400 K $ <T_{\rm ff} < $ 20000 K, this term can be 
approximated as a linear trend to within a factor of 2. Finally, we note 
$1< g < 2$ for these wavelengths and temperatures. Thus, the overall free-free
emissivity scales roughly as $j_{\rm ff} \propto T_{\rm ff}^{-1.5}$. This means the 
free-free emissivity rises as the temperature drops (mainly from the increasing 
density of the cooling gas), at least until the gas cools further (and the 
exponential term steepens) or the gas recombines.  
This strong inverse dependence on $T_{\rm ff}$ suggests that free-free emission
from cooler gas might be expected to be more prominent than that of the hotter 
gas at these wavelengths.
However, the proper assessment of the relative contributions requires a 
detailed evaluation of the mass and density of the cooling gas at temperature $T_{\rm ff}$ as
a function of time: $M(T_{\rm ff},t)$, $n(T_{\rm ff},t)$, which is far
beyond the scope of this paper.

To summarise, we argue that a  free-free model could reproduce the observations if the emission is self-absorbed, either at 13~GHz or at 40~GHz. In what follows, we will analyze both cases. The data used for the free-free models are the same as the ones used for the warm carbon model (i.e. Tables~\Ref{tab:data7200} and \Ref{tab:data7400}).

\subsection{Results for the self-absorbed model}
\label{sct:results_ff}
The resulting fits of the SED for days 7200 and 7400 for the self-absorbed free-free at 13~GHz model are shown in Figures~\ref{fig:sed_ffcut13}, 
%\& \ref{fig:sed_ffcut40}, 
and the parameters which characterize the components of the SED for these days 7200 with this value of the cut-off frequency, are given in Table~\Ref{tab:parameters_ffcut13}. We then assume that for the whole period 6000 -- 8000 days the value of the temperature and of the mass of the cold dust component are the mean of the values derived for days 7200 and 7400: i.e. $M_{\rm Cold~Dust,13}$ = 0.17~$\pm$0.03~$M_\sun$ and $T_{\rm Cold~Dust,13}$ = 24~$\pm$0.1~K for the cut-off at 13~GHz.

The correlations between the parameters that characterize the free-free emission self-absorbed at 13~GHz are shown in Figure~\ref{fig:correlation_ff13} (they would be very similar for the 40~GHz cut-off).

Figure~8 of \cite{Kangas2023} shows that the near-IR evolution of the flux density is similar to that of the mid-IR. In that figure, the authors show that the {\it Spitzer} continuum at 3.6 and 4.5~\micron\ follows the same trend as their near-IR broad band data. Table~5 of these authors give flux density for day 7227 of  $\sim$0.85~$\pm~0.04$~mJy in the J-band (1.22~\micron), $\sim$0.35~$\pm~0.01$~mJy in the H band (1.65~\micron) and $\sim$0.57~$\pm~0.02$~mJy in the K$_s$-band (2.15~\micron). An important check of the free-free model is therefore that the predicted NIR flux is below these fluxes. We find for the same day and with our free-free model, flux densities of 0.02, 0.07 and 0.28~mJy, respectively for the three filters. That means that there is no inconsistency between these set of data. Furthermore, our free-free model must be compatible with a later peak and decrease in the near-IR emission as well. Since the {\it Spitzer} data used for building our model were obtained before the peak emission, it is most probable that the trends shown in the present work in mass, temperature, spectral index, etc. were to change after day $\sim$9000.

We then compute the fits for the 9 {\it Spitzer} days. Results of this series of fits are shown in Figure~\Ref{fig:sed_9d_ffcut13}.

We then perform the same analysis with a cut-off frequency at 40~GHz. Figure~\Ref{fig:sed_ffcut40} show our fits for the days 7200 and 7400 with a free-free emission self-absorbed at that frequency and Table~\Ref{tab:parameters_ffcut40} gives the parameters that define each component of the SED.
Our first conclusion, then, is that the three models fit reasonably (and equally) well the data, as can be seen in Figures~\Ref{fig:sed_wc}, \ref{fig:sed_ffcut13} and \ref{fig:sed_ffcut40}  .

In order to constrain the Cold Dust, we fix the mean values obtained for these days: $M_{\rm Cold~Dust,40}$ = 0.17~$\pm$0.03~M$_\sun$ and $T_{\rm Cold~Dust,40}$ = 23~$\pm$0.1~K for the 40~GHz cut-off. We then fit the {\it Spitzer} data obtained between days 6000 and 8000 under the 40~GHz self-absorbed free-free model. Our results are shown in Figure~\Ref{fig:sed_9d_ffcut40}.

\section{Time evolution  of the various components of the SED}
\label{sct:evolution}
Figures~\Ref{fig:individual-9d-wc-fits}, \ref{fig:sed_9d_ffcut13} and \ref{fig:sed_9d_ffcut40} give us the parameters of the various components of the SED for each date when SNR~1987A has been observed by {\it Spitzer} and thus their evolutions (during the analyzed period).

\subsection{The Warm Carbon Dust}

Figure~\ref{fig:evolution_warmcarbon} shows the evolution of the warm carbon dust parameters. Our fit indicates that the mass of the warm carbon dust has been increasing linearly during the period from $\sim~1 \times 10^{-8}$ to $\sim~5 \times 10^{-8}$ M$_\sun$ (the slope being $2.1 \times 10^{-11}$~M$_\sun$ day$^{-1}$). The temperature seems to have remained constant at about 474~K, although it cannot be excluded that, without the uncertainties, it has slightly decreased (from $\sim$500~K to $\sim$450~K, assuming that the first data  point at day 6136 is an outlier). 

We note also that during the analyzed period the increase of the mass of the warm carbon dust is tied to that of the silicates dust by the relationship (see Figure~\ref{fig:masses_wc_Si}):
\begin{equation}
M_{\rm Warm~Carbon} = 5.6 \times 10^{-3} M_{\rm Silicates}
\end{equation}

\subsection{The Free-Free Emission}
\label{sct:free-free}

Figure~\Ref{fig:evolution_ff} shows the evolution of the ionization coefficient $(1 - \xi)$ and of the temperature: 
the temperature of the ionized gas seems to have been constant (at $\sim$1370~K), although a slight decrease cannot be excluded, while the ionization coefficient was clearly increasing linearly (from $\sim$5\% to $\sim$10\%; the slope being $2.4 ~\times 10^{-3}$ percent day$^{-1}$). That makes sense since a lower temperature ties in with a higher ionization fraction, such that the [3 - 6]~\micron\ emission is not changed (to the first order).  

\subsection{The Silicates Dust}

Figure~\ref{fig:evolution_silicates} shows the time evolution of the silicates dust parameters under the three hypothesis (warm carbon dust, free-free emission self-absorbed at 13~GHz and free-free emission self-absorbed at 40~GHz). Note that in the wavelength domain that is common to the free-free emission and the silicates dust, the cut-off frequencies (13 and 40 GHz) do not affect either. Therefore the fits of both components, hence the value of the parameters, are the same for the two cut-off frequencies.  The parameters defining the silicates dust follow the very same trend in the three models: the silicates dust mass increases linearly with time while the temperature remains constant.
We find:
\begin{equation}
Slope({\rm Warm~Carbon}) = 3.6 \times 10^{-9} M_\sun\ {\rm day}^{-1}
\end{equation}
\begin{equation}
Slope({\rm free-free})_{\rm 13GHz} = 2.5 \times 10^{-9} M_\sun\ {\rm day}^{-1}
\end{equation}
\begin{equation}
Slope({\rm free-free})_{\rm 40GHz} = 2.6 \times 10^{-9} M_\sun\ {\rm day}^{-1}
\end{equation}
Further, in the free-free hypothesis the mass is slightly lower than in the warm carbon case, and the temperatures are roughly similar at 197 -- 200~K.

\subsection{The Radio Synchrotron Spectral Index}

\cite{Staveley2007} and \cite{Zanardo2010} observed a gradual flattening of the synchrotron spectrum from a spectral index $\alpha = -1.0 ~(S_\nu \propto \nu^\alpha$)
at day 3000, to {$\alpha = -0.75$} at day 7300. Our results, shown in Figure~\Ref{fig:evolution_synchrotron_alpha}  instead implies that in the 13~GHz self-absorbed free-free model the spectral index increases (becomes harder) 
from -0.85 at day 6500 to -0.75 at day 7300, then flattens around this value until day $\sim$7700, and finally seems to decrease slightly until day $\sim$8000. Thus, the index in the case 13~GHz behaves roughly as reported by \cite{Staveley2007} and \cite{Zanardo2010} at least until day $\sim$7700. 
On the other hand, it is slightly higher in the warm carbon dust model than in the 40~GHz  free-free model, but in both cases it is notably steeper than in the 13~GHz model {$\alpha$} = [$\sim$-0.78, $\sim$-0.68]. It is noteworthy that it follows the same evolution trend in the the three models. 

The difference between the models is a natural consequence of the 13~GHz self-absorbed free-free contributing at long wavelengths (whereas the 40~GHz and the warm carbon models do not). Clearly, extra emission from free-free means reduces (steepens) the synchrotron to compensate. 

\cite{Staveley2007} and \cite{Zanardo2010} argue also that the spectral index 
could reach the {\it ``canonical''} value for certain shell-type SNR of $\alpha = -0.53$ by 2018 (day $\sim11500$) or so. Figure~\ref{fig:evolution_synchrotron_alpha} indicates clearly that this value was not reached yet in any of the 3 models. Although it is difficult to make long-term forecasts, it seems doubtful that this {\it ``canonical''} value could be reached on the predicted day, if we judge by the values of $\alpha$ obtained with our attempt to adjust the SED at day 10,000 ($\alpha = -0.80$ for the free-free model and $\alpha = -0.76$ for the warm carbon model; see Appendix). 
We note, however, that only 50 out of the 294 remnants listed in \cite{Green2019}'s catalog of Galactic SNRs have a spectral index $\alpha$ = (-0.5, -0.6), and a simple power law is not always adequate to describe the spectra \citep{Dubner2015}. \cite{Urosevic2014} and \cite{Loru2019} note that the spectral indices are steepening in SNRs where the particle acceleration mechanisms are not efficient, which could be the case for the 13~GHz free-free model. However, these authors argue, on the basis of the Diffuse Shock Mechanism (DSA) theory, that the {\it ``canonical''} value for young SNRs is around $\alpha$ = -0.7, which would favour the warm carbon dust and the 40~GHz free-free hypothesis.

To our knowledge, observations or theoretical studies of variations with time over relatively short time scales ($\sim$yr) of the radio synchrotron spectral index in SNRs are very sparse, although their spatial variations have been extensively studied - see for instance \citealp{Furst1988,Cassam2007,Williams2018,Zanardo2013} and references therein). \cite{Zanardo2013} in particular present a map of the spectral index values for SNR~1987A. They show that these values are mainly in the range [-0.3, -0.8], but can be as low as -1 and even lower in the center and the South-East region of the remnant. We stress here that the values derived in the present work are relative to an {\it averaged} synchrotron emission which prevents a study of the spatial variations, and cannot therefore be compared to those obtained by these authors.

As far as we know, only \cite{Kimani2016} and \cite{Nayana2022} study the variations with time of the spectral indices. The former argue that it shows no signs of evolution and remains steep at $\alpha$ = -1  for SN~2008iz in M82 (unlike that of SN 1993J which started flattening at day $\approx$ 970), and the latter show that the SN~2016gkg index displays some variation around $\alpha$ = -1, explained by the fact that the environment is often non-uniform, with variations in density, magnetic fields, and other properties. Figure~\Ref{fig:evolution_synchrotron_alpha} doesn't show any of these variations for the
three models. It displays instead a gradual clear hardening in the three models, and a possible flattening at day $\approx$7300, similar to (although much later than) SN~1993J.   

In conclusion, the analysis of the spectral index does not allow to separate the three hypotheses. 

\section{The Light Curves}
\label{sct:lc_theoretical}

We use the evolution of the parameters of the SED components to compute the overall SED evolution $F(\lambda, t)$ over the period of days 6000 -- 8000, hence to build the theoretical light curves for each of the {\it Spitzer} wavelength. We then compare our results with the measured data. 

Figure~\Ref{fig:lc_models} illustrates the fact that a free-free model fits the light curves as well as, or even better than, the warm carbon component. 
We have associated $\Delta$ values (Eq. \ref{eq:Delta}) to our fits. We believe that this parameter represents better their quality than the canonical $\chi^2$'s.  
Intuitively $\Delta$ represents the rms error ($\sigma$) of the fit in units normalized to the $\sigma$ of the data. $\Delta >> 1$ reflects a poor model fit, while $\Delta < 1$ indicates an overfitting.

As we have $j = 5$ {\it Spitzer} light curves we compute:
\begin{equation}
\Delta^2 = \left(\sum_{j=1}^{5} \Delta_j^2\right)/5. 
\end{equation}
Then: 
\begin{equation}
\chi^2_{\rm Warm~Carbon} = 117.
\end{equation}
\begin{equation}
\chi^2_{\rm free-free_{13GHz}} = 61 
\end{equation}
\begin{equation}
\chi^2_{\rm free-free_{40GHz}} = 85
\end{equation}
and:
\begin{equation}
\Delta_{\rm Warm~Carbon} = 3. 
\end{equation}
\begin{equation}
\Delta_{\rm free-free_{13GHz}} = 1.8 
\end{equation}
\begin{equation}
\Delta_{\rm free-free_{40GHz}} = 1.9
\end{equation}
which shows that a fine analysis of the light curves clearly favors the self-absorbed free-free model, in particular the one with a cut-off frequency of 13~GHz. 

Note that one may assume that there are unaccounted systematic errors present that are not reflected in the formal errors on the data. 

\section{Conclusion}
\label{sct:conclusion}
We showed that free-free emission could perfectly reproduce the observed excess emission at the shortest near-IR wavelengths (3 -- 5~\micron) of the SNR~1987A spectrum. The mechanism for heating the gas that produces this emission is not clear. In a particularly precise case of self-absorbed emission, the products of collisional heating processes could be compatible with the physical parameters of the ER, as revealed by recent fine analyses of emission lines. The cut-off frequency is set by the total mass of the emitting region. It remains that the low temperature of the ionized gas is at the limit of the required condition.

Nonetheless, given the fact that free-free emission reproduces the observations so well, we cannot rule out that an unknown general mechanism would be able to explain its origin. We note, in particular, that \cite{Sun2025} suggest that the X-ray emission may be subject to absorption from the hot ionized foreground gas. We have not explored the possibility of such an external contribution, which goes beyond the present work. 

We believe that the possibility that free-free radiation could explain the excess emission at short infrared wavelengths should be explored for the periods (days 7500 - 9000, and days 9000 - 12000). Unfortunately, the lack of data in the later time period to estimate the mass and temperature of the silicates dust (e.g. from 8 to 30~\micron) does not allow this approach, unless we are able to correctly estimate the evolution of this dust component for these dates.  

Indeed, our results would need also to be discussed in light of the {\it James Webb Space Telescope} (JWST) observations. SN 1987A has been first observed at day 12927 \citep{Jones2023, Larsson2023, Bouchet2024} and \cite{Larsson2023} report that the NIRSpec data show the same excess emission at the shortest IR wavelengths as the one discovered by {\it Spitzer} that is discussed here. Other observations of SNR~1987A followed and will be continued. However, the epoch of those observations corresponds to the end of the phase 3 in the hydrodynamical model of \cite{Woltjer1972} and \cite{Reynolds2008}, when the shell is cooling and dust destruction begins, and the beginning of phase 4, when the expansion velocity of the remnant becomes comparable to the random motions of the interstellar gas (at that point the shell begins to lose its identity and merge into the interstellar medium, see~\S\Ref{sct:lightcurves}).

The physical processes involved at the time of the JWST observations, past, present and future, are therefore quite different from those taking place during the period considered in this work, when the blast wave has entered the main body of the ER while the ejecta continue their free expansion. We think then that there is no point in extrapolating our proposed free-free model for checking its consistency with the JWST results. 

Moreover, it is noteworthy to recall that {\it Spitzer} observations were essentially unresolved and the data refer always to the ``total''  flux density (the only concern was the contamination of Stars 2 and 3, while these stars are excluded directly in JWST data). Furthermore, most of the JWST observations of SNR~1987A concern spatial sub-selections to look at specific parts of the ER or ejecta. This makes comparisons between {\it Spitzer} and JWST data often difficult and sometimes hazardous. 

Finally, we wish the work presented here to be a plea for analyzing data obtained with the JWST at much more recent times with the same approach as the one presented here. Under the condition, however, that ALMA and ATCA continue their observations at agreed time, in order to evaluate each individual component of the SED at the same epoch.

\section*{Acknowledgments}

PB would like to emphasize that the first idea that inspired this work arose from a discussion with John Danziger, who left us unfortunately last July 21, 2025 (RIP). The authors therefore wish to dedicate this article to his memory.  
PB is also grateful to Eli Dwek for enriching discussions and to an anonymous referee, extremely conscientious and meticulous, who has considerably improved our first manuscript. 
This work is based on observations made with the NASA {\it Spitzer} Space Telescope.
The data were obtained from the Mikulski Archive for Space Telescopes at the Space Telescope Science Institute, which is operated by the Association of Universities for Research in Astronomy, Inc., under NASA contract NAS 5-03127 for JWST. This publication makes use also of data products from the Wide-field Infrared Survey Explorer, which is a joint project of the University of California, Los Angeles, and the Jet Propulsion Laboratory/California Institute of Technology, funded by the National Aeronautics and Space Administration. Work by R.G.A. was supported by NASA under award number 80GSFC24M0006.

\facilities{
The IRAC and MIPS\footnote{\url{https://irsa.ipac.caltech.edu/data/SPITZER/docs/}} instruments onboard the {\it Spitzer} Space Telescope [NASA]; the AKARI (ASTRO-F) [Japan Aerospace Exploration Agency] satellite\footnote{\url{https://www.isas.jaxa.jp/en/missions/spacecraft/past/akari}}; the Australia Telescope Compact Array\footnote{\url{https://www.narrabri.atnf.csiro.au/}} (ATCA) operated by the Commonwealth Scientific and Industrial Research Organisation [CSIRO]; the {\em Herschel} Infrared Space Telescope\footnote{\url{https://www.herschel.caltech.edu/}}[European Space Agency]; the Atacama Pathfinder Experiment (APEX)\footnote{\url{http://www.apex-telescope.org/ns/}} [European Southern Observatory (ESO)] and the Swedish-ESO Submillimetre Telescope\footnote{\url{https://www.eso.org/public/denmark/teles-instr/lasilla/swedish/}} (SEST) operated by the Swedish National Facility for Radio Astronomy}. 

\software{
GDL\footnote{\url{https://github.com/gnudatalanguage/gdl}} \citep{Park2022,coulais2025gdl11smartgreen}, MPFIT\footnote{\url{https://pages.physics.wisc.edu/~craigm/idl/fitting.html}}$^{,}$\footnote
{\url{https://packages.debian.org/sid/gdl-mpfit}} 
\citep{Markwardt2009}, IDL Astro Lib\footnote{\url{https://packages.debian.org/sid/gdl-astrolib}} \citep{Landsman1995} }

\bibliography{Biblio/add,Biblio/biblio}
\bibliographystyle{aasjournal}

\section{Appendix}

\begin{figure*}
\begin{center}
\includegraphics[width=8.5cm]{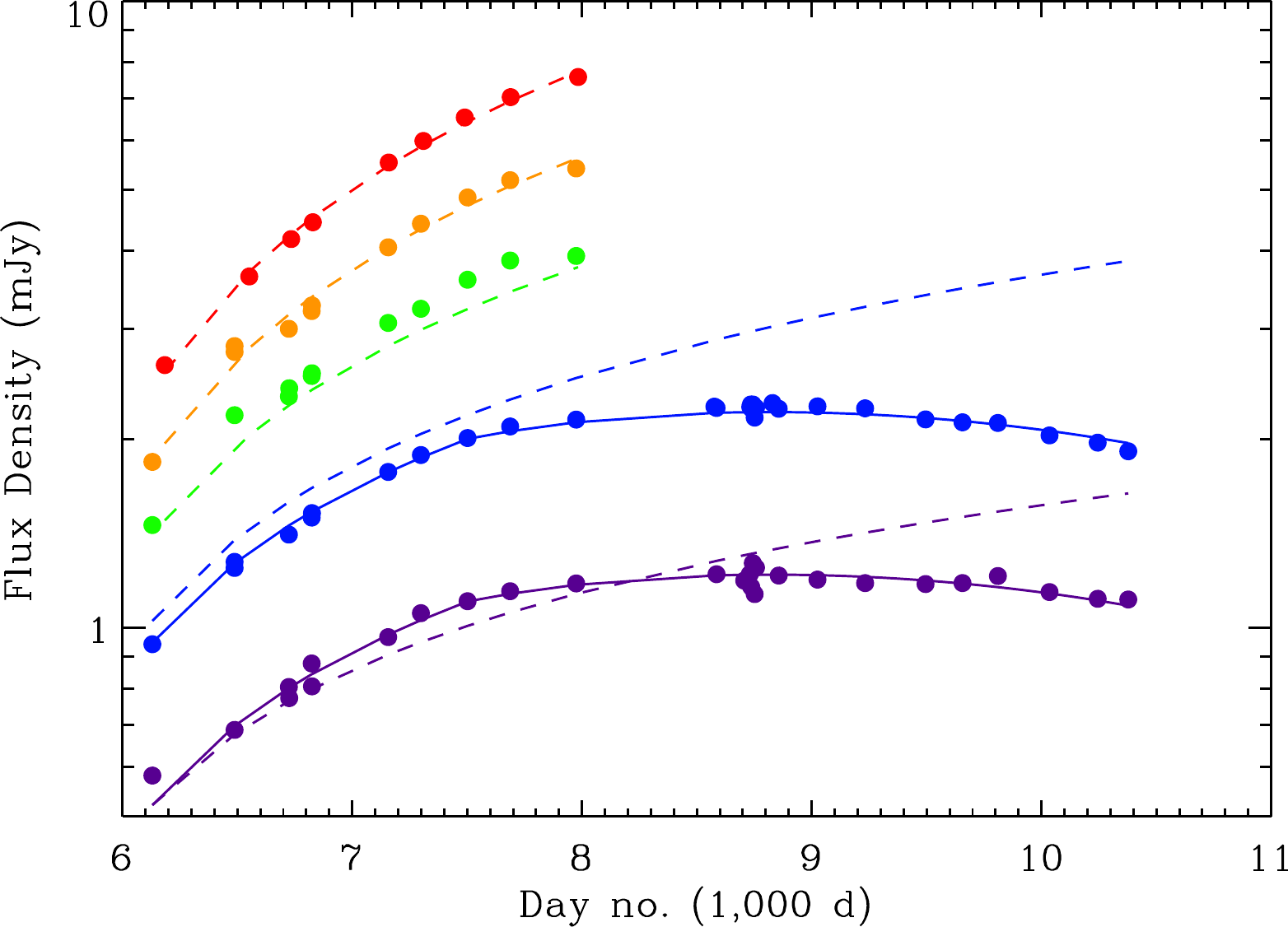} % RGA
 \end{center}
  \caption{Light curves of SN 1987A from {\it Spitzer} observations prior to day 10500 
  \citep[Fig. 8 from][]{Arendt2016}. Purple, blue, green, orange, red = 3.6, 4.5, 5.8, 8, 24 $\mu$m. The dashed lines represent a model of the evolution as given by Eq. (7) of that work.}
 \label{fig:lc_IRAC} 
\end{figure*}

\begin{figure*}
\begin{center}
   \includegraphics[width=8.5cm]{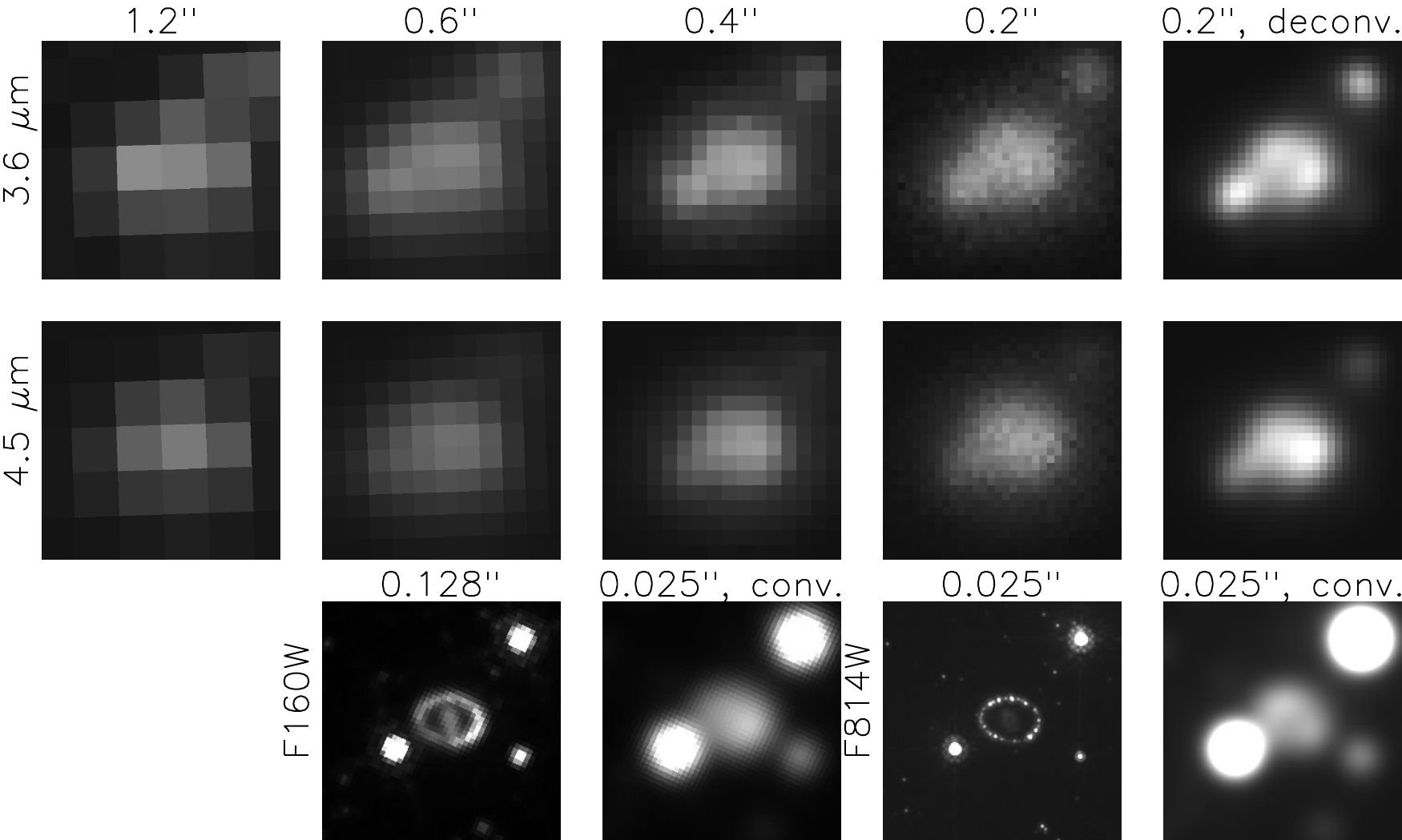} %RGA
\end{center}
  \caption{$6\farcs4 \times 6\farcs4$ IRAC images of SNR 1987A. 
  The first image in the top row is a single IRAC 3.6 $\mu$m 
exposure with the detector's $1\farcs2$ pixels. The following images show
the improved resolution enabled by mosaicking all IRAC observations 
(including post-cryogenic data, i.e. day 6000-12000), collected at 
a wide variety of sub-pixel dither offsets and roll angles. The last image 
shows the application of a deconvolution procedure to the $0\farcs2$ mosaic.
The second rows shows the same for 4.5 $\mu$m data.
The bottom row shows 2 shorter wavelength HST images at their original resolutions and convolved to simulate the resolution of the deconvolved IRAC images. Adapted from \cite{Arendt2020}.}
  \label{fig:imageshort}
\end{figure*}

\begin{figure*}
\centering
\includegraphics[width=8.5cm]{ 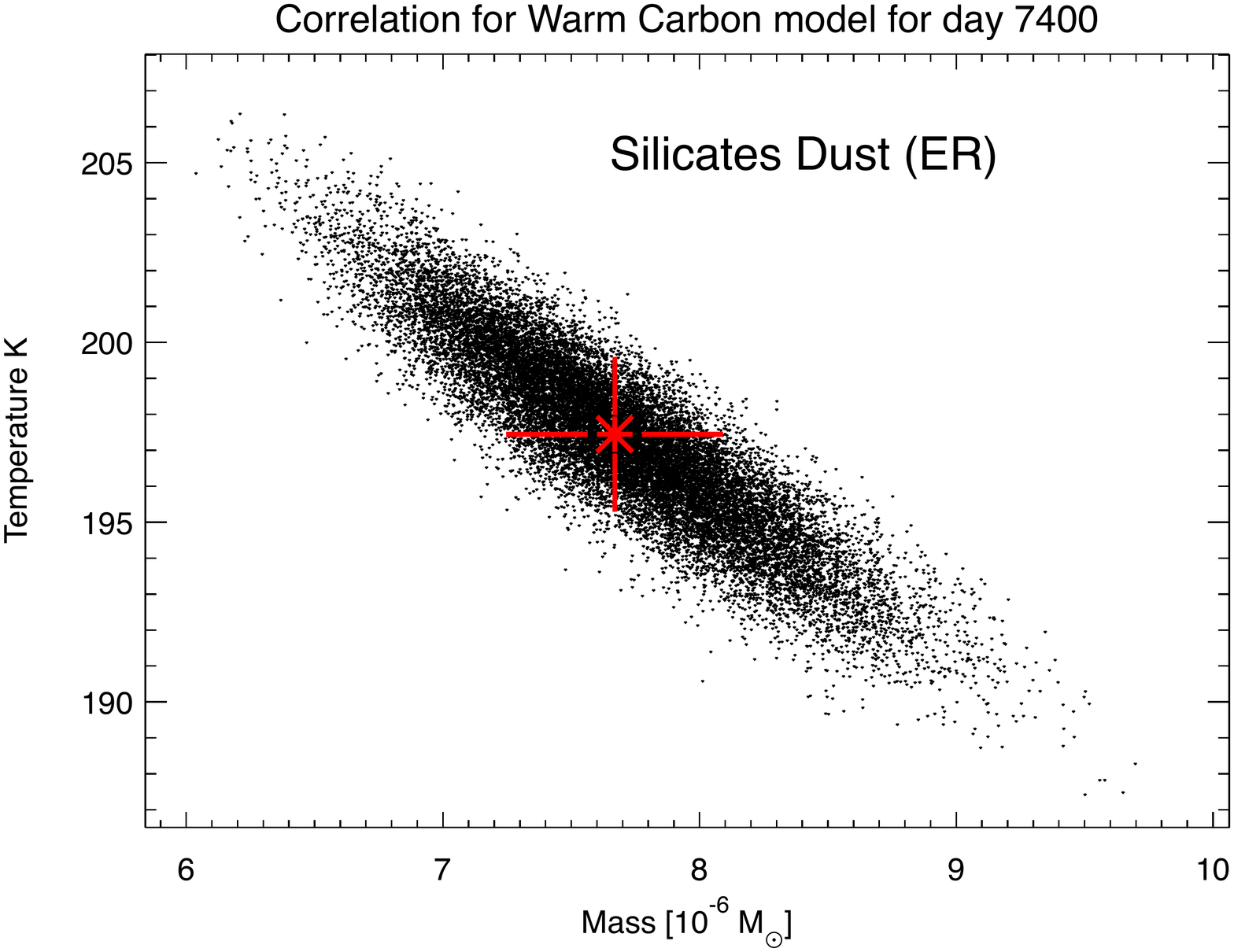}
\includegraphics[width=8.5cm]{ 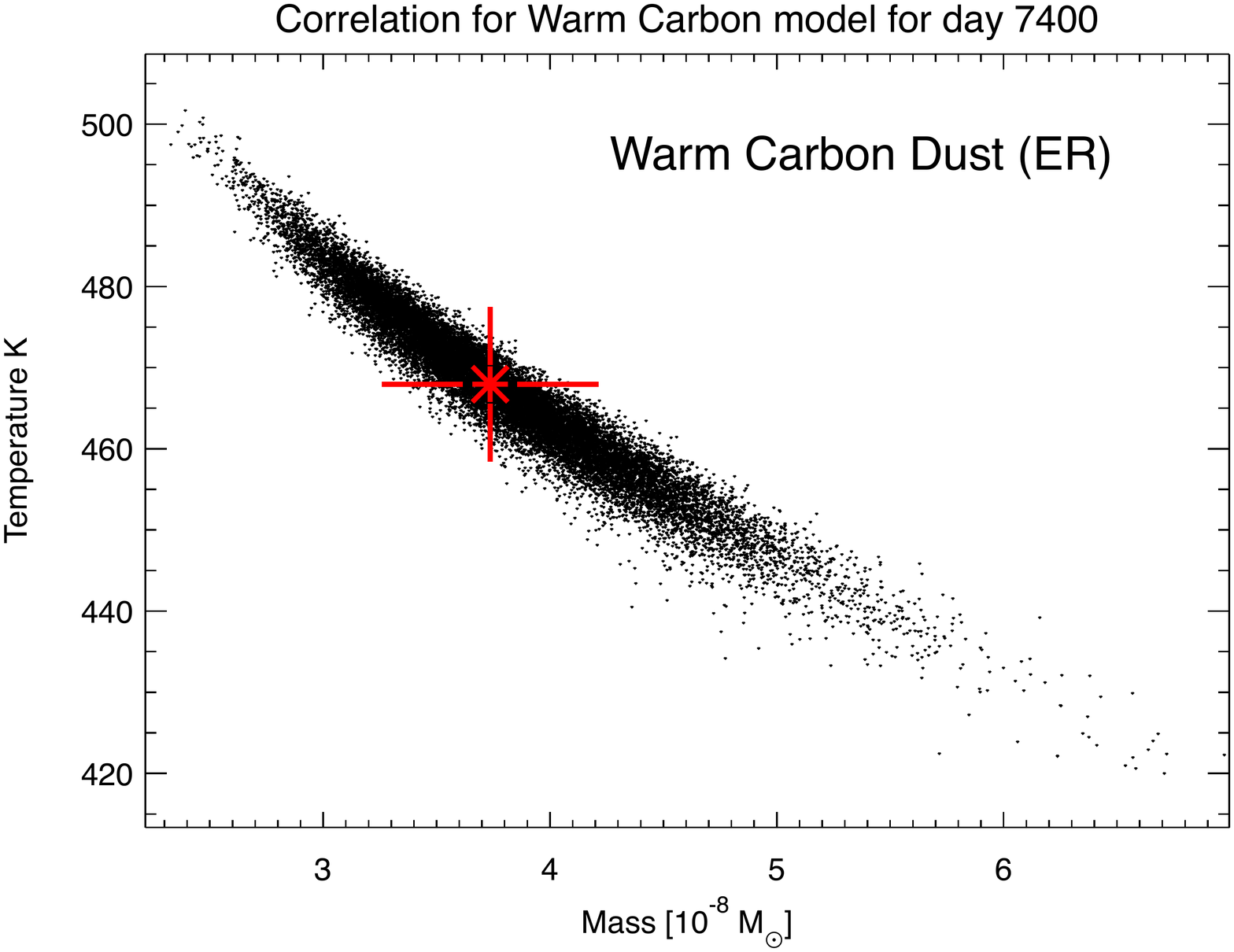}
\includegraphics[width=8.5cm]{ 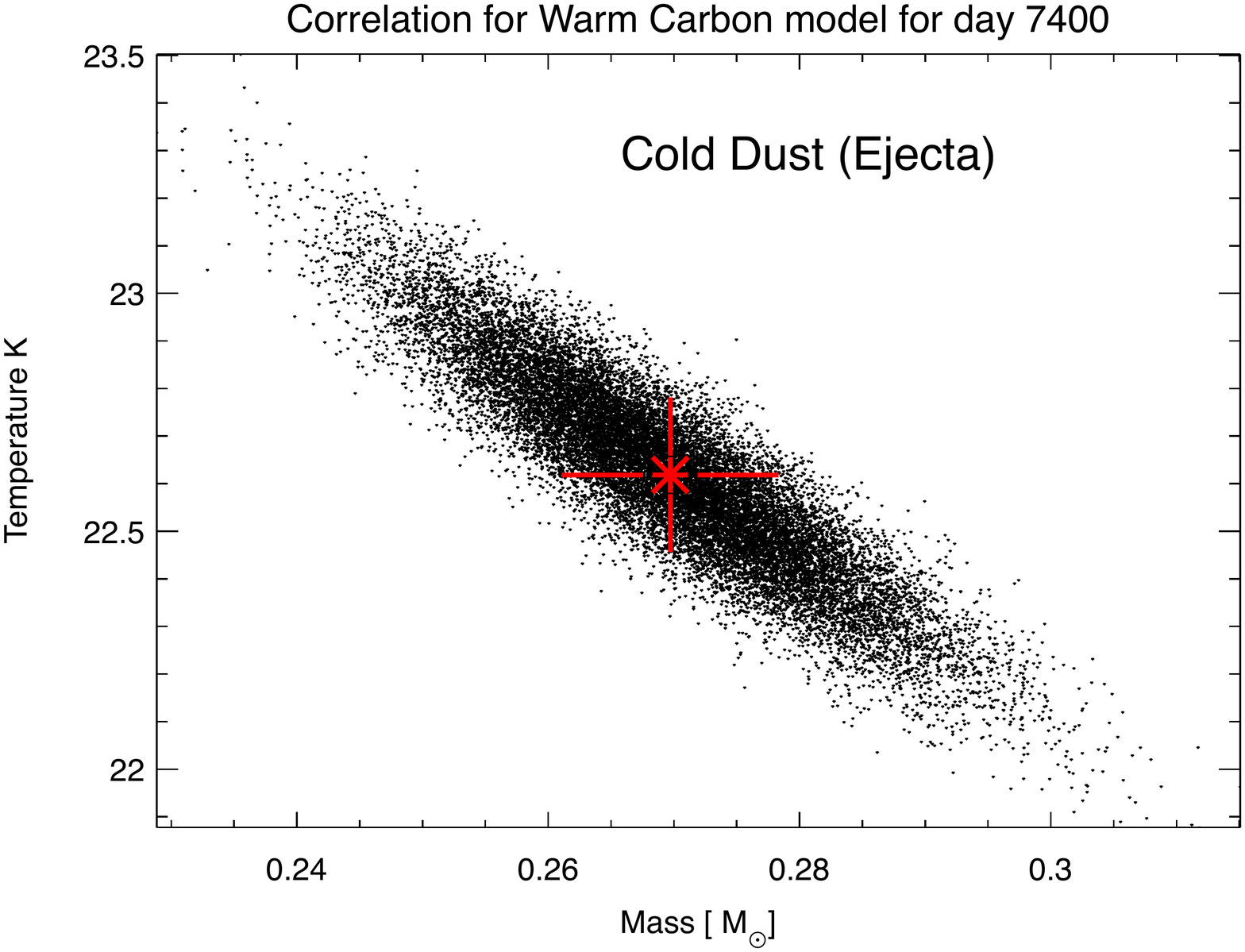} 
%\end{center}
  \caption{The correlations between both parameters of each component of the SED at day 7400 derived from our fitting procedure for the warm carbon case. The selected couple of parameters in each case are indicated by the red cross.}
  \label{fig:correlation_wc}
\end{figure*}

\begin{figure*}
\centering
\includegraphics[width=8.5cm]
{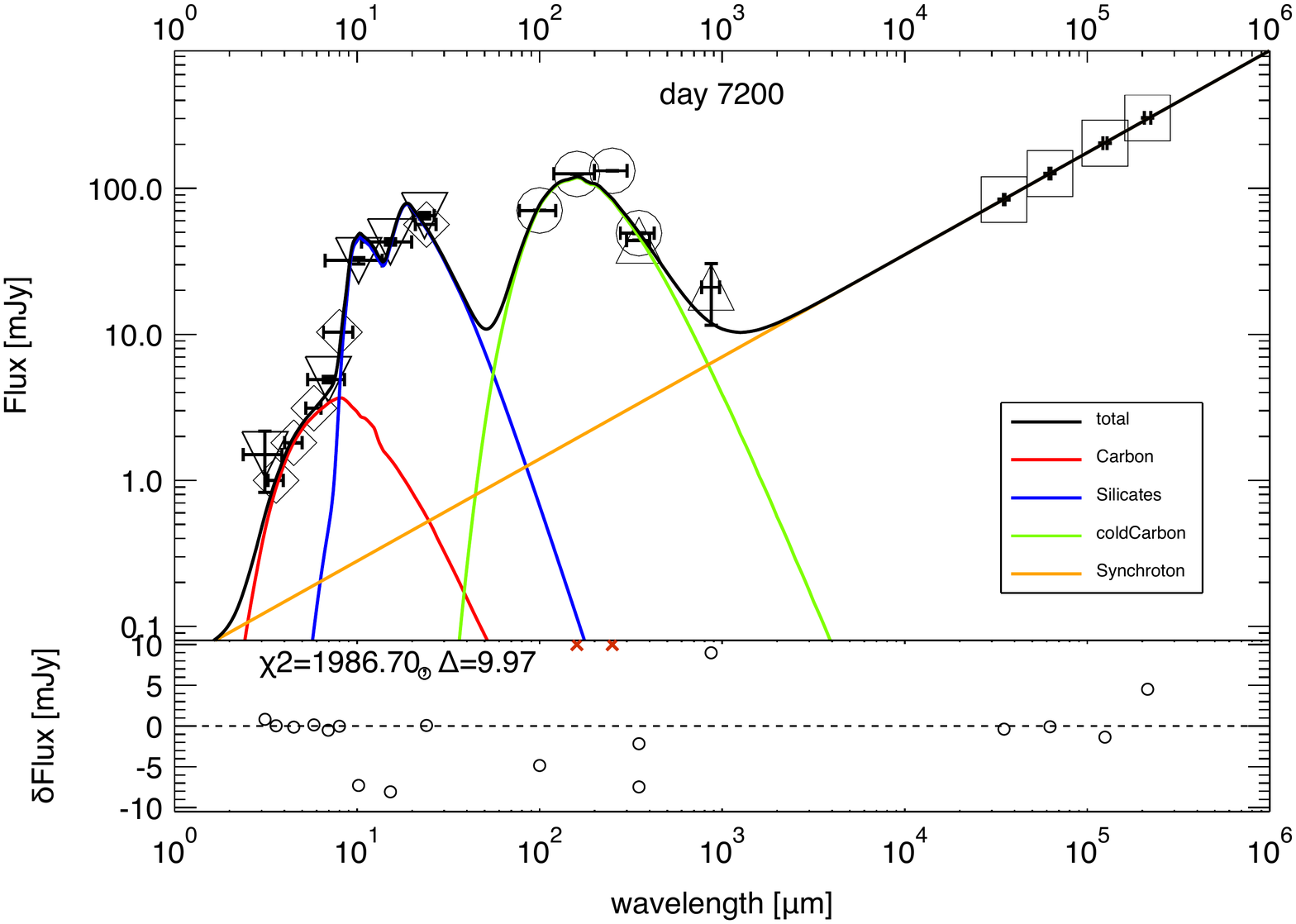}
\includegraphics[width=8.5cm]
{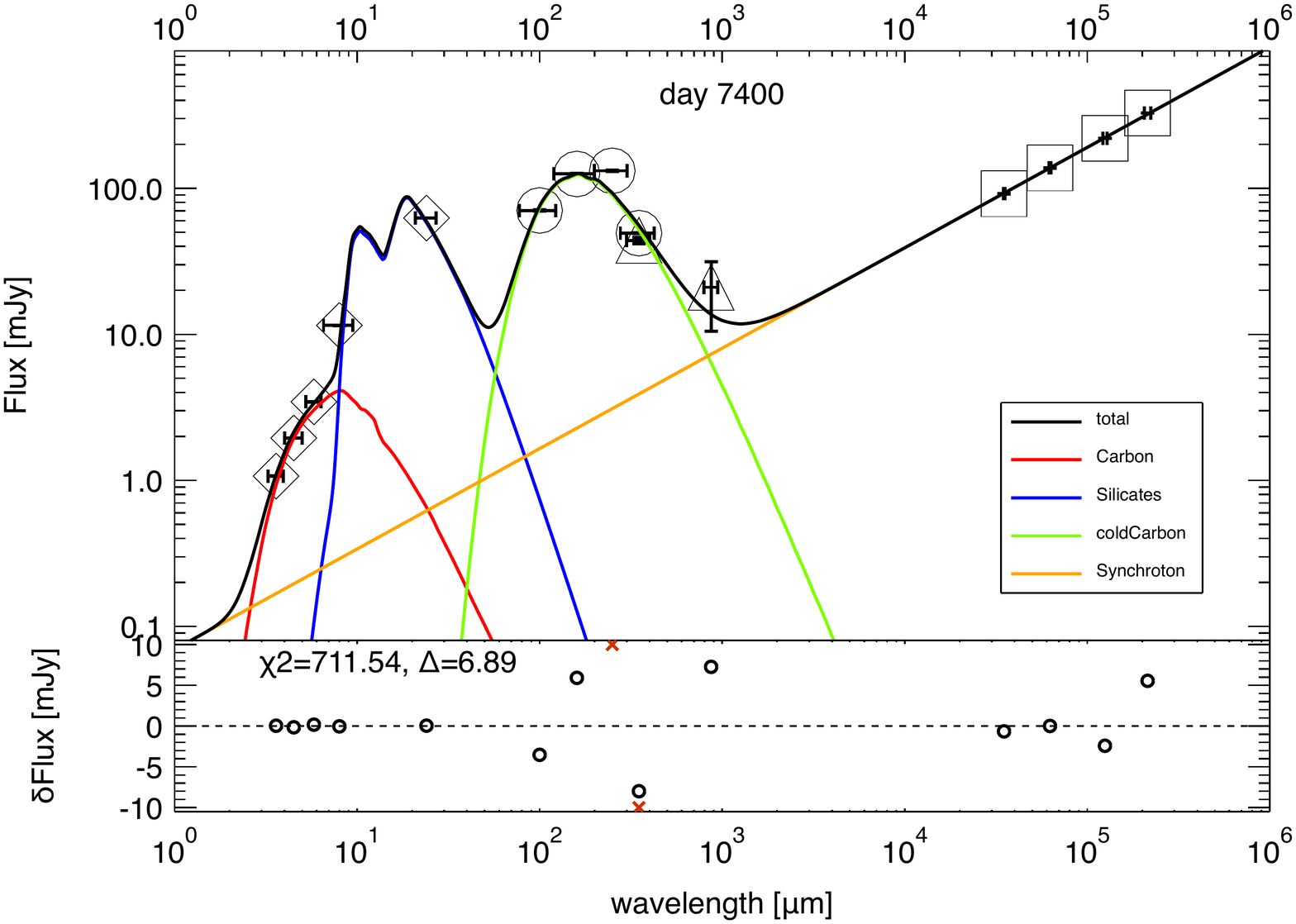}
  \caption{SED of SNR~1987A at days 7200 (top) and 7400 (bottom) with our resulting fits. The short wavelengths range has been fitted with a warm carbon
  dust component. See text for the caution to be taken regarding the cold dust component fit. The measurement errors ($\sigma$) and the width of the filters are given by the height and width of the cross associated to each data point. $\chi^2$ and $\Delta$ as defined by Eq.~(\Ref{eq:chi2}) and Eq.~(\Ref{eq:Delta}) characterize the goodness of our fits. The symbols for the telescopes used are given in Table~\Ref{tab:telused} and the color code is: red = warm carbon dust; blue = silicates dust; green = cold dust; yellow = radio synchrotron; black = sum of all the components. The lower panel below the SED displays $\delta(Flux) = Flux(measured) - Flux(fit)$. In that panel, red data points means that they lie outside the axis range. All the following figures representing the SED of SNR~1987A with our fits obey the same colors and legends.} 
  \label{fig:sed_wc}
\end{figure*}

\begin{figure*}
\begin{center}
 \includegraphics[width=5.5cm]{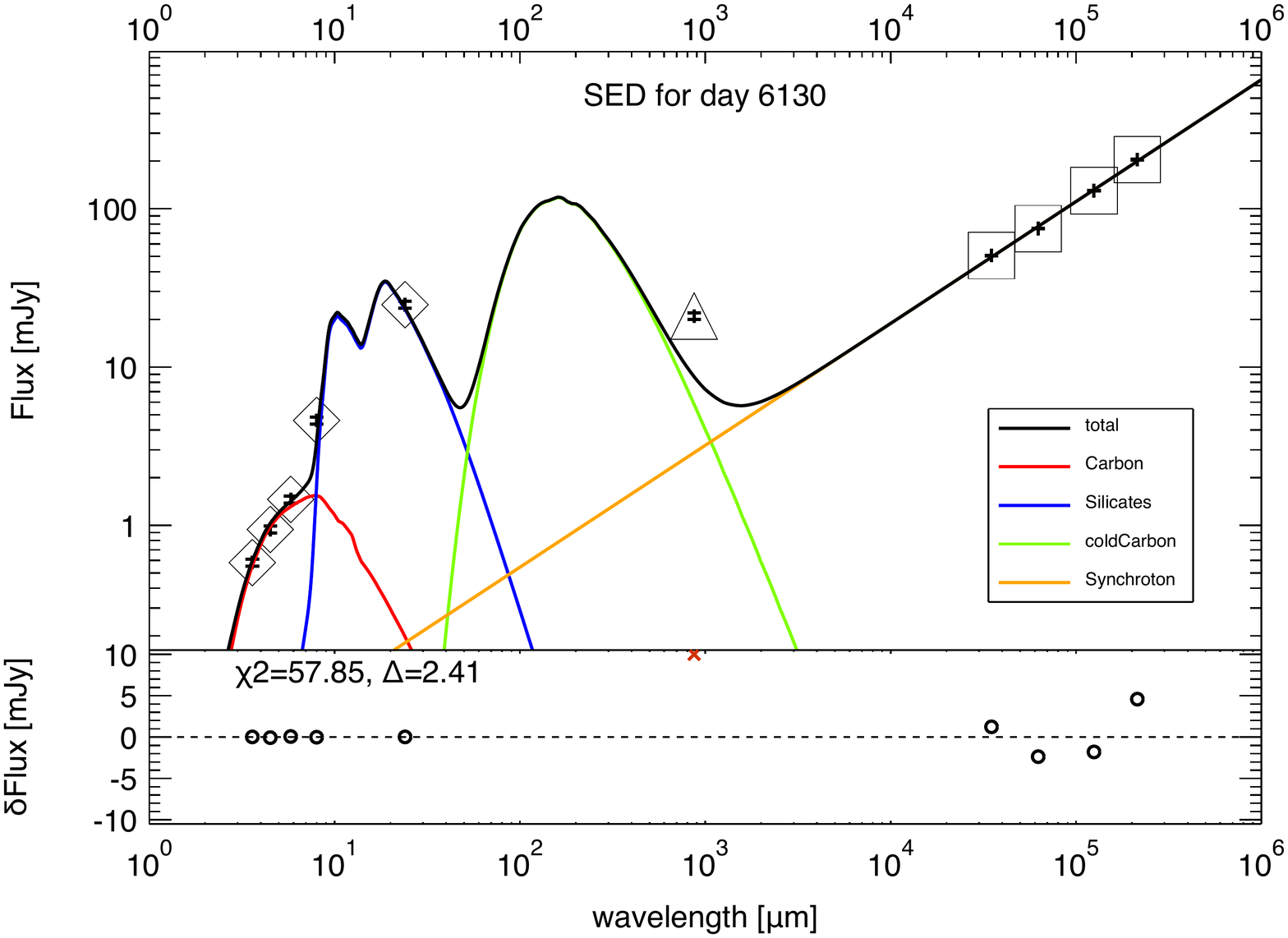 }
 \includegraphics[width=5.5cm]{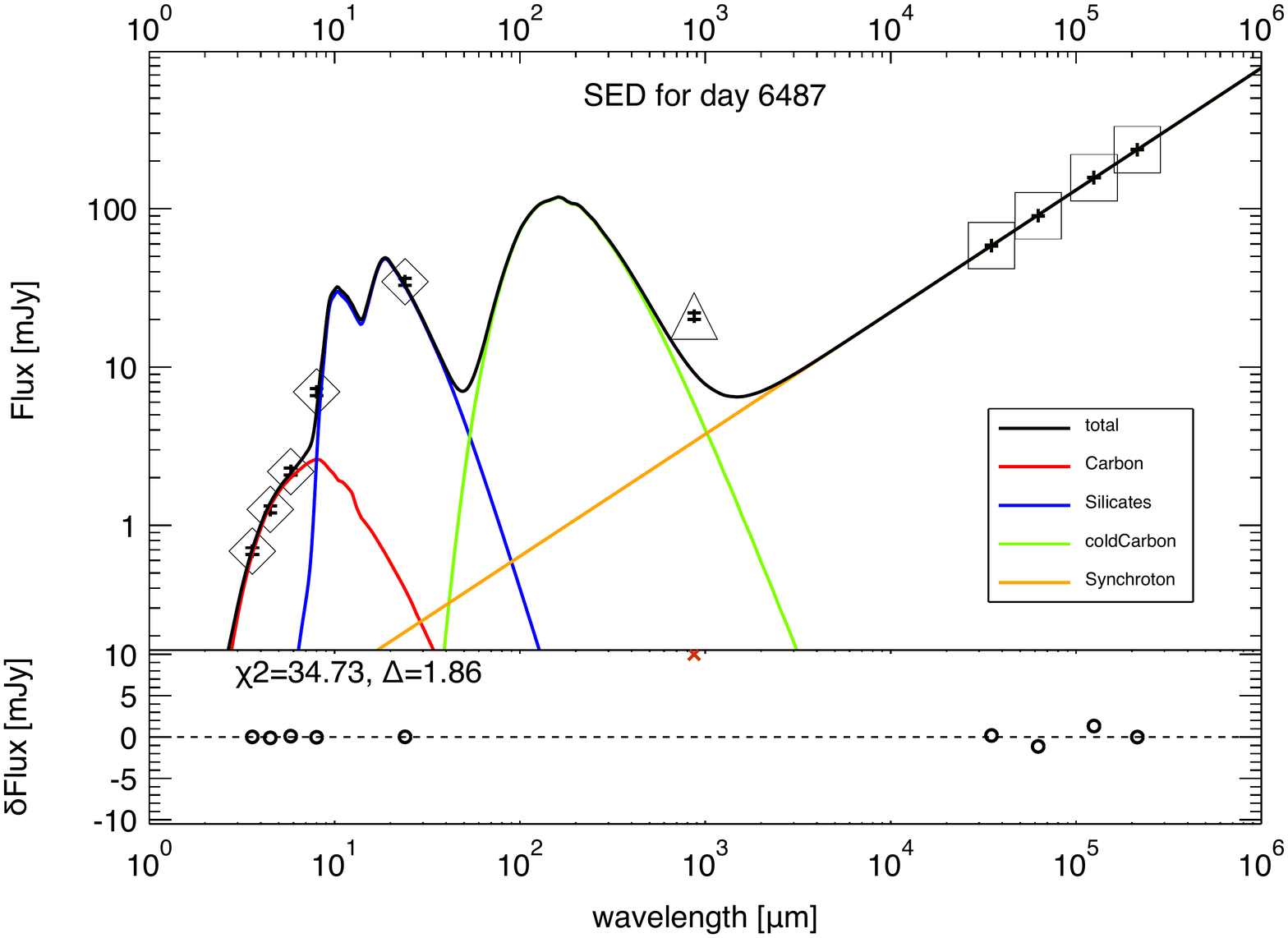 }
 \includegraphics[width=5.5cm]{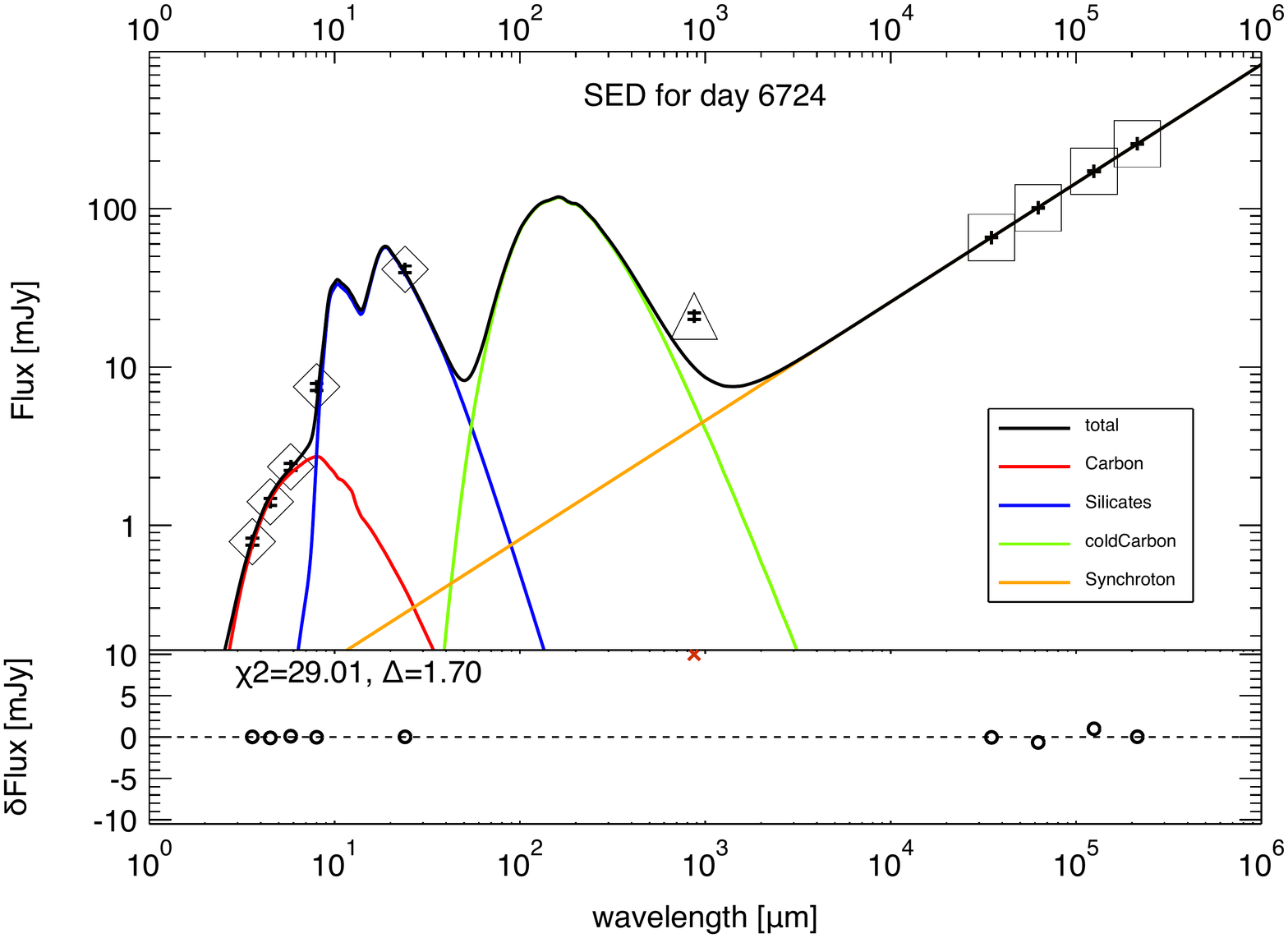}
 \includegraphics[width=5.5cm]{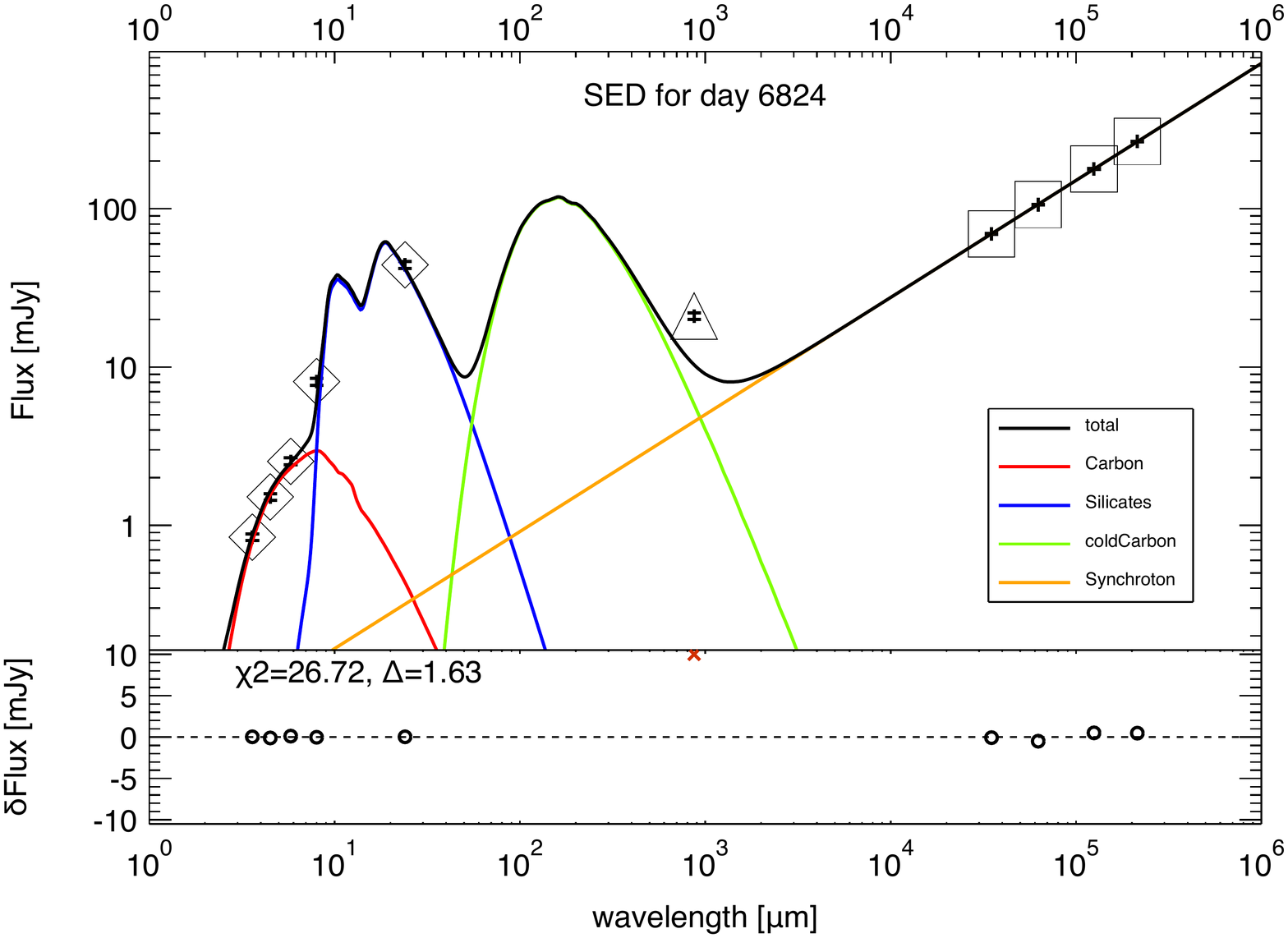}
 \includegraphics[width=5.5cm]{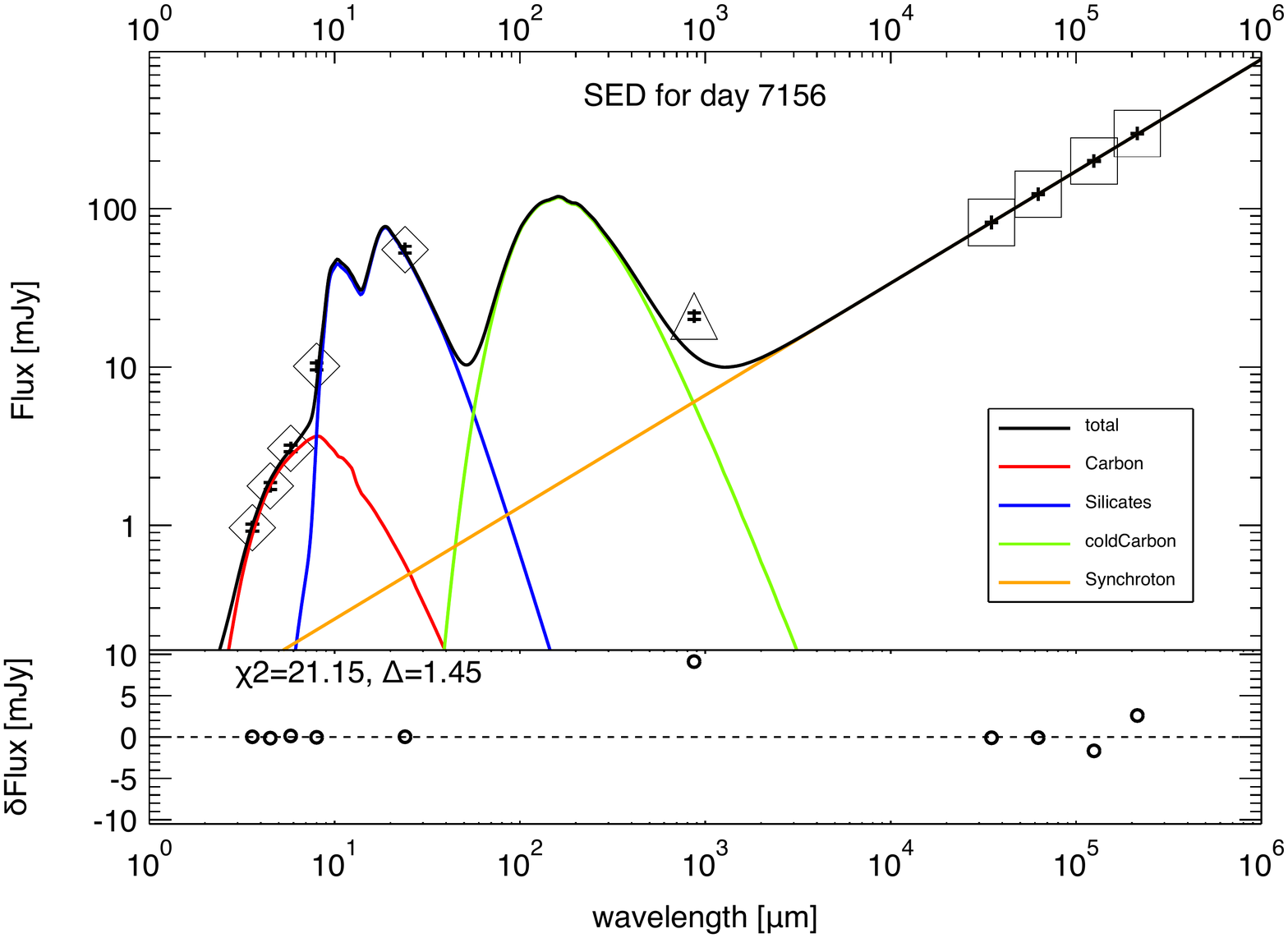}
 \includegraphics[width=5.5cm]{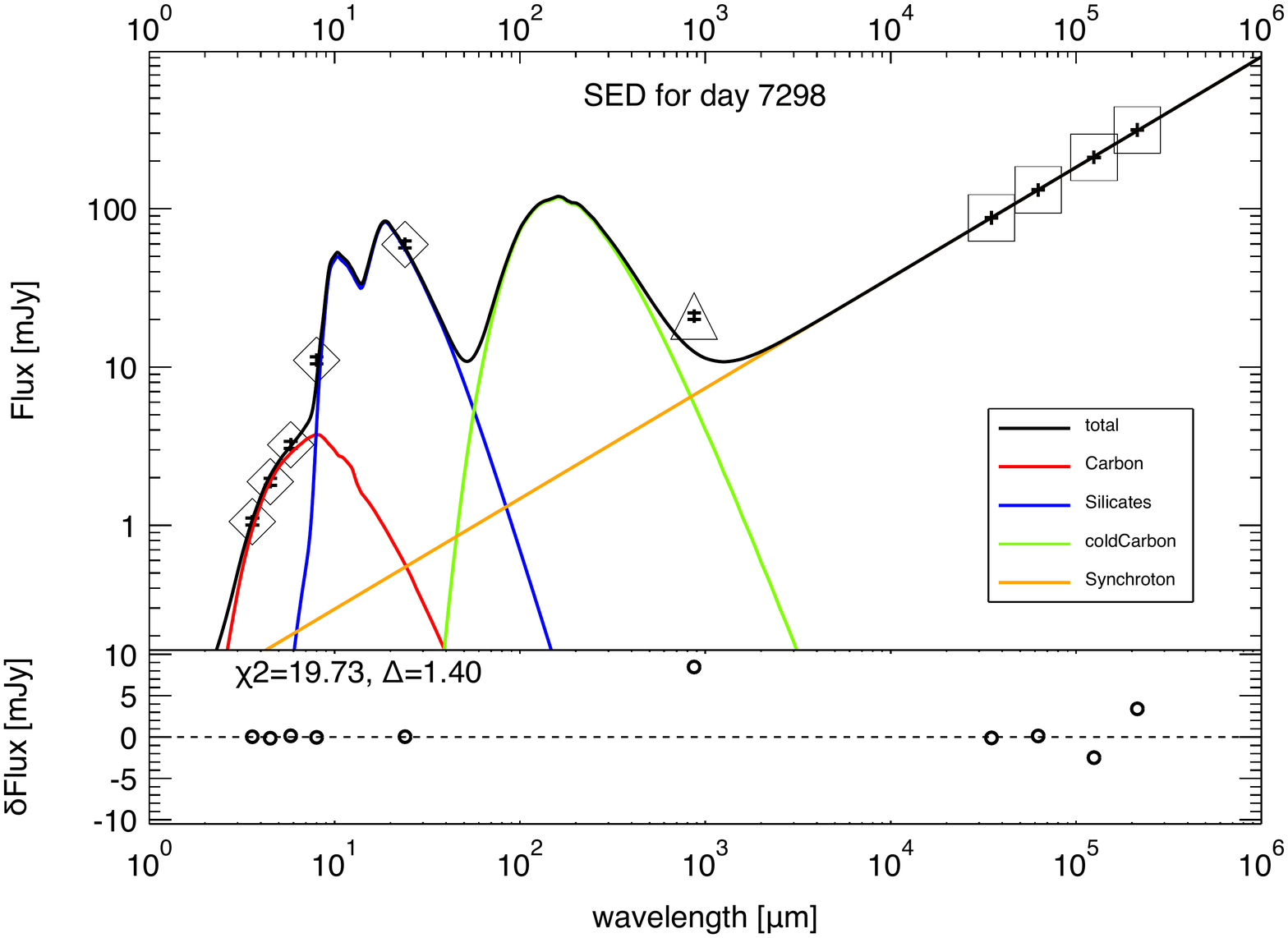}
 \includegraphics[width=5.5cm]{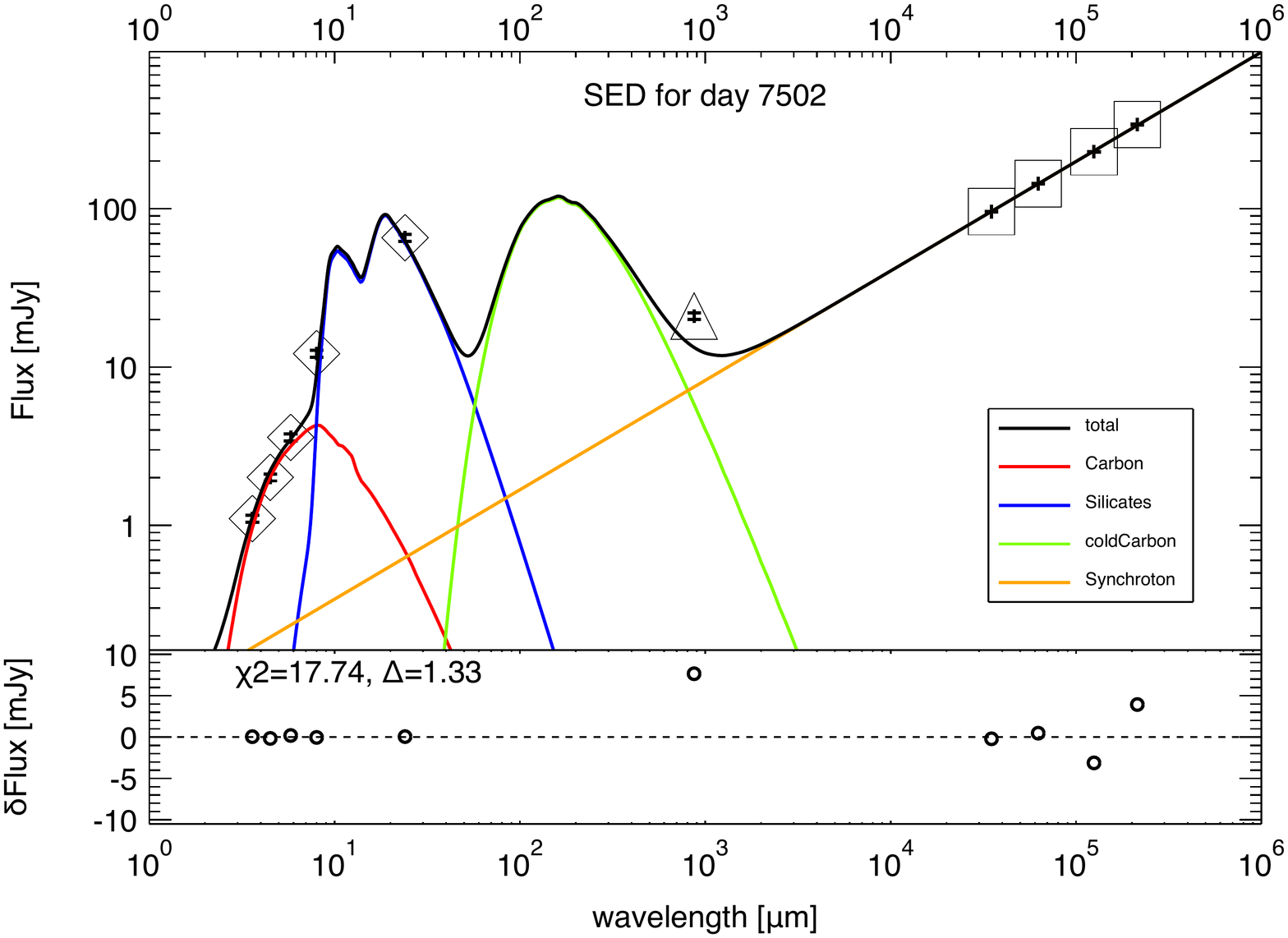}
 \includegraphics[width=5.5cm]{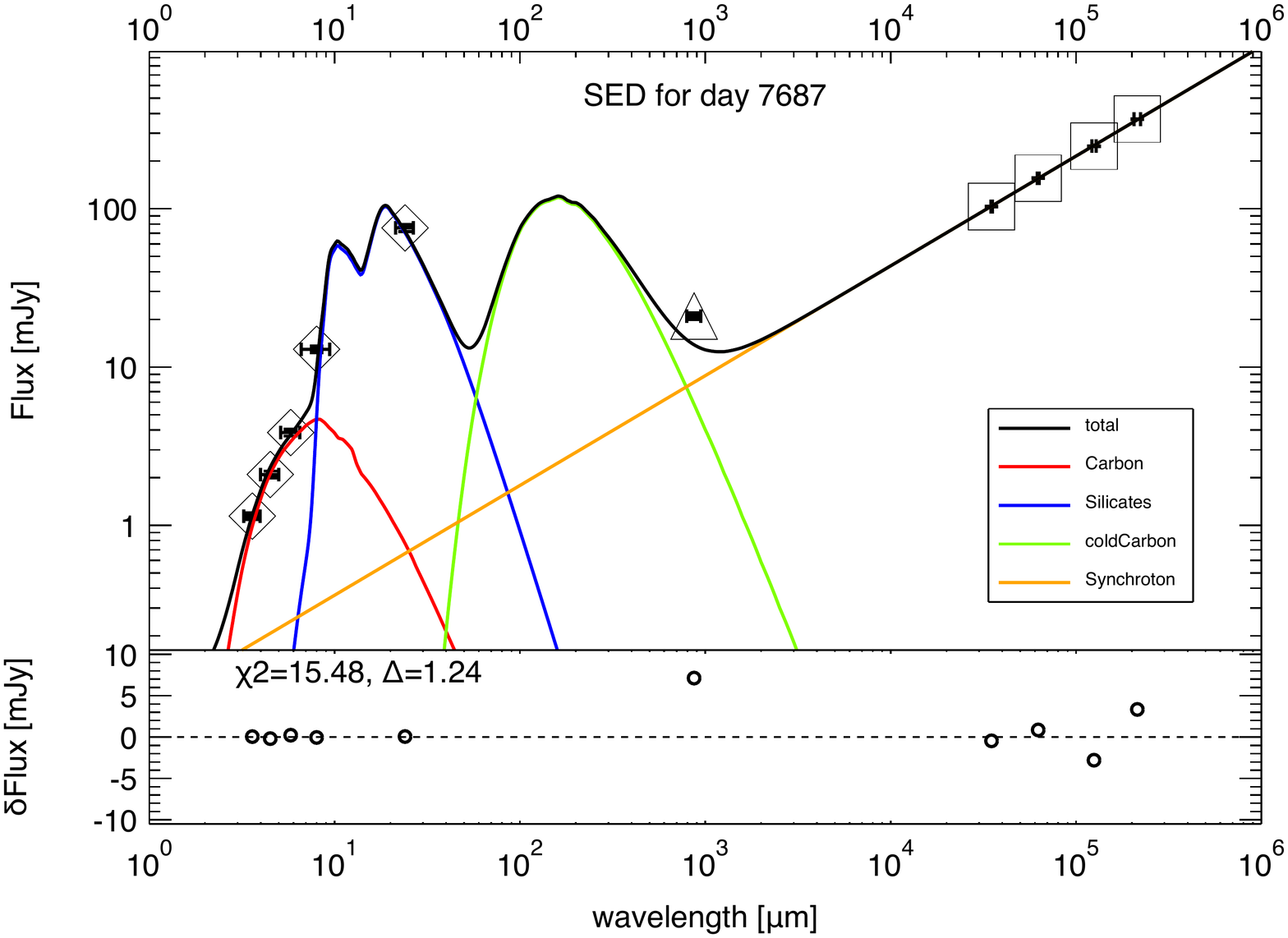}
 \includegraphics[width=5.5cm]{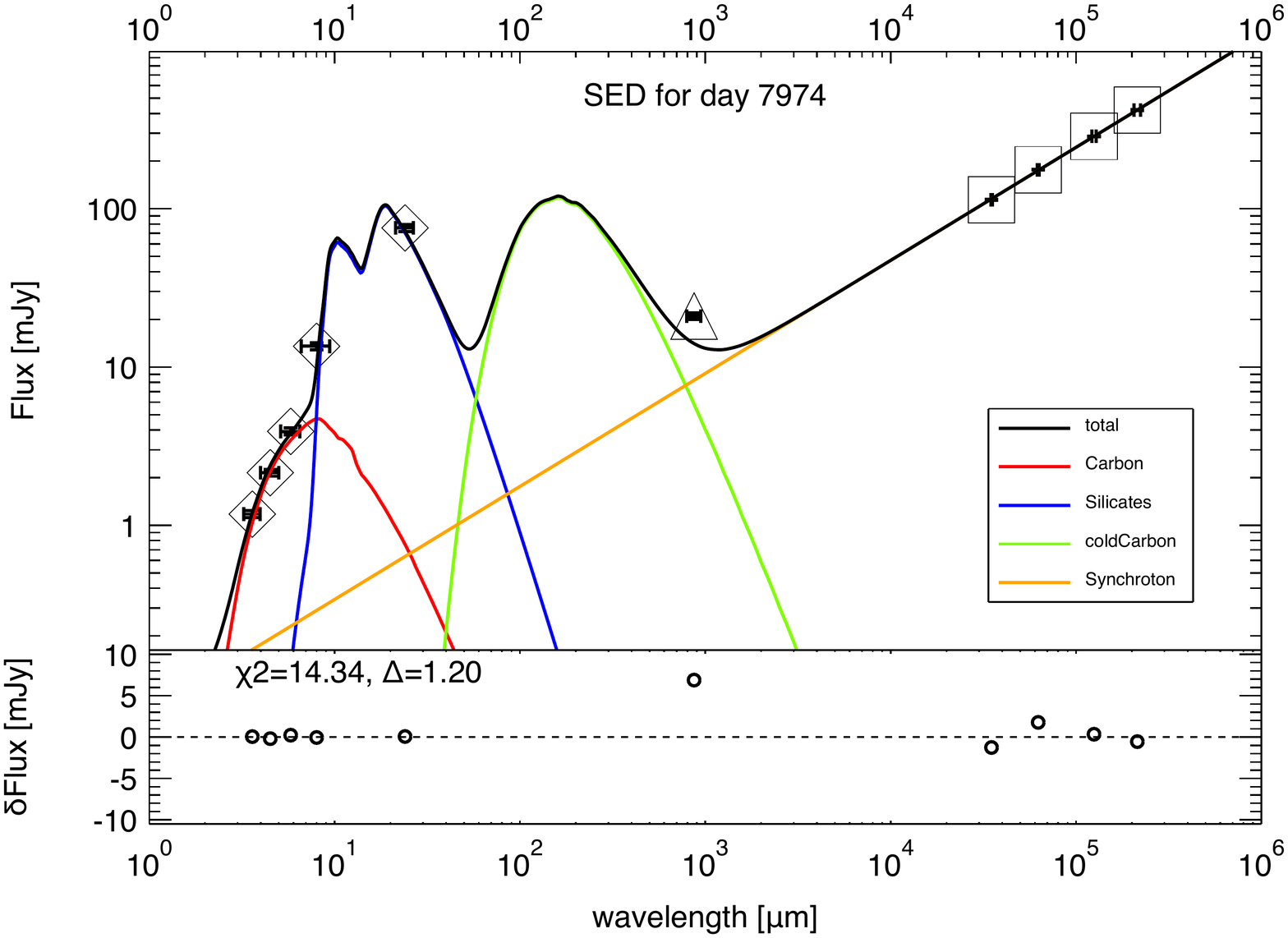}
 \includegraphics[width=5.5cm]{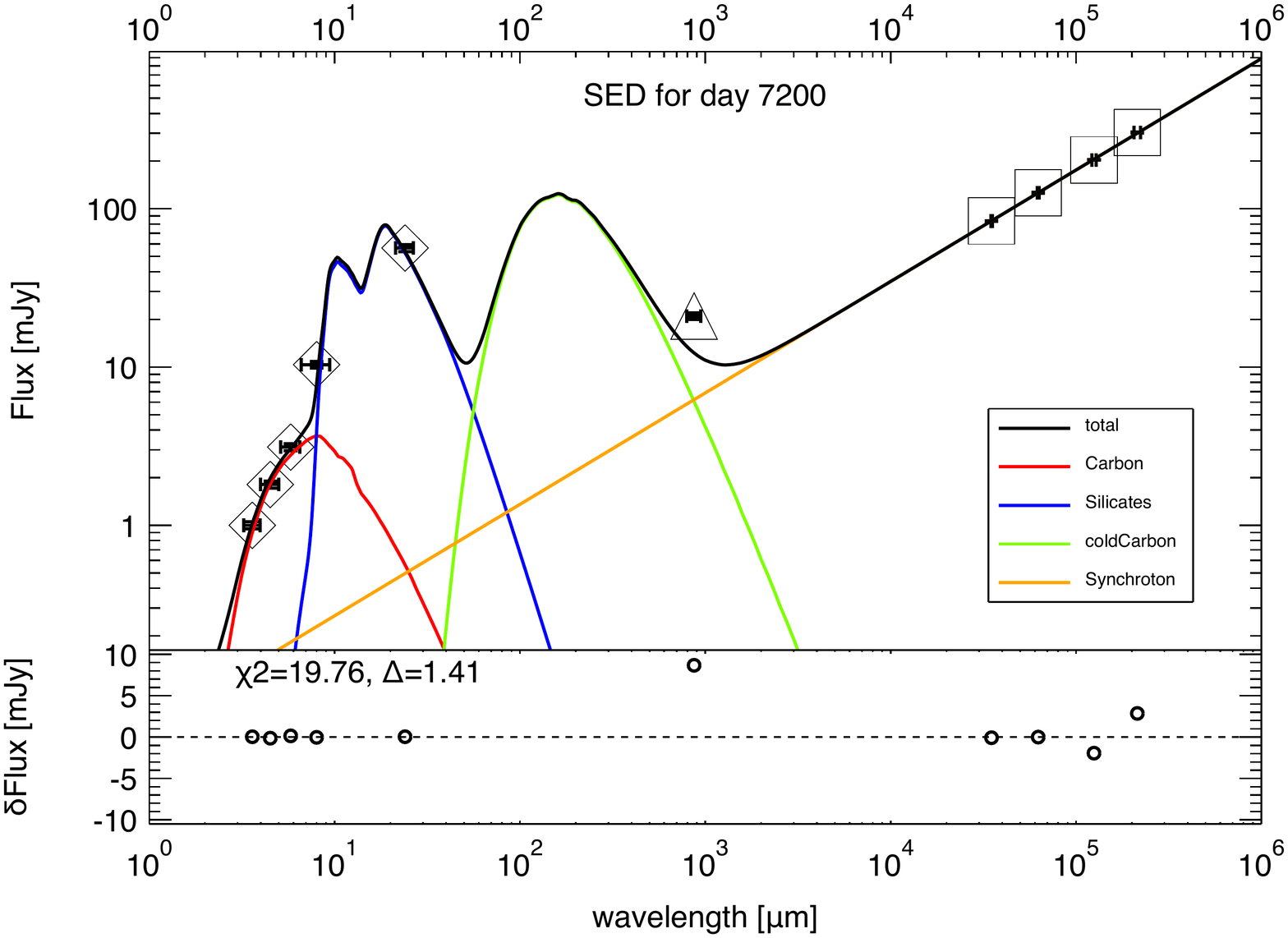}
 \includegraphics[width=5.5cm]{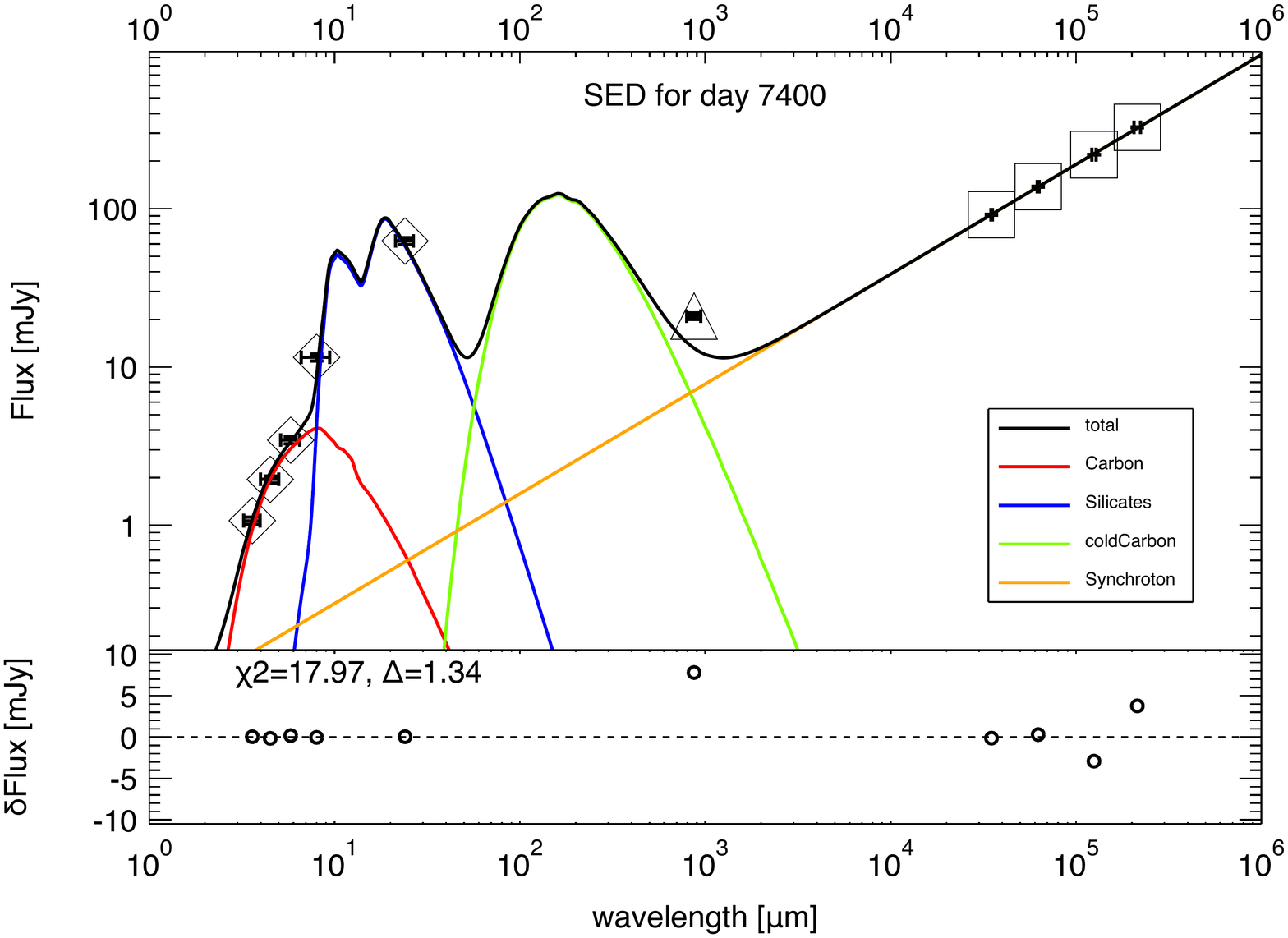}
\end{center}
  \caption{SED of SNR~1987A for the {\it Spitzer} data acquired during the period 6000 -- 8000 days with a warm carbon dust component to account for the excess emission at near-IR wavelengths. Symbols and colors
  follow the conventions used in Figure \ref{fig:sed_wc}. The two bottom panels are the interpolated data for the two epochs analysed in more detail, shown for comparison.} 
  \label{fig:individual-9d-wc-fits}
\end{figure*}

\begin{figure}
\begin{center}
\includegraphics[width=8.5cm]
{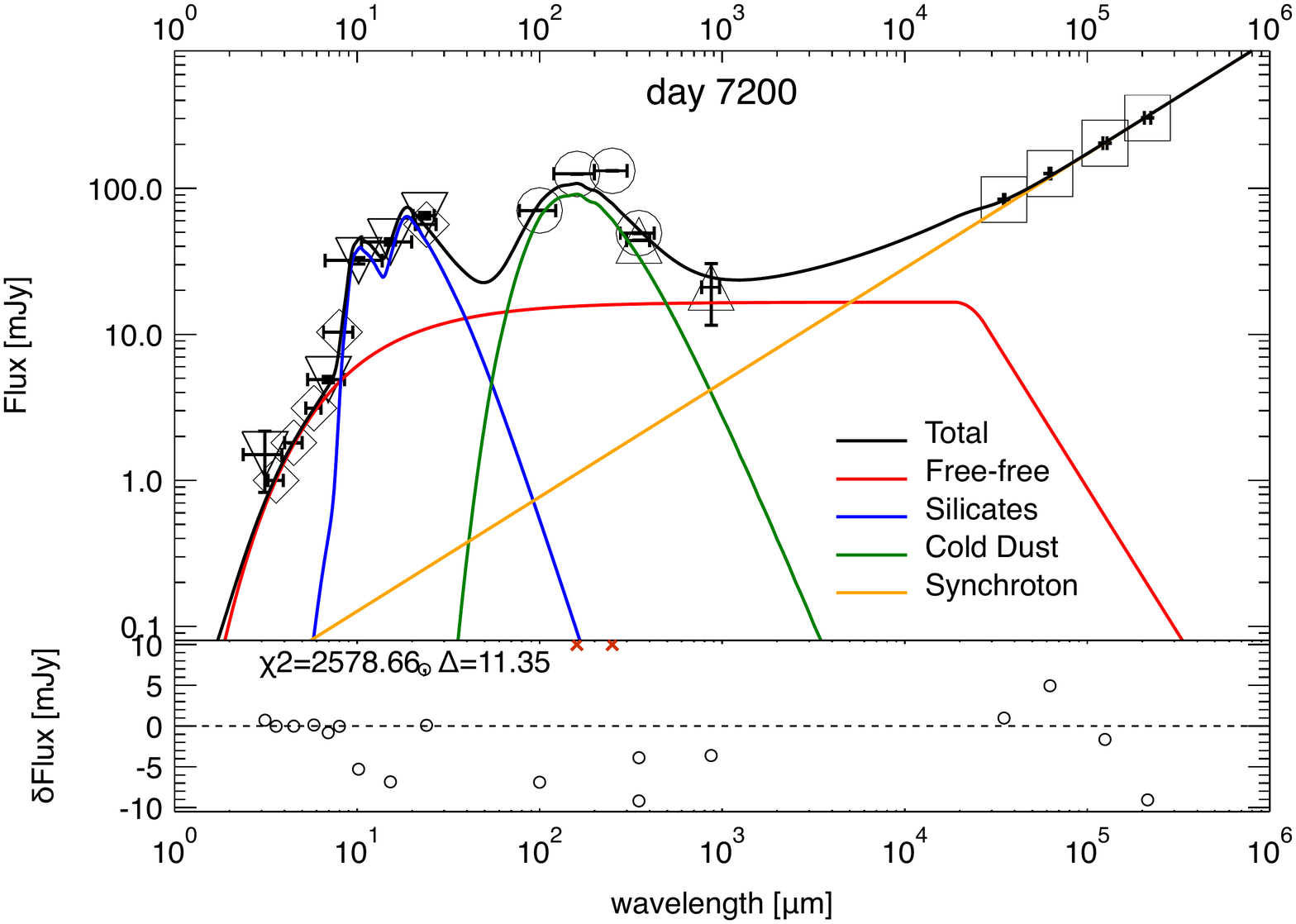}
\includegraphics[width=8.5cm]
{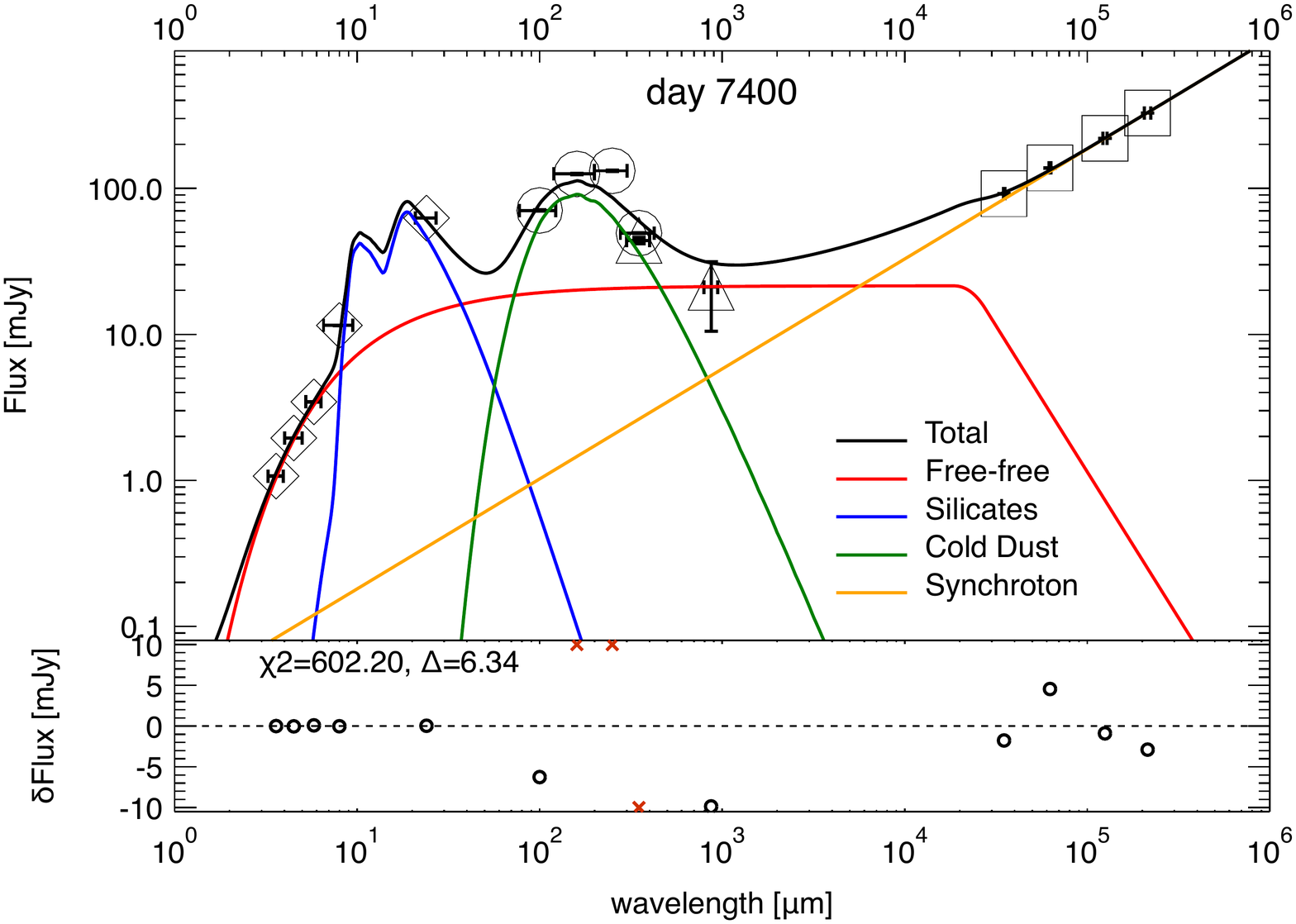} 
\end{center}
  \caption{SED of SNR~1987A at days 7200 (top) and 7400 (bottom) with our resulting fits. The short wavelengths range has been fitted with a free-free emission self-absorbed at 13~GHz. See text for the caution to be taken regarding the cold dust component fit. The measurement errors ($\sigma$) and the width of the filters are as in Figure~\Ref{fig:sed_wc} and the parameters $\chi^2$ and $\Delta$ are defined by Eq.~(\Ref{eq:chi2}) and Eq.~(\Ref{eq:Delta}) that characterize the goodness of our fits; the symbols for the telescopes used are given in Table~\Ref{tab:telused} and the color code is as in Figure~\Ref{fig:sed_wc}.} 
  \label{fig:sed_ffcut13}
\end{figure}

\begin{figure}
\begin{center}
\includegraphics[width=8.5cm]
{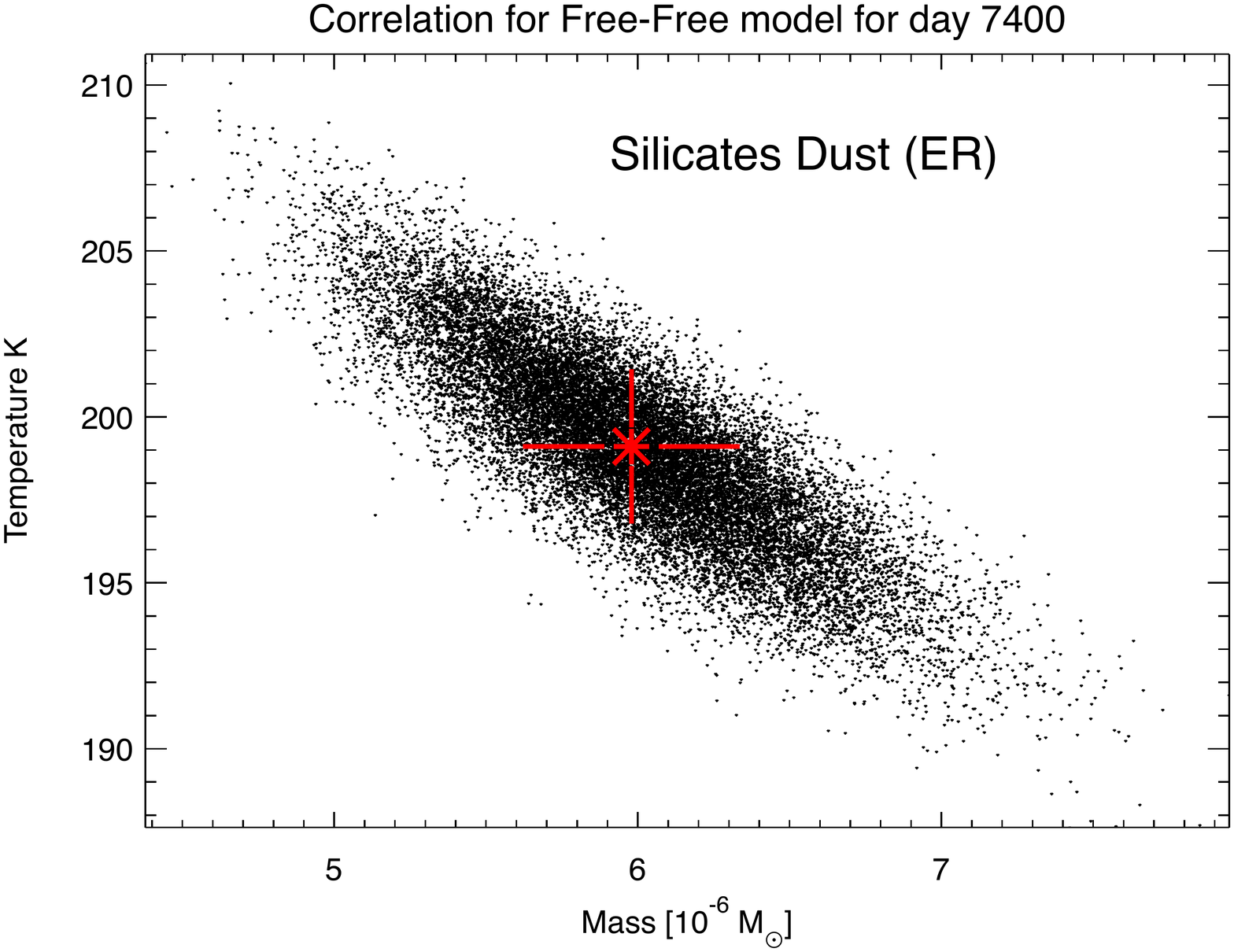} 
\includegraphics[width=8.5cm]{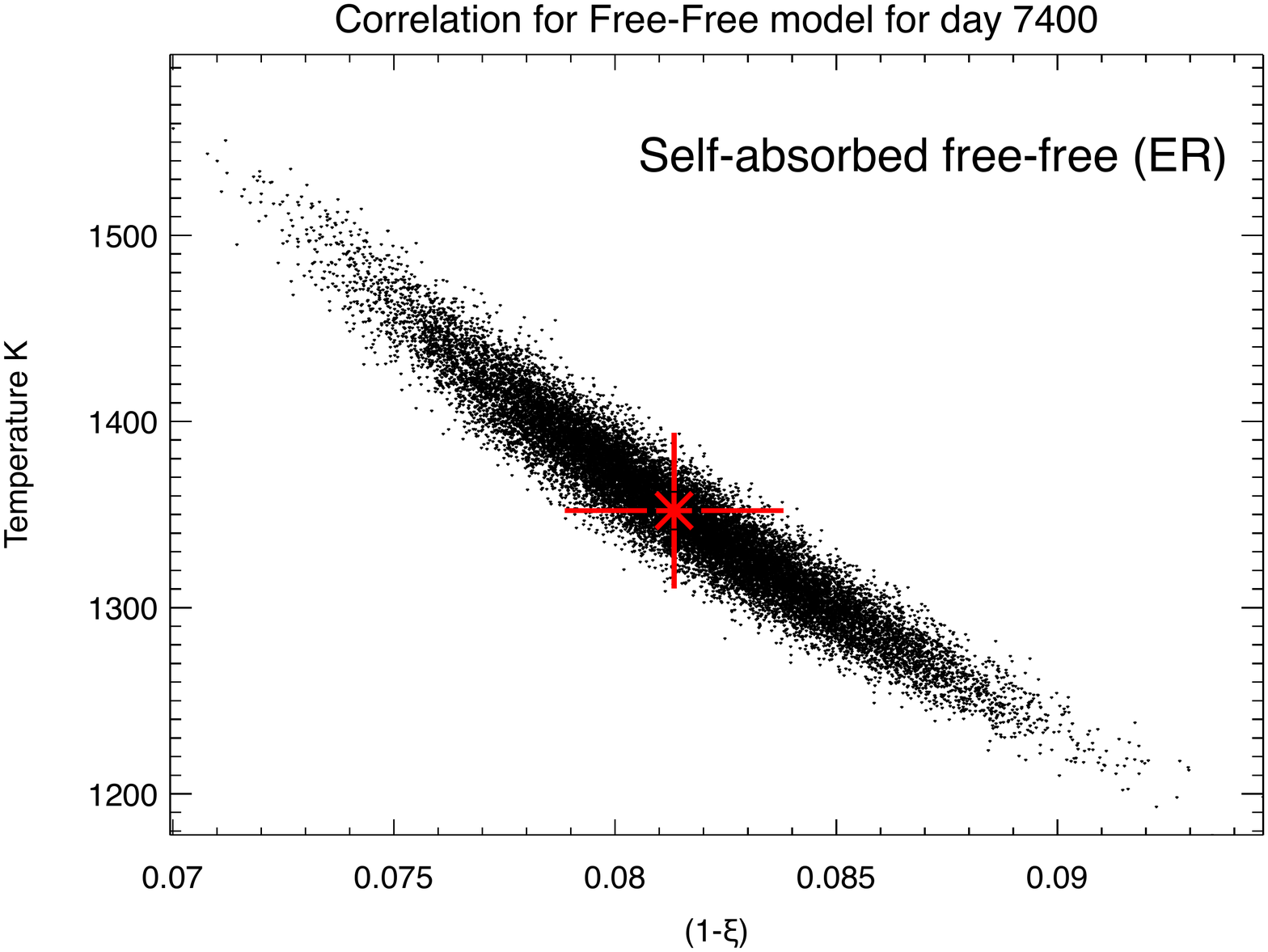} 
\includegraphics[width=8.5cm]
{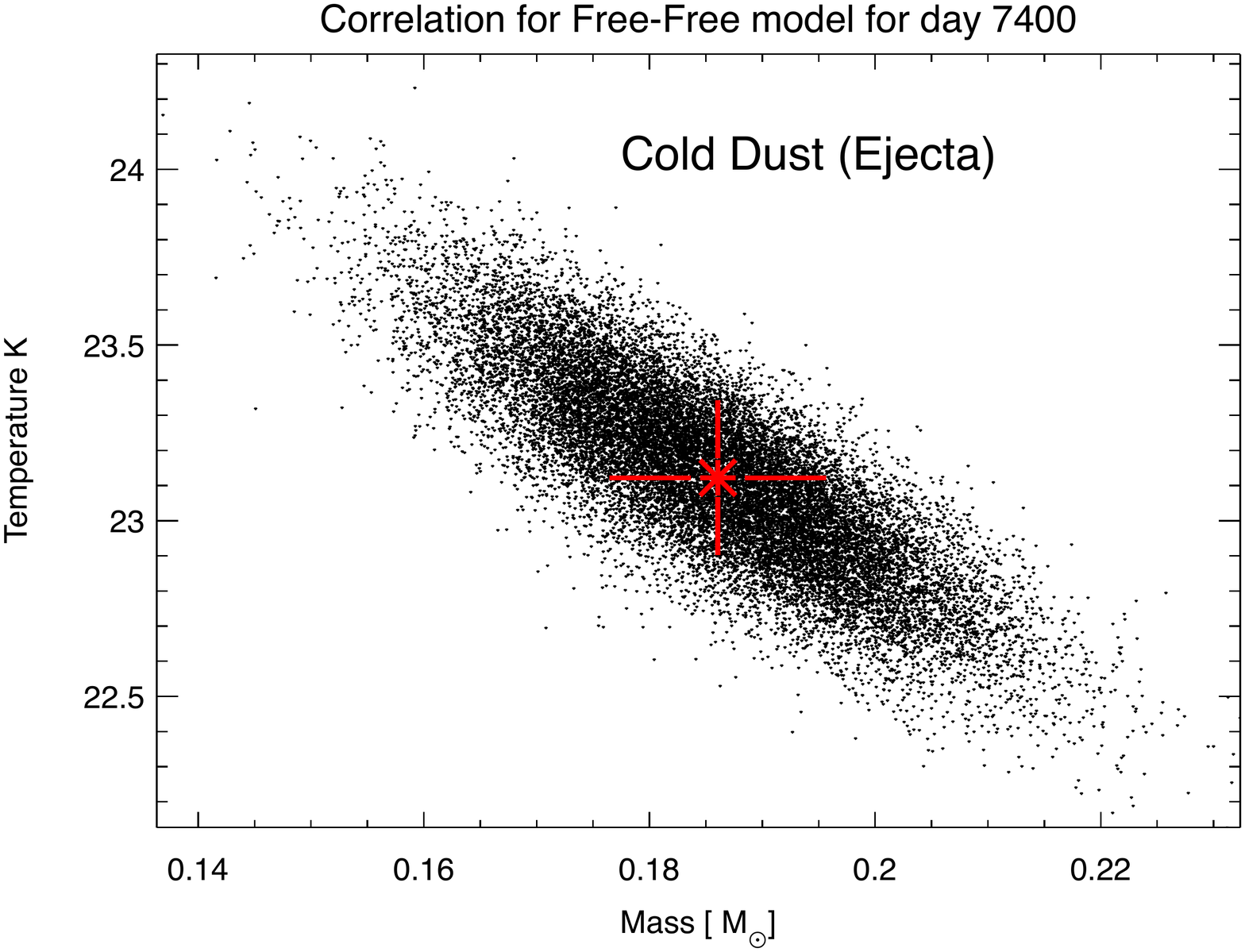}
\end{center}
  \caption{Same as Figure~\Ref{fig:correlation_wc} for the free-free model self-absorbed at 13~GHz at day 7400.}
\label{fig:correlation_ff13} 
\end{figure}

\begin{figure*}
\begin{center}
\includegraphics[width=5.5cm]{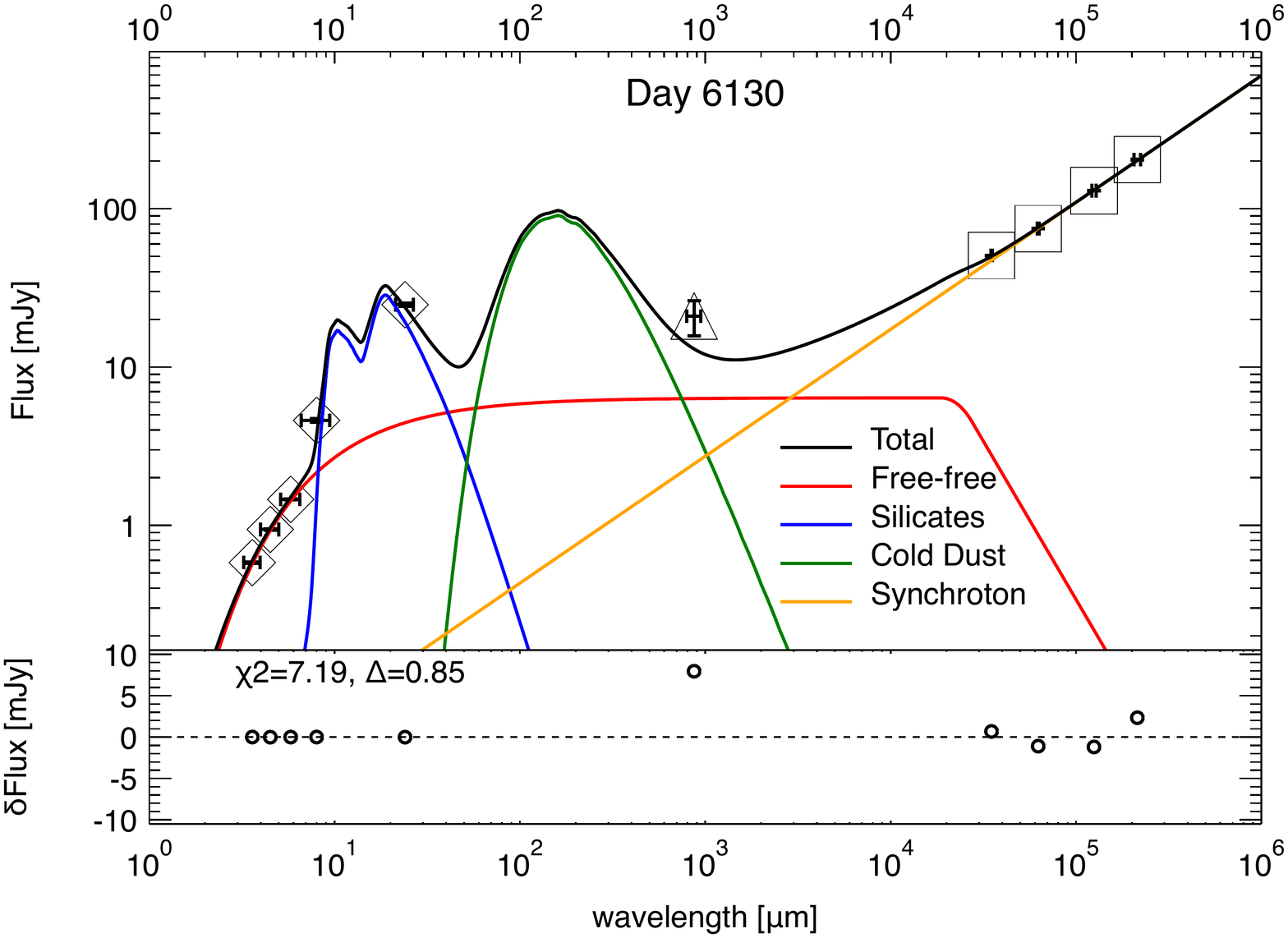 }
\includegraphics[width=5.5cm]{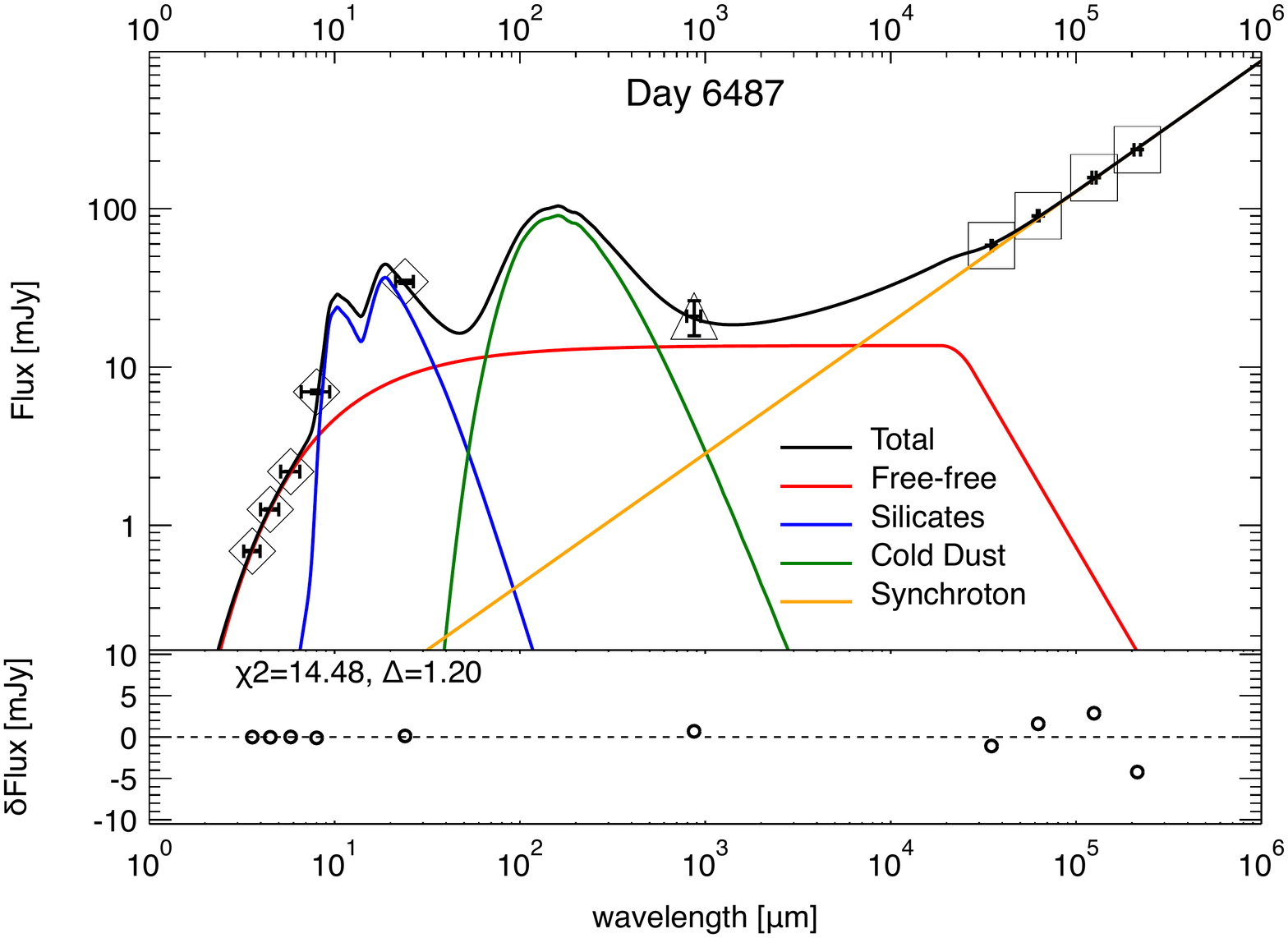 }
\includegraphics[width=5.5cm]{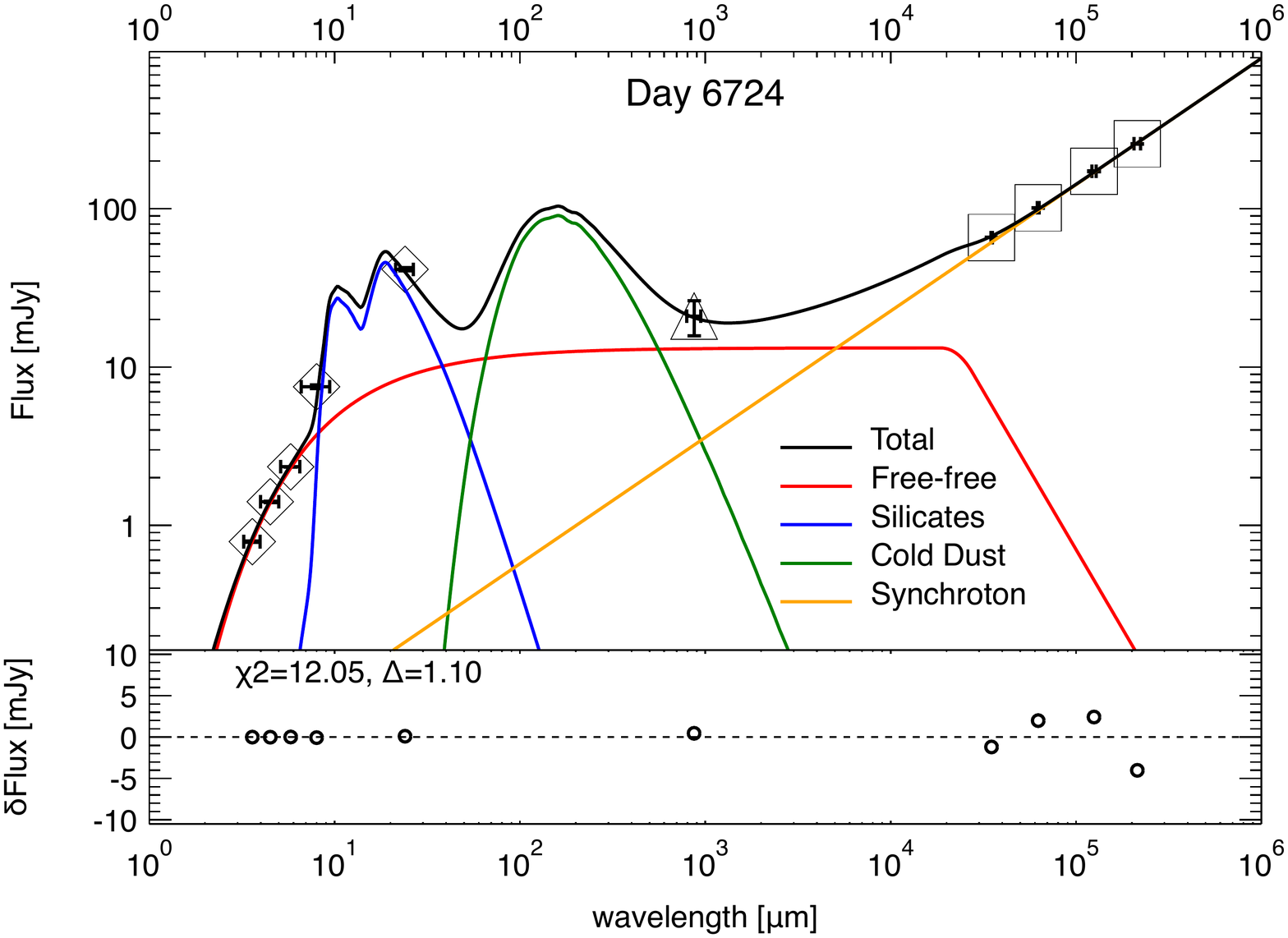 }
\includegraphics[width=5.5cm]{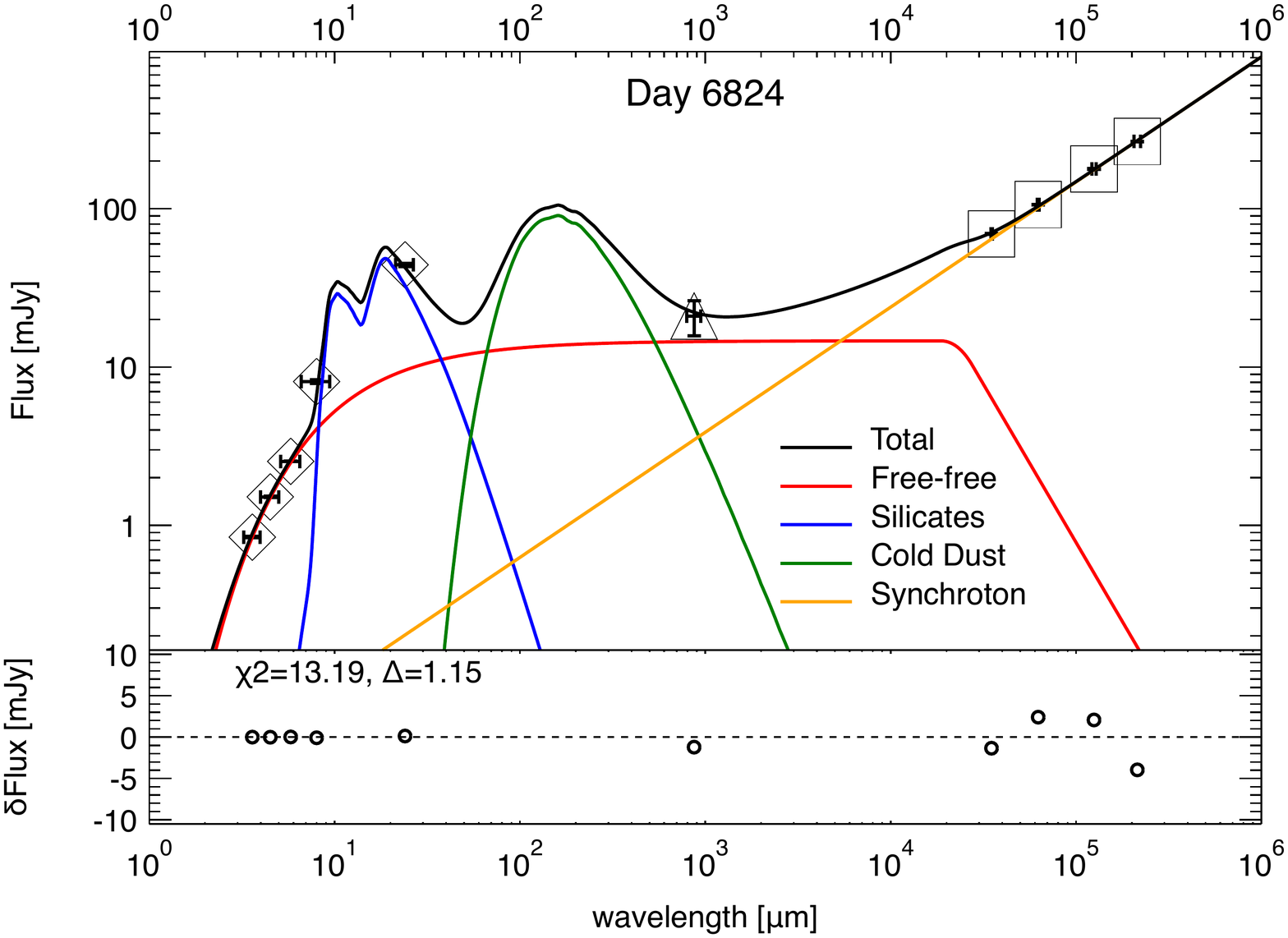 }
\includegraphics[width=5.5cm]{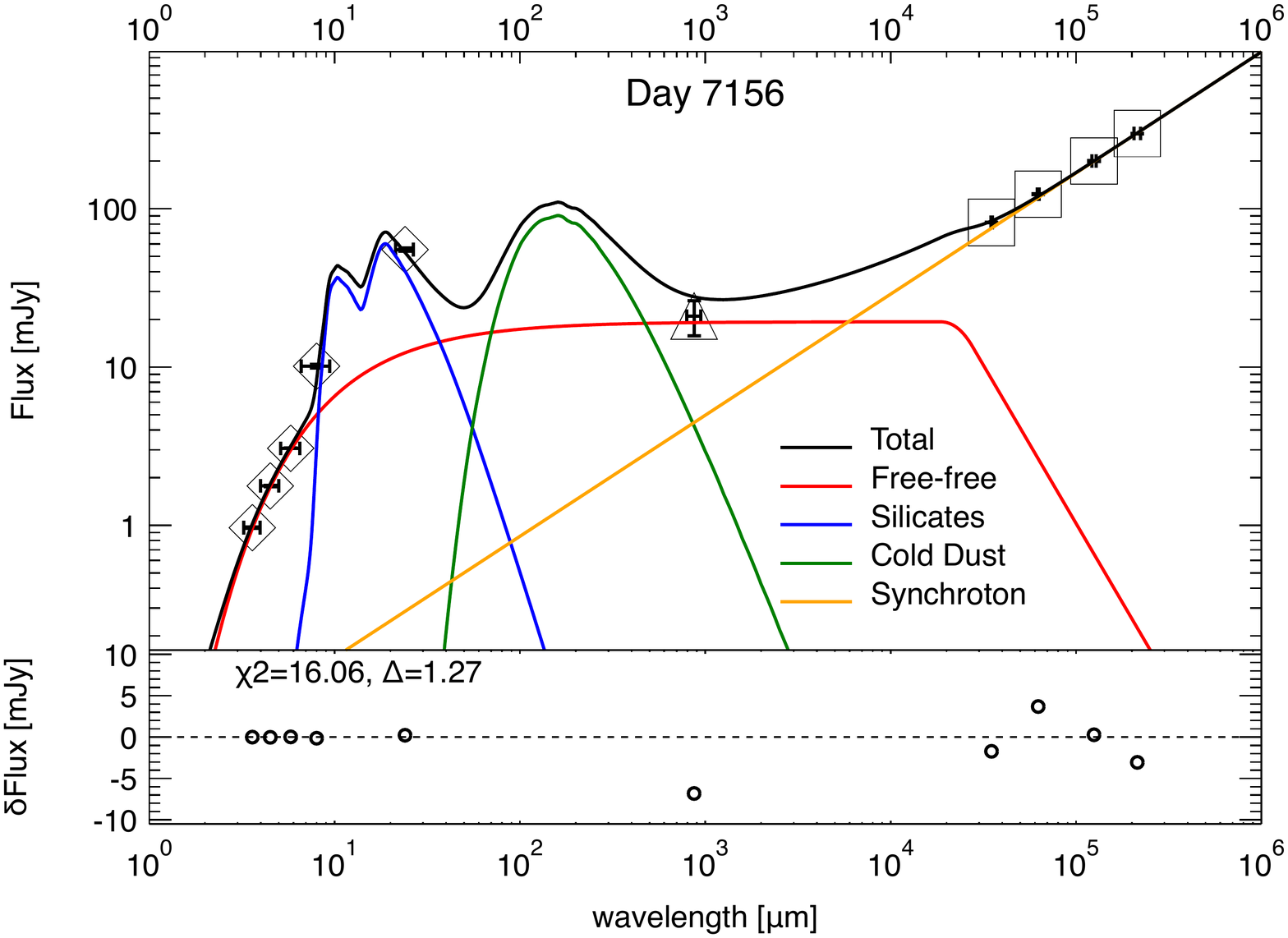 }
\includegraphics[width=5.5cm]{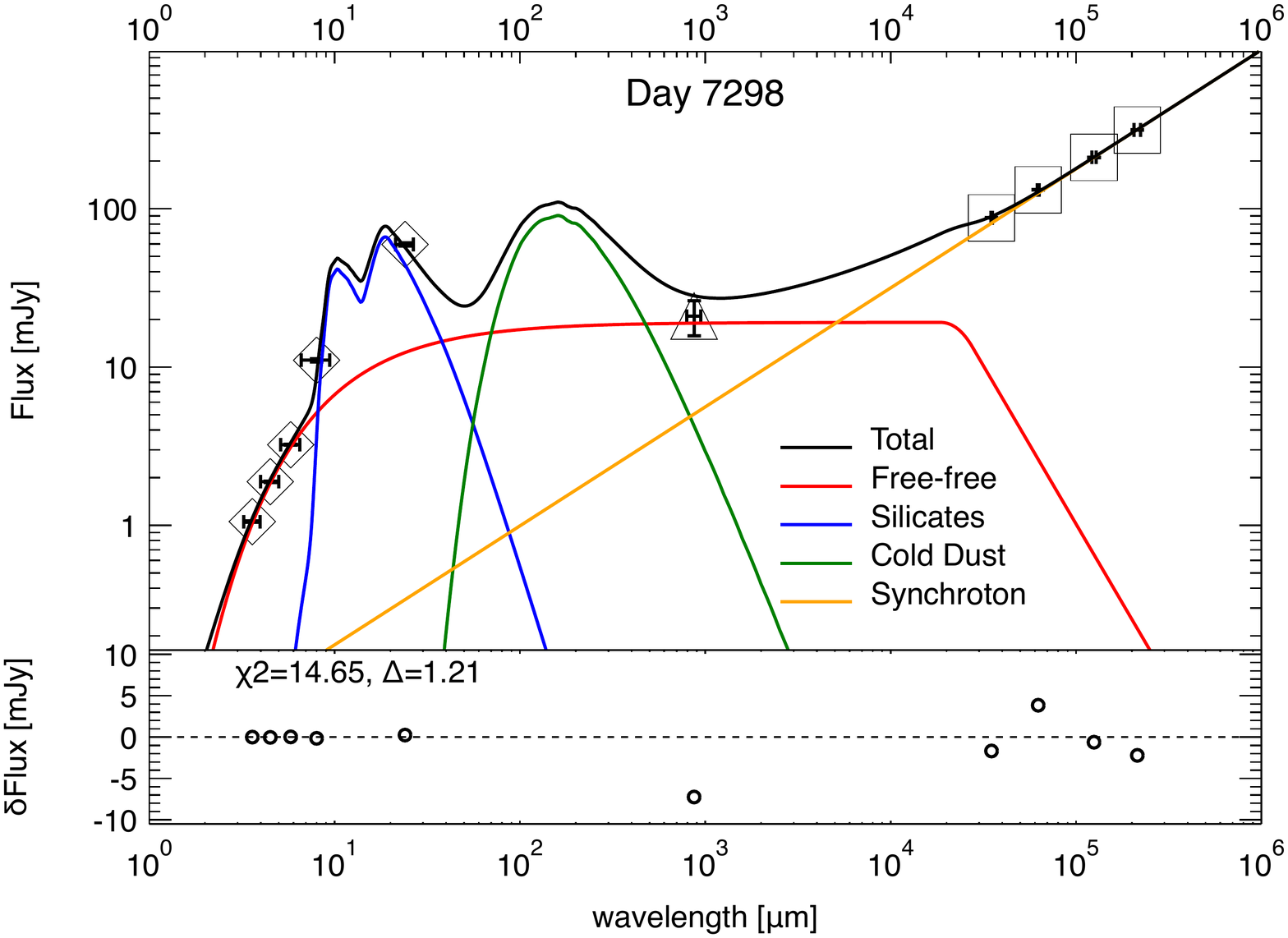 }
\includegraphics[width=5.5cm]{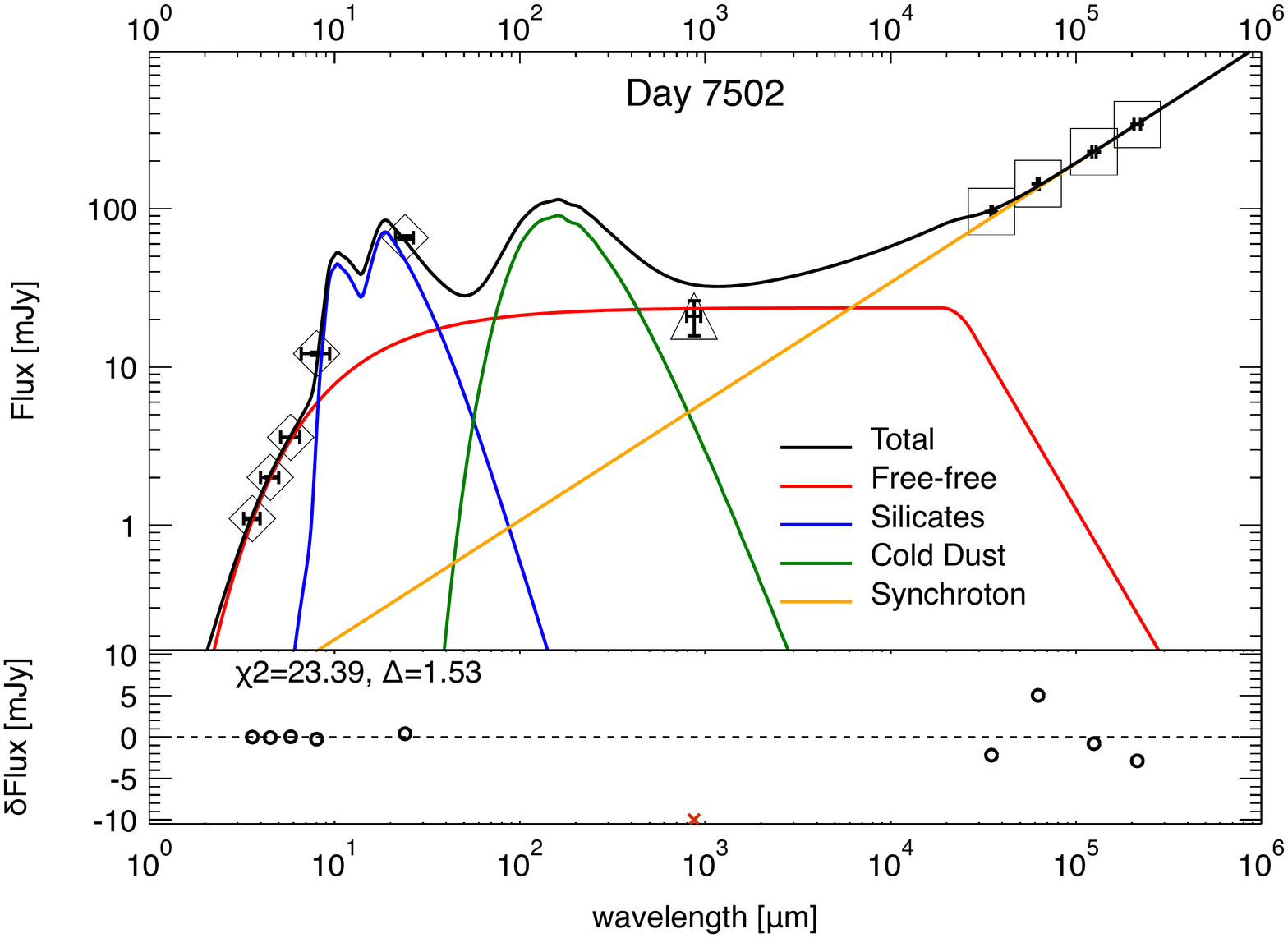 }
\includegraphics[width=5.5cm]{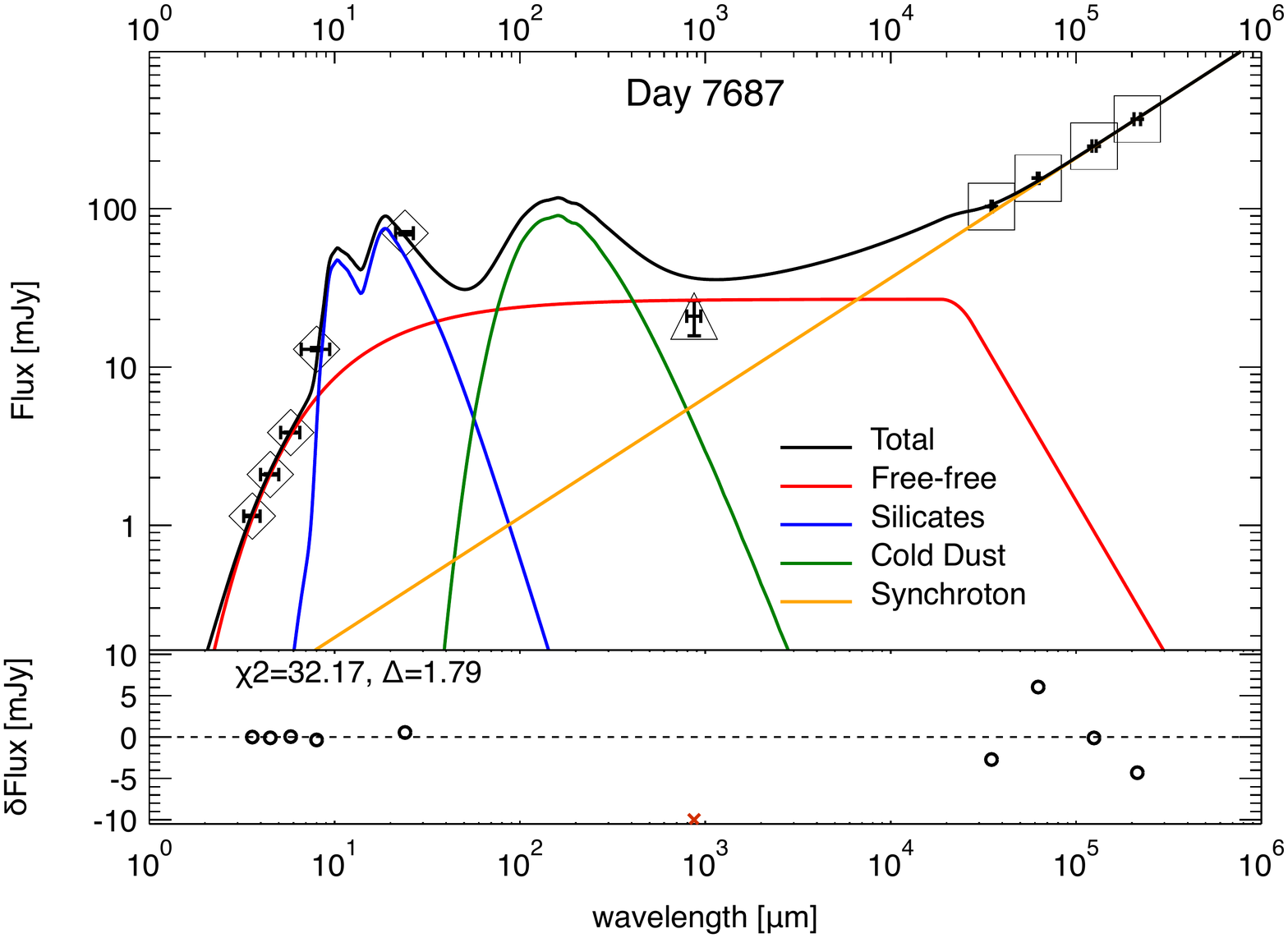 }
\includegraphics[width=5.5cm]{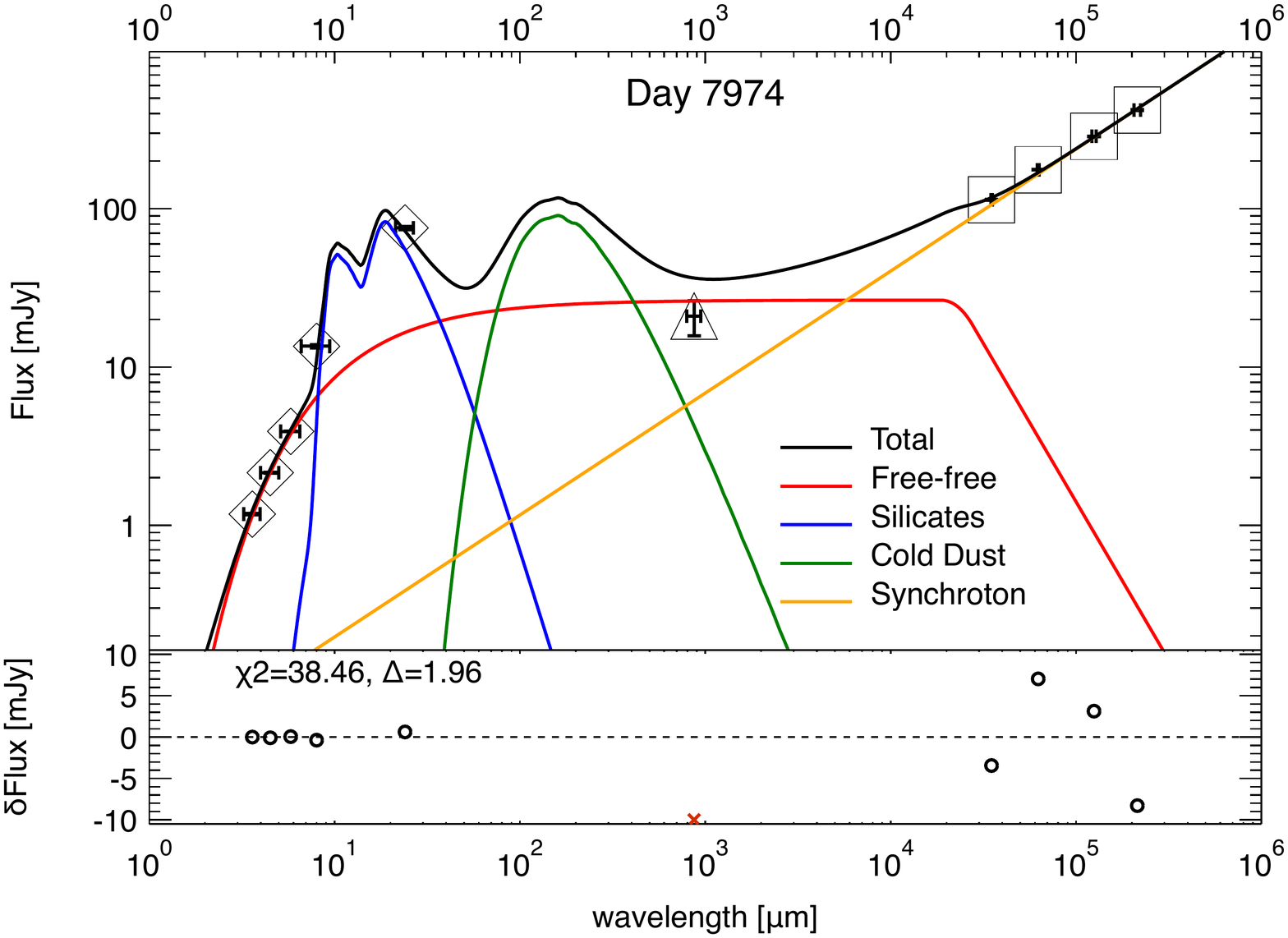 }
\includegraphics[width=5.5cm]{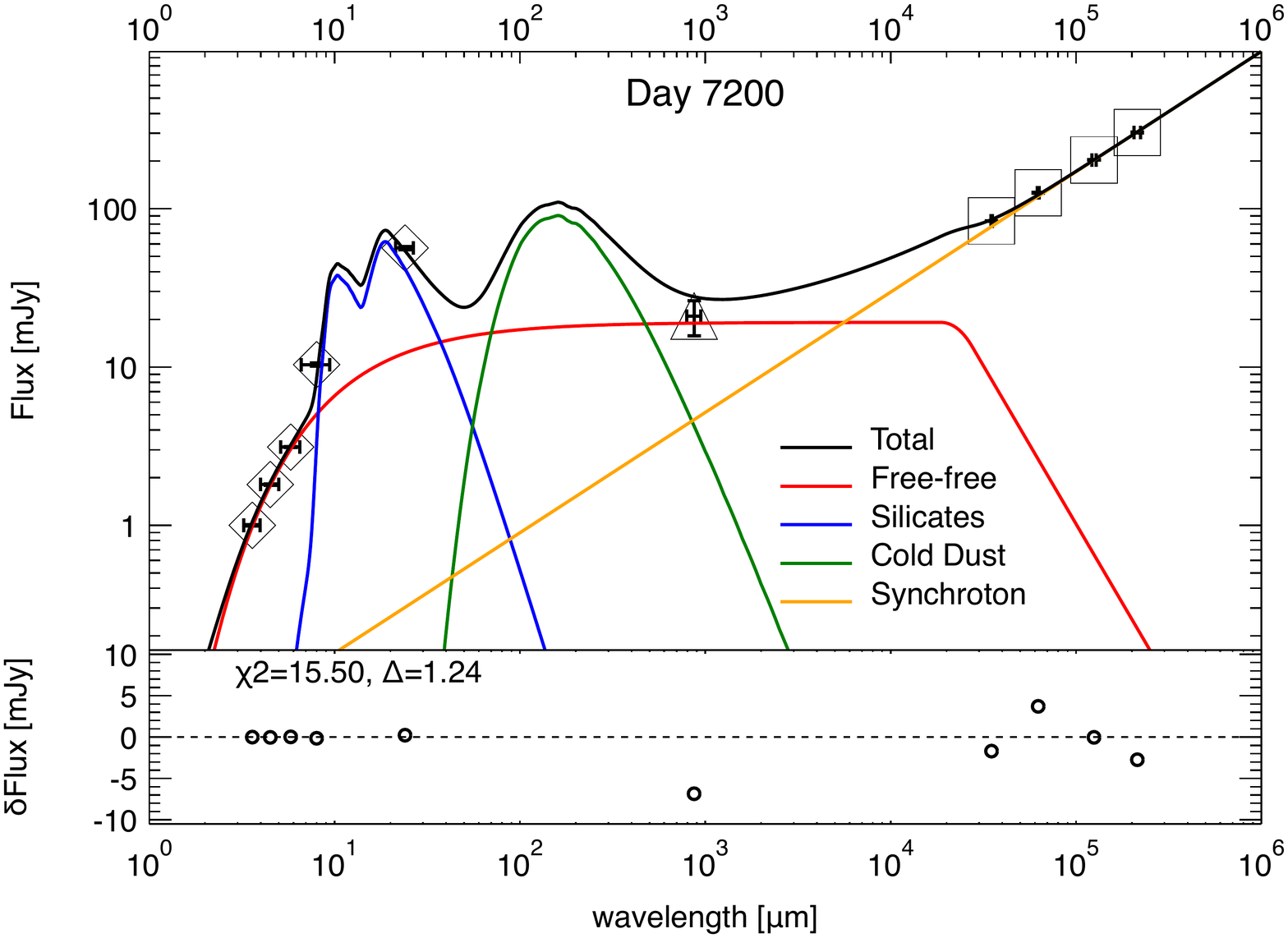 }
\includegraphics[width=5.5cm]{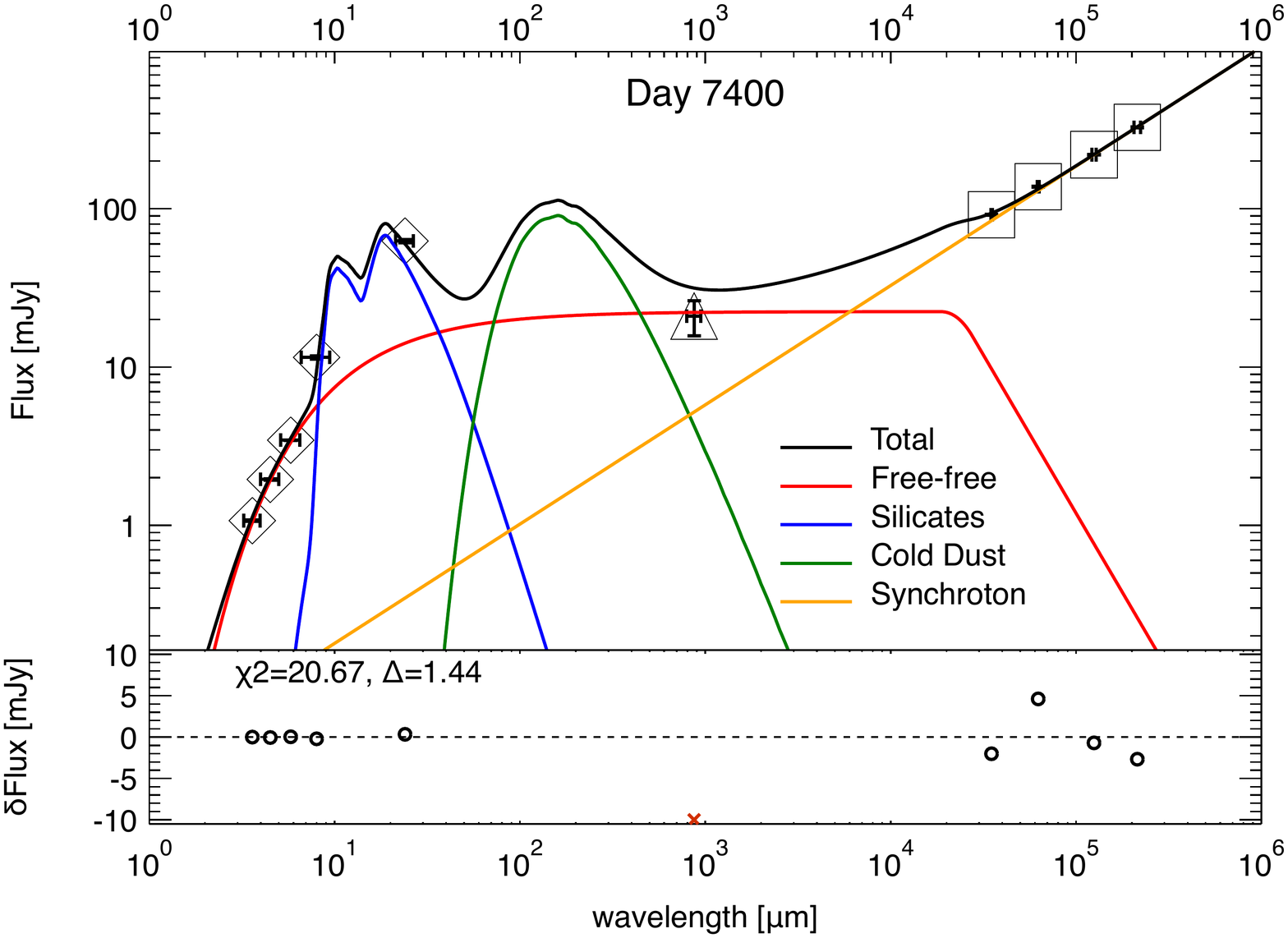 }
\end{center}
  \caption{Same as Figure~\Ref{fig:individual-9d-wc-fits} with a free-free radiation optically thick at 13~GHz instead of a warm carbon dust; symbols as in Table~\Ref{tab:telused}.The two bottom panels are the interpolated data for the two epochs analysed in more detail, shown for comparison.} 
  \label{fig:sed_9d_ffcut13}
\end{figure*}

\begin{figure*}
\begin{center}
\includegraphics[width=8.5cm]
{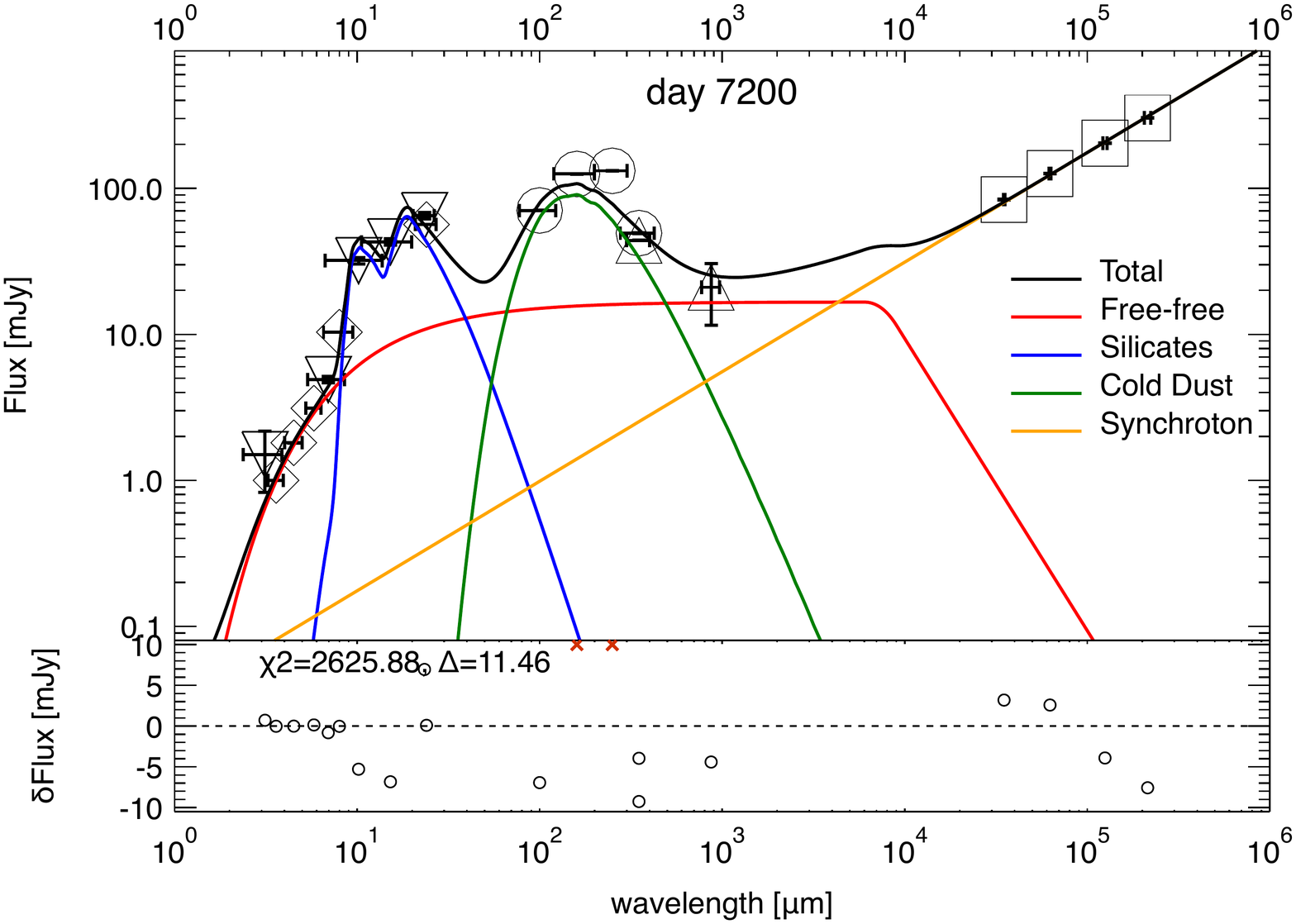}
\includegraphics[width=8.5cm]
{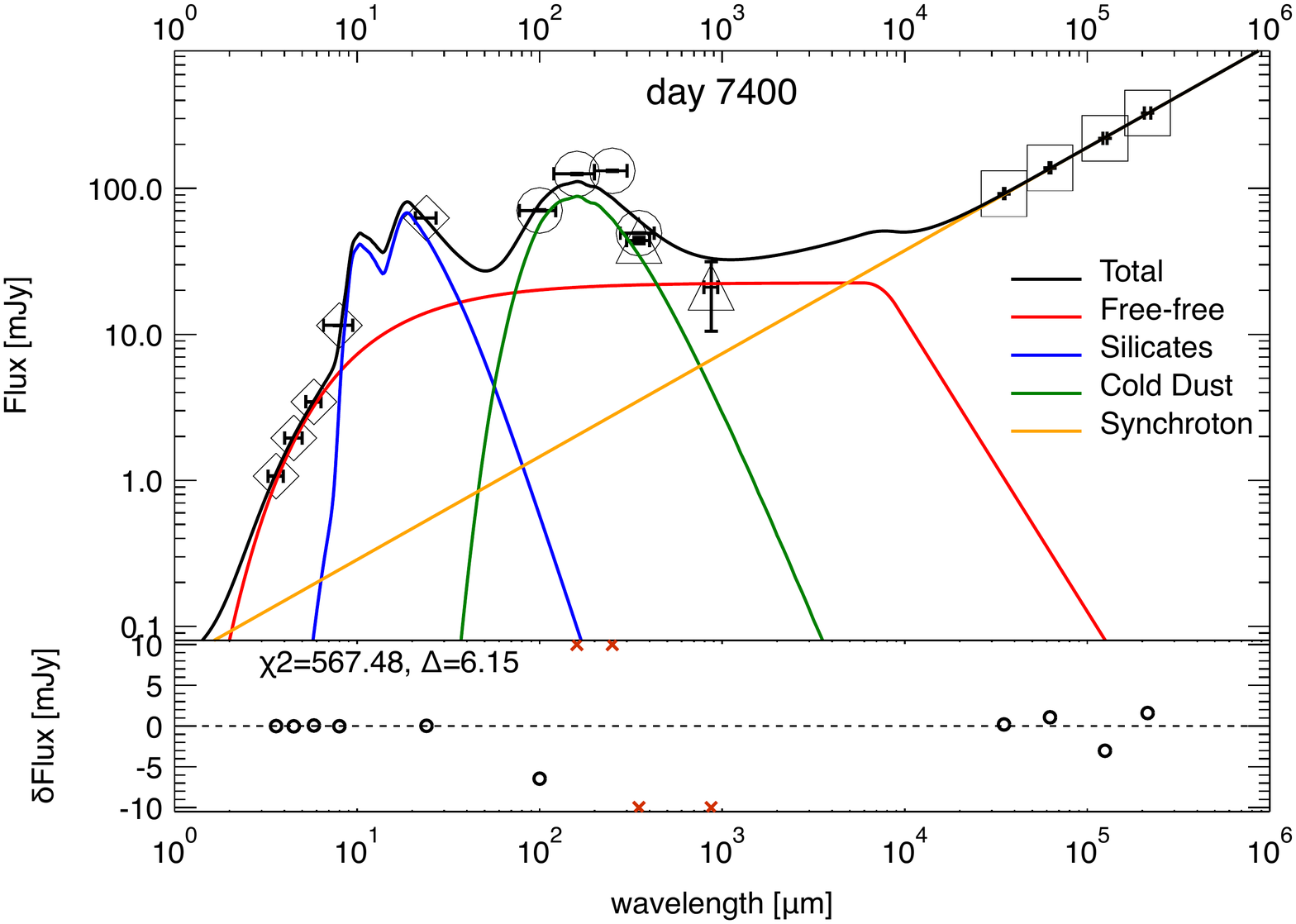} 
\end{center}
  \caption{Same as Figure~\ref{fig:sed_ffcut13} with a free-free emission self-absorbed at 40~GHz. Symbols as in Table~\Ref{tab:telused}. The two bottom panels are the interpolated data for the two epochs analysed in more detail, shown for comparison.} 
  \label{fig:sed_ffcut40}
\end{figure*}

\begin{figure*}
\begin{center}
\includegraphics[width=5.5cm]{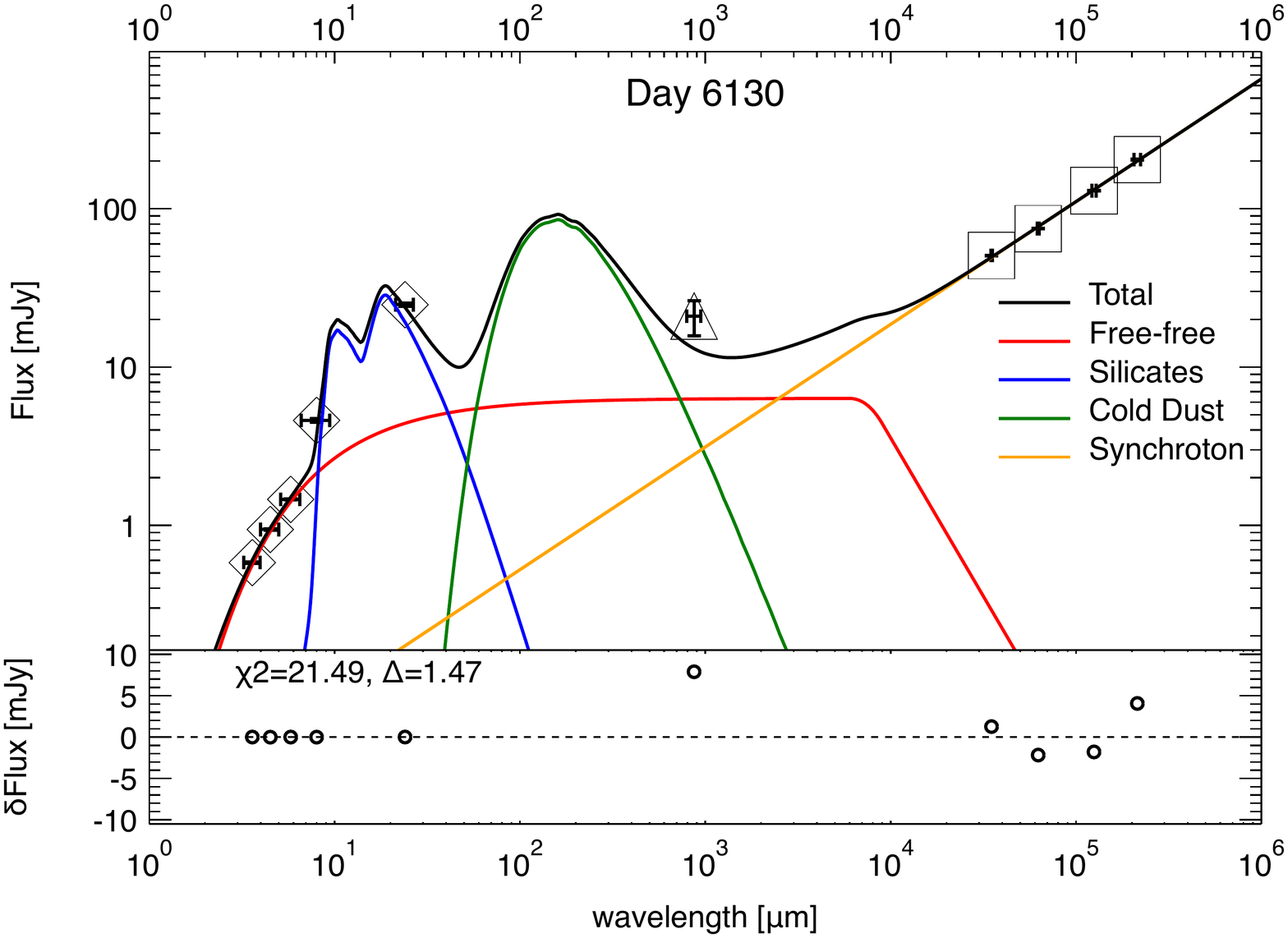 }
\includegraphics[width=5.5cm]{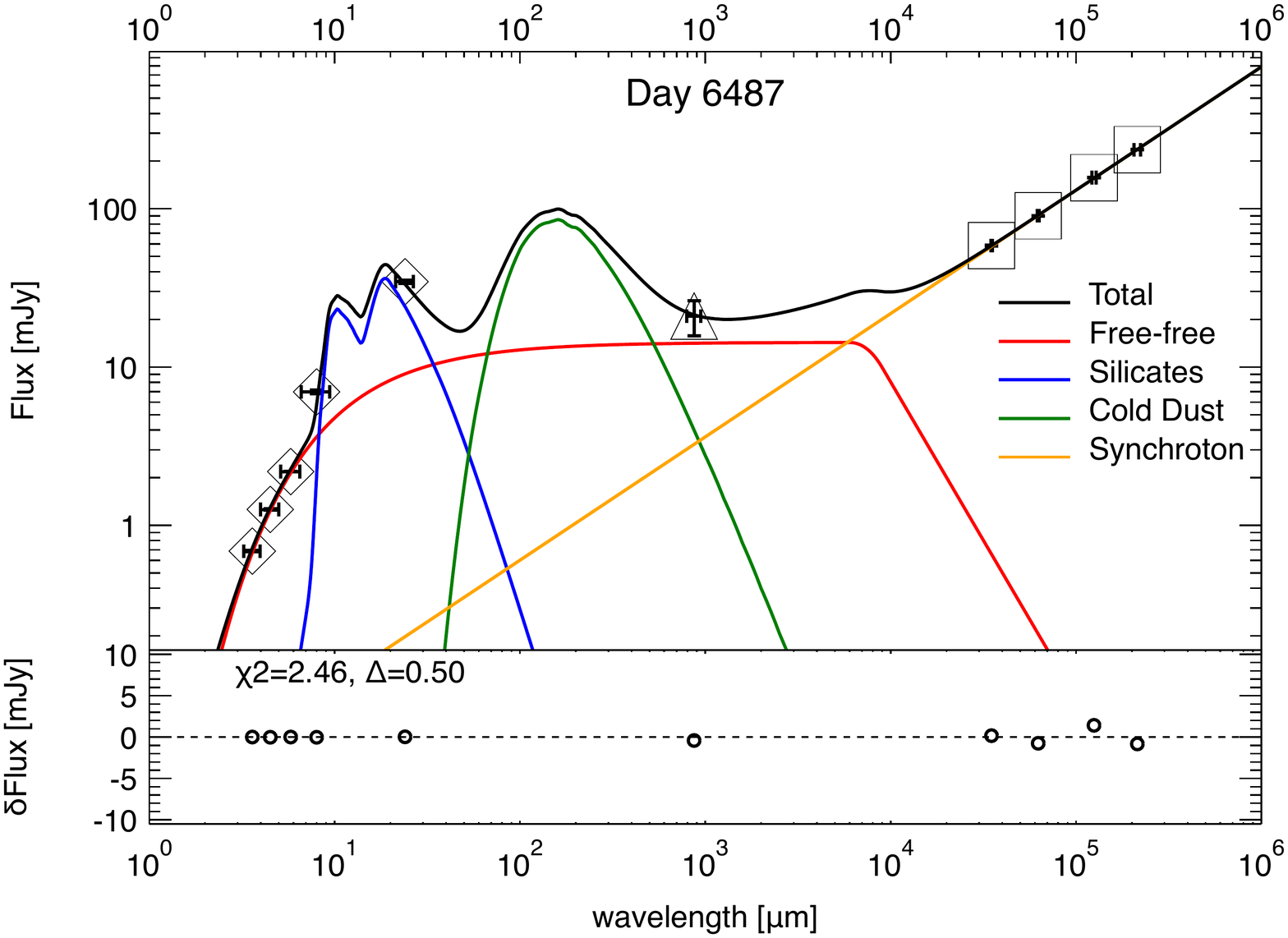 }
\includegraphics[width=5.5cm]{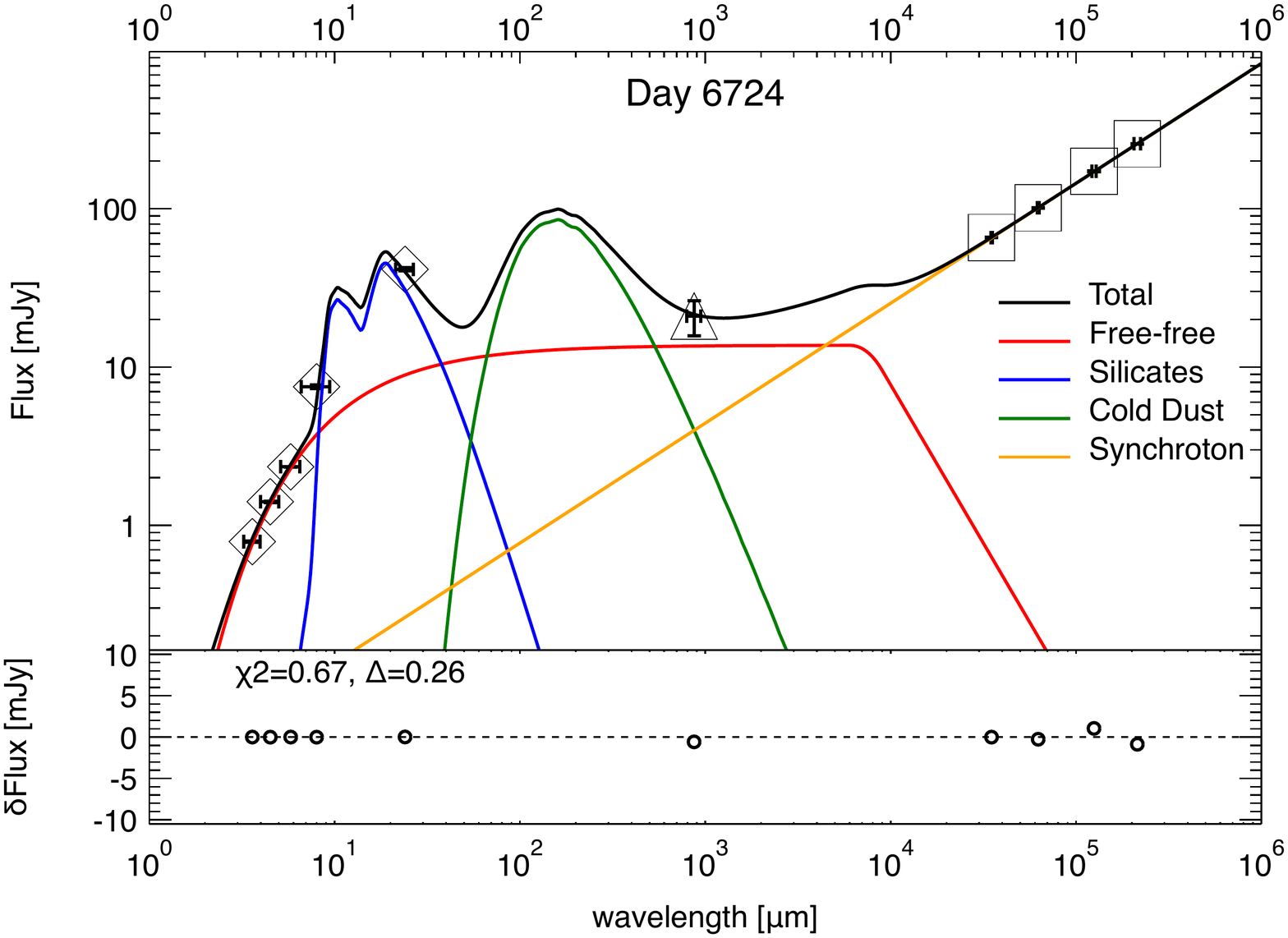 }
\includegraphics[width=5.5cm]{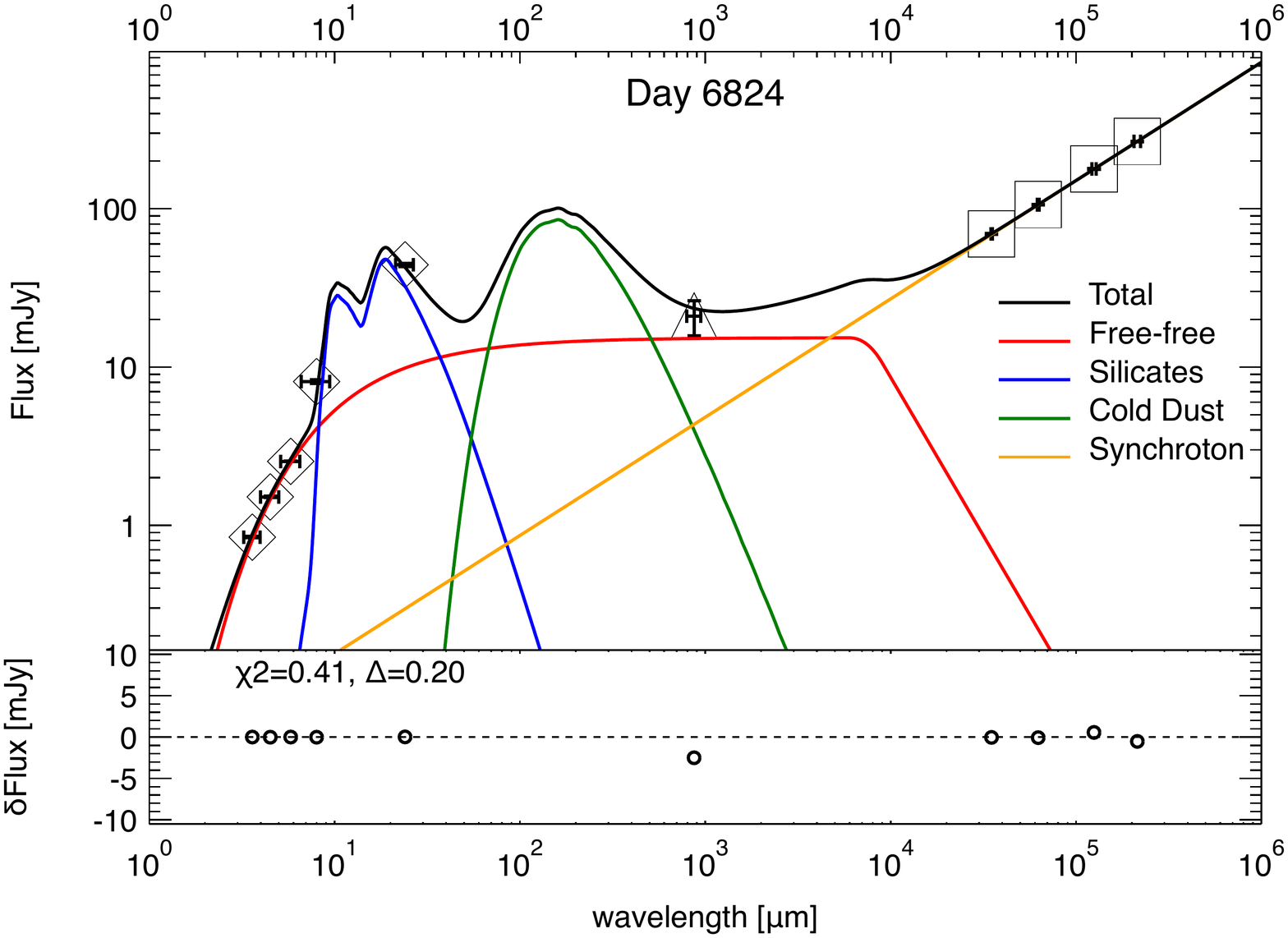 }
\includegraphics[width=5.5cm]{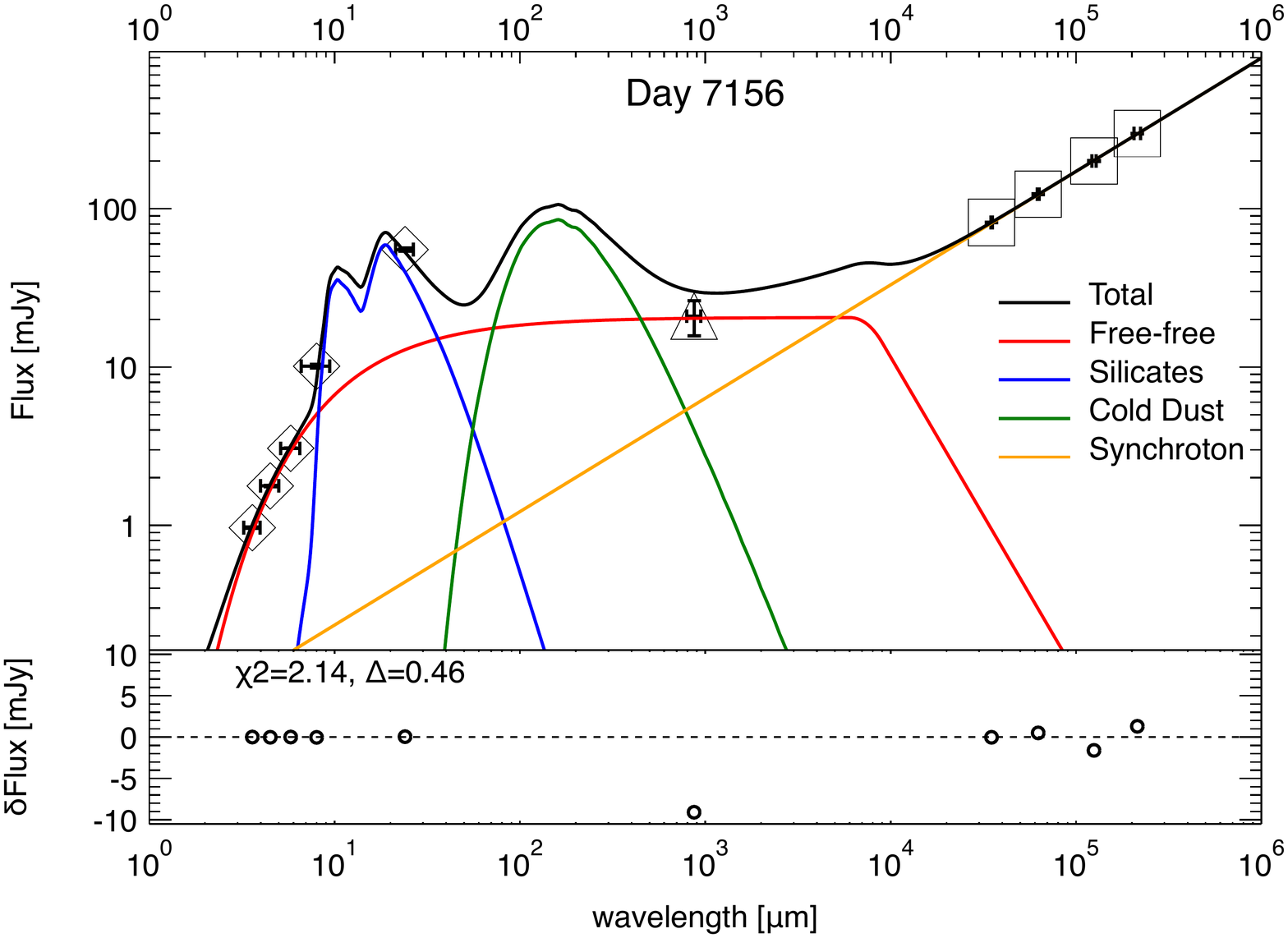 }
\includegraphics[width=5.5cm]{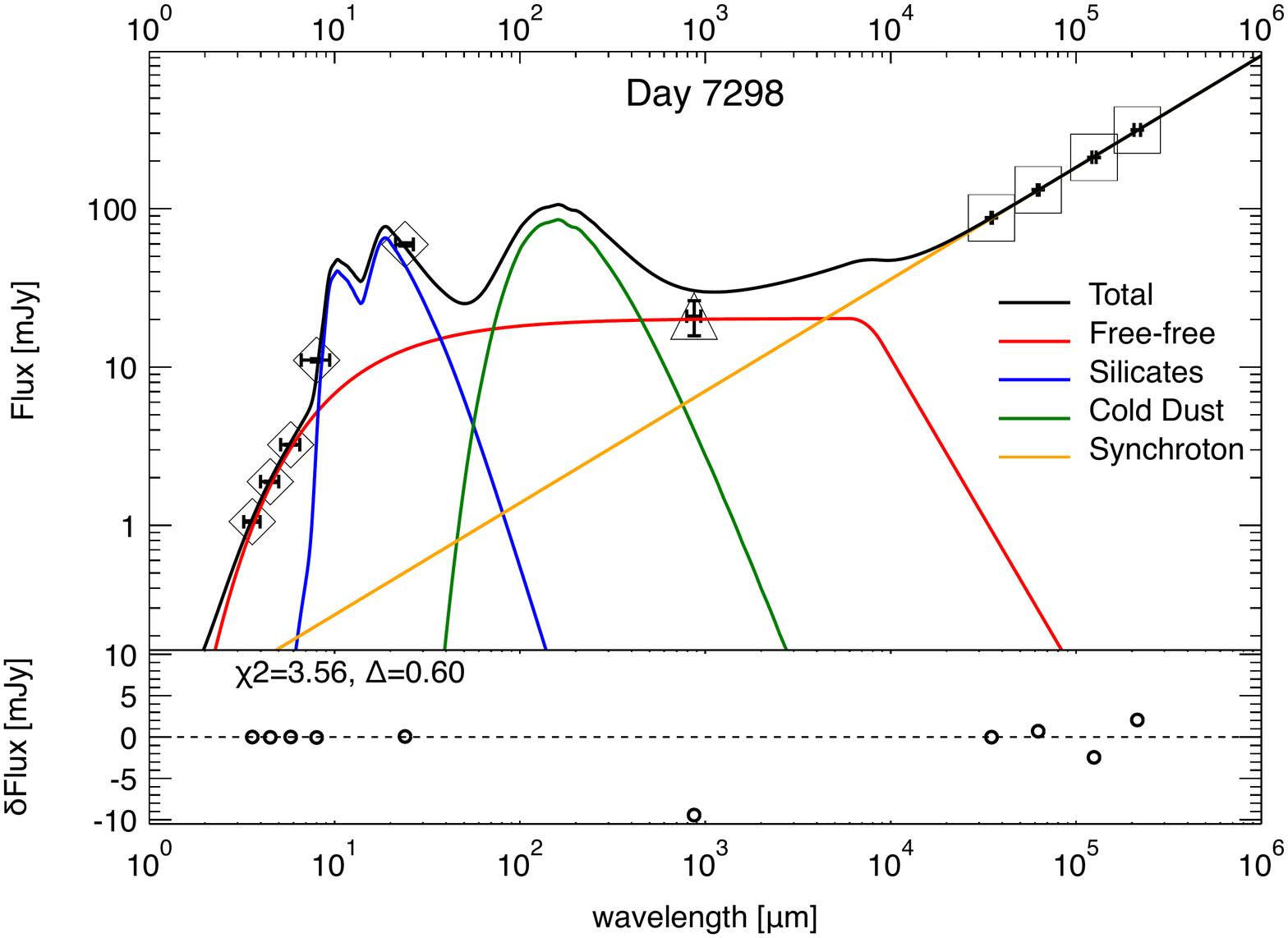 }
\includegraphics[width=5.5cm]{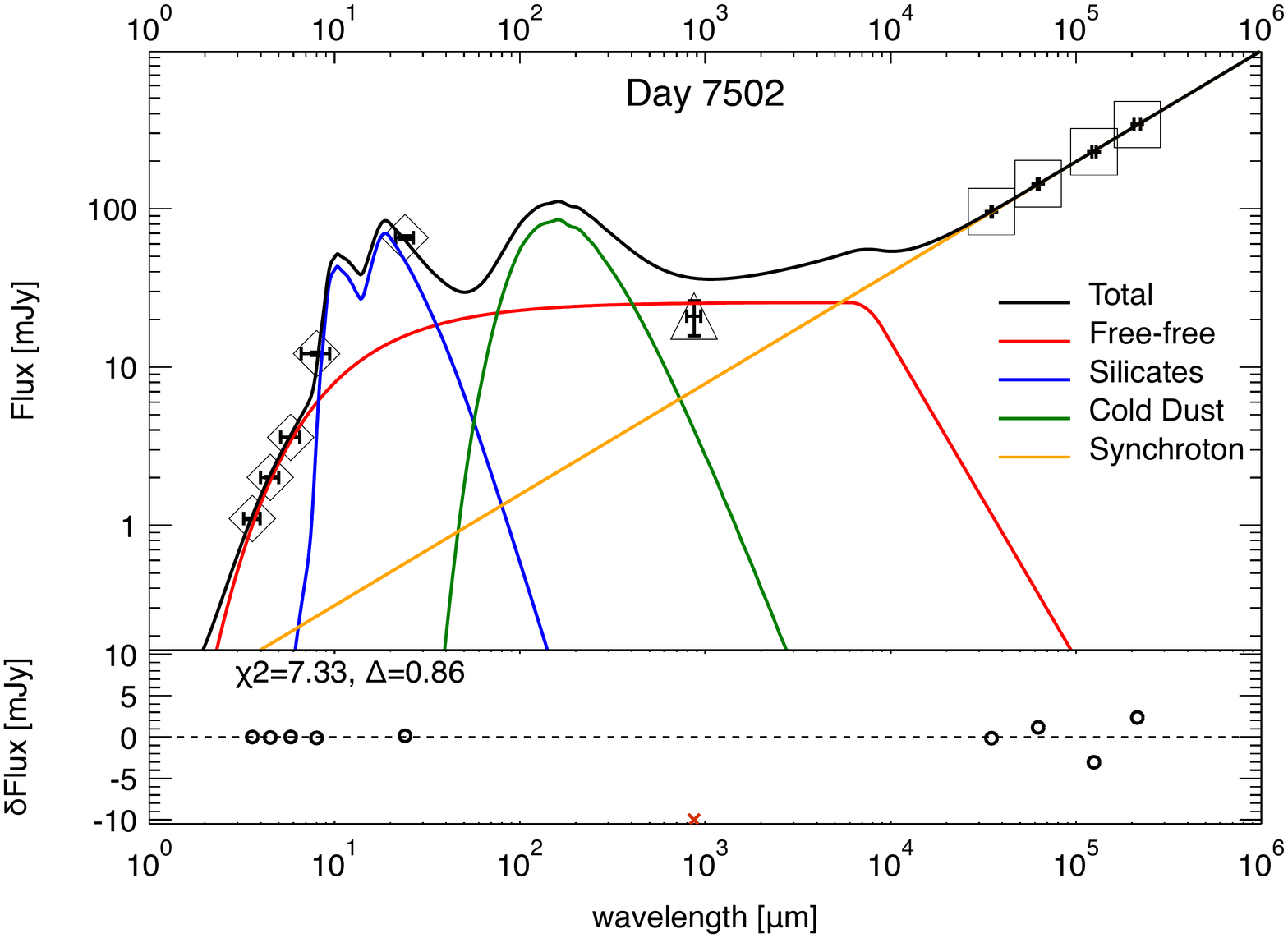 }
\includegraphics[width=5.5cm]{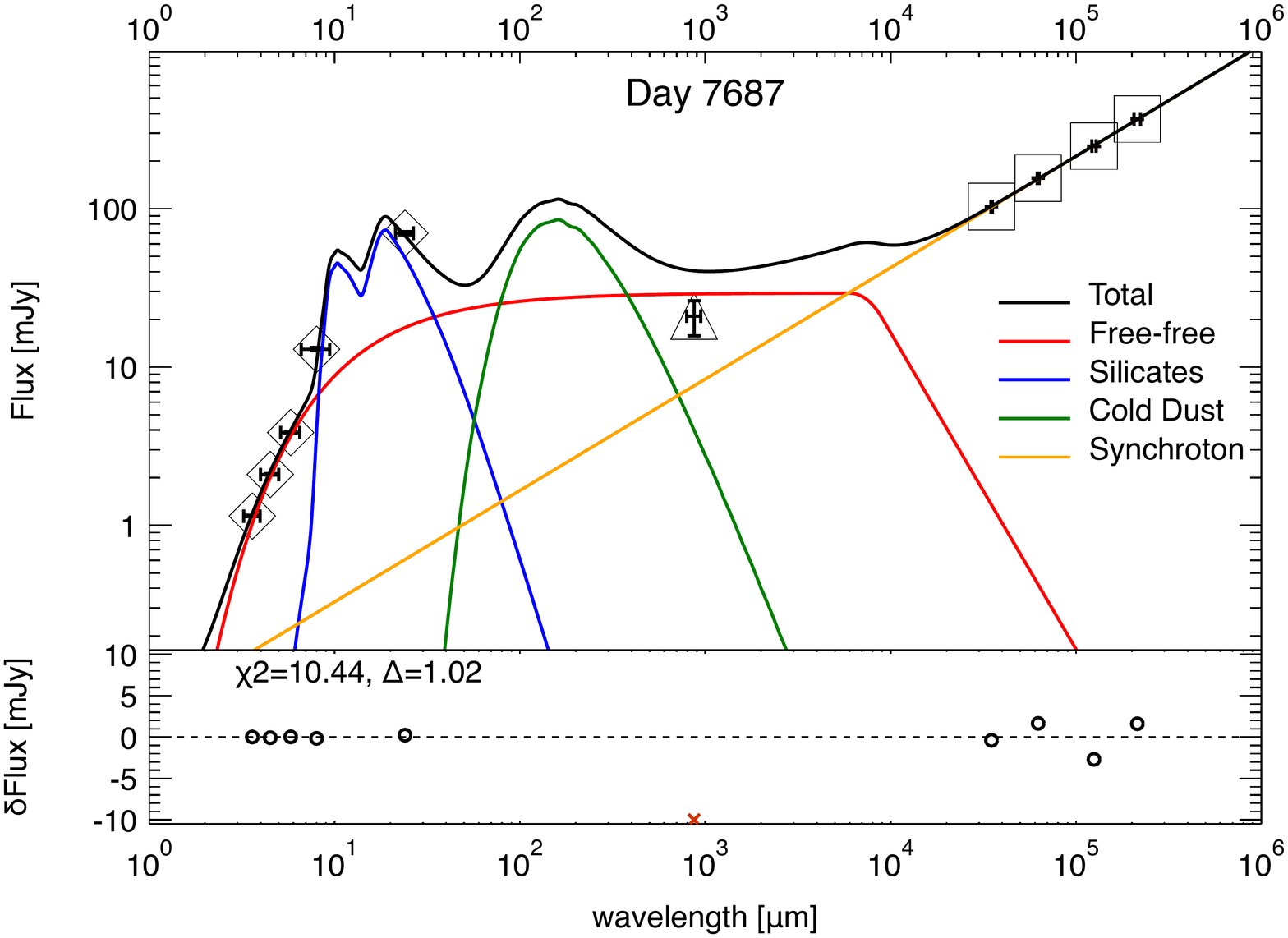 }
\includegraphics[width=5.5cm]{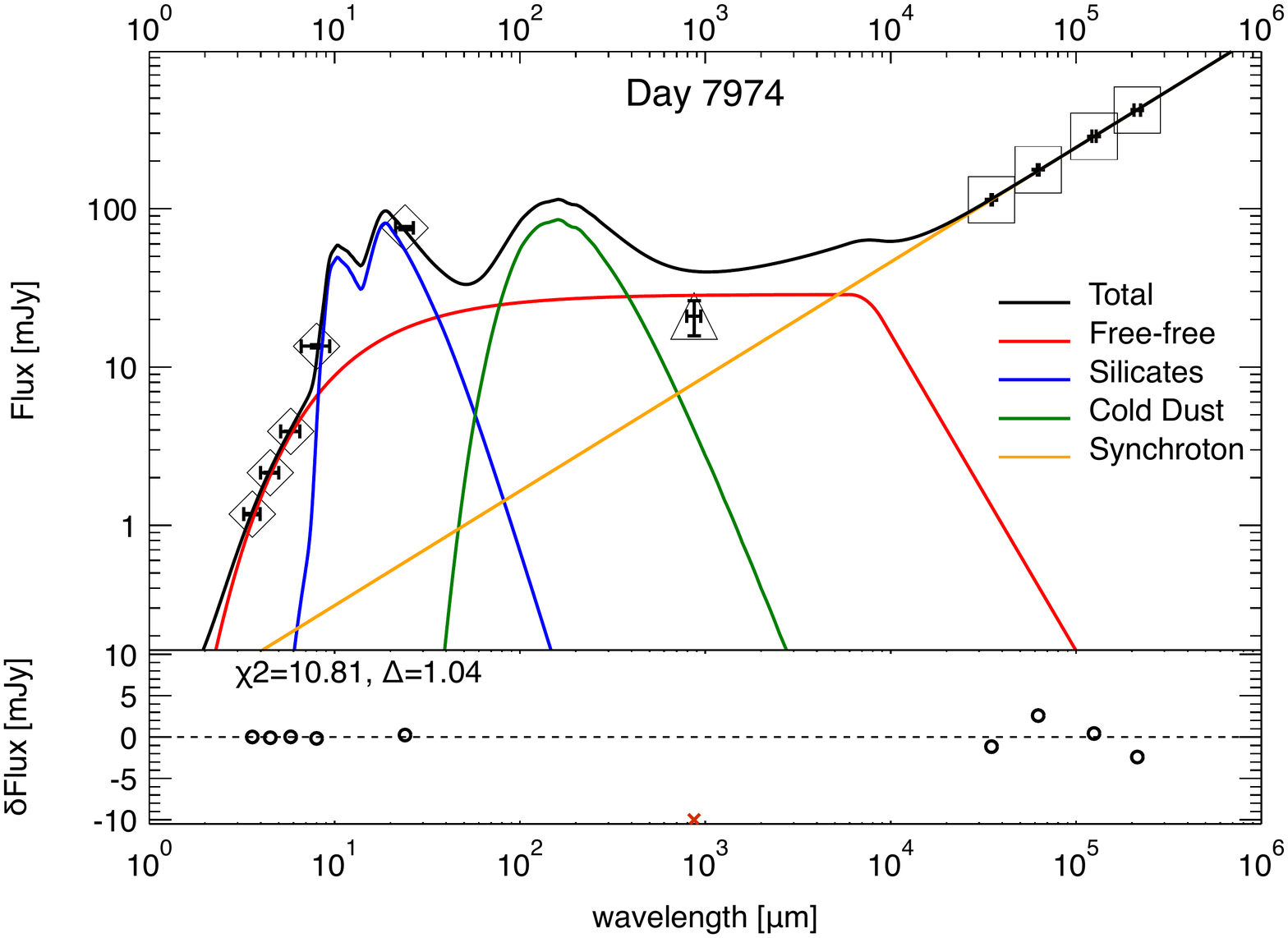 }
\includegraphics[width=5.5cm]{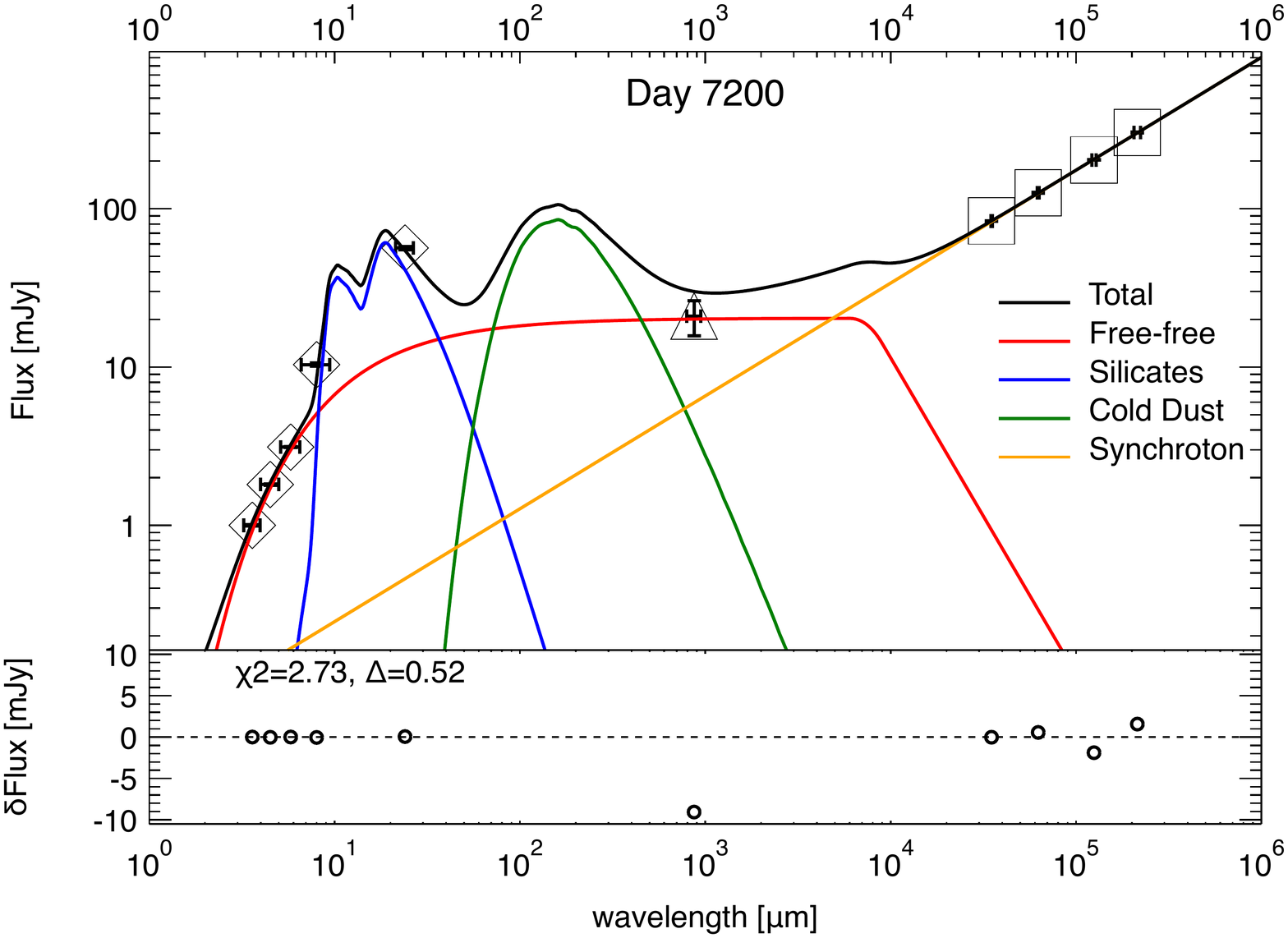 }
\includegraphics[width=5.5cm]{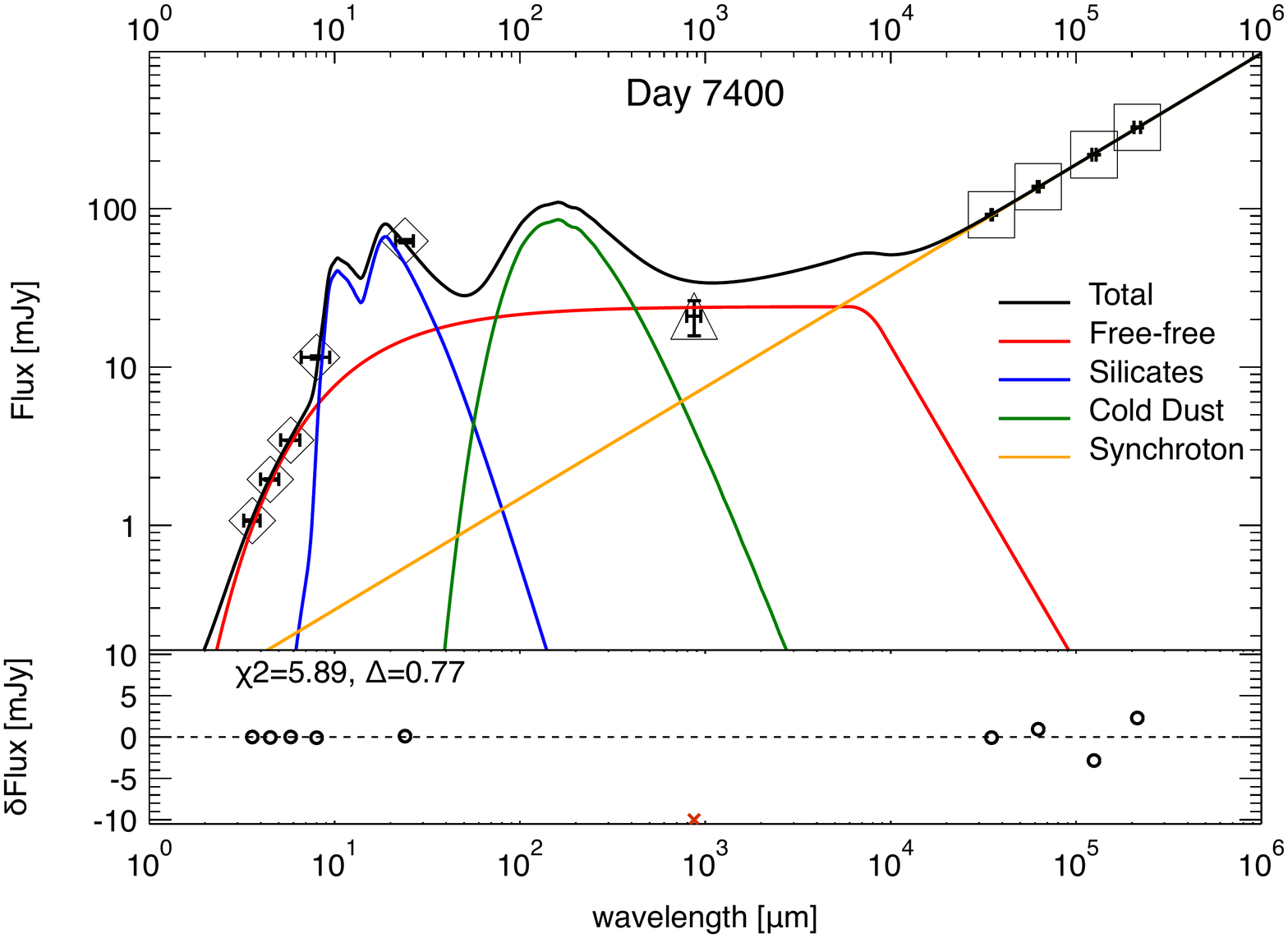 }
\end{center}
  \caption{Same as Figure~\Ref{fig:sed_9d_ffcut13} with a free-free radiation optically thick at 40~GHz; symbols as in Table~\Ref{tab:telused}.} 
  \label{fig:sed_9d_ffcut40}
\end{figure*}

\begin{figure*}
\begin{center}
  \includegraphics[width=8.5cm]{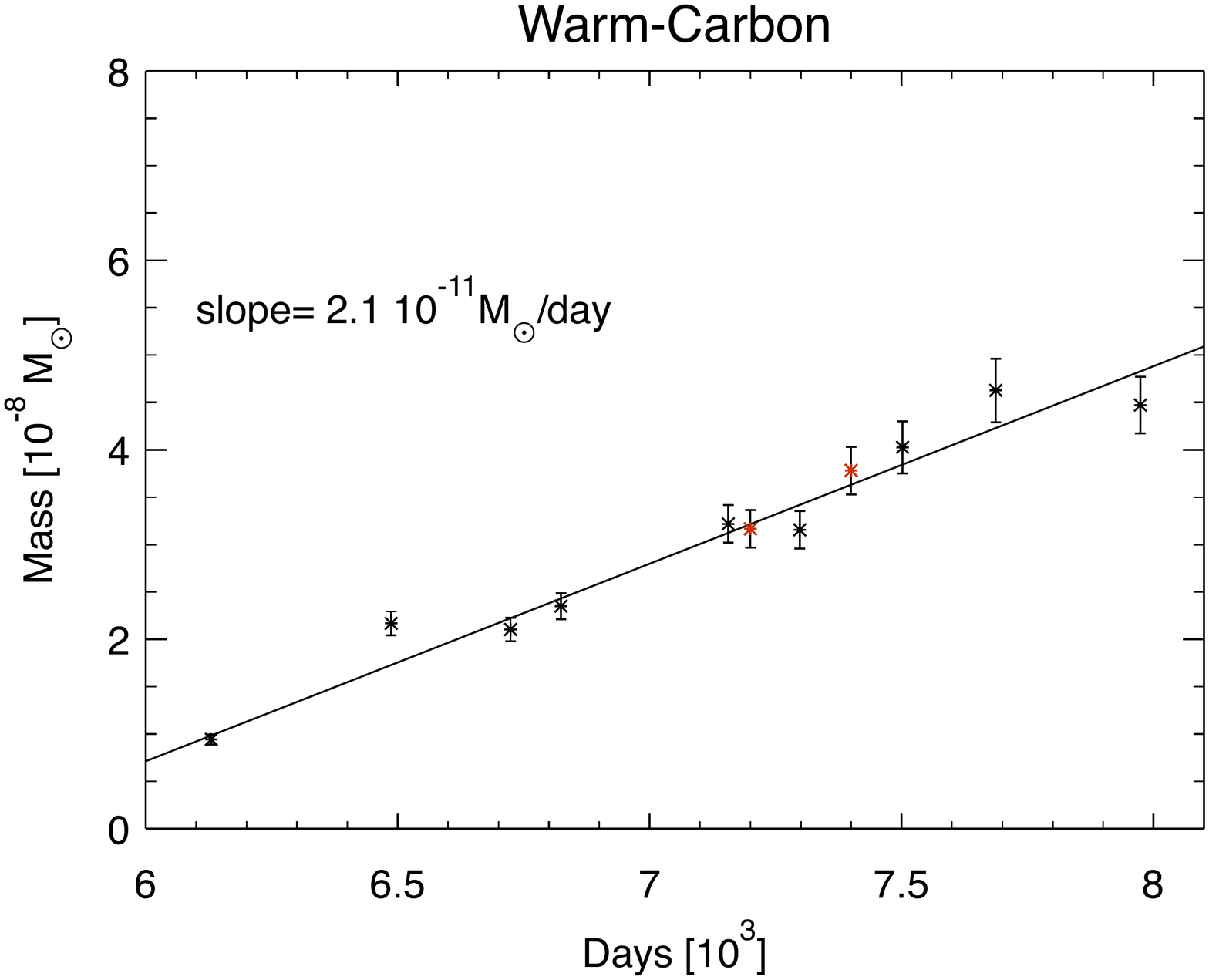}
  \includegraphics[width=8.5cm]{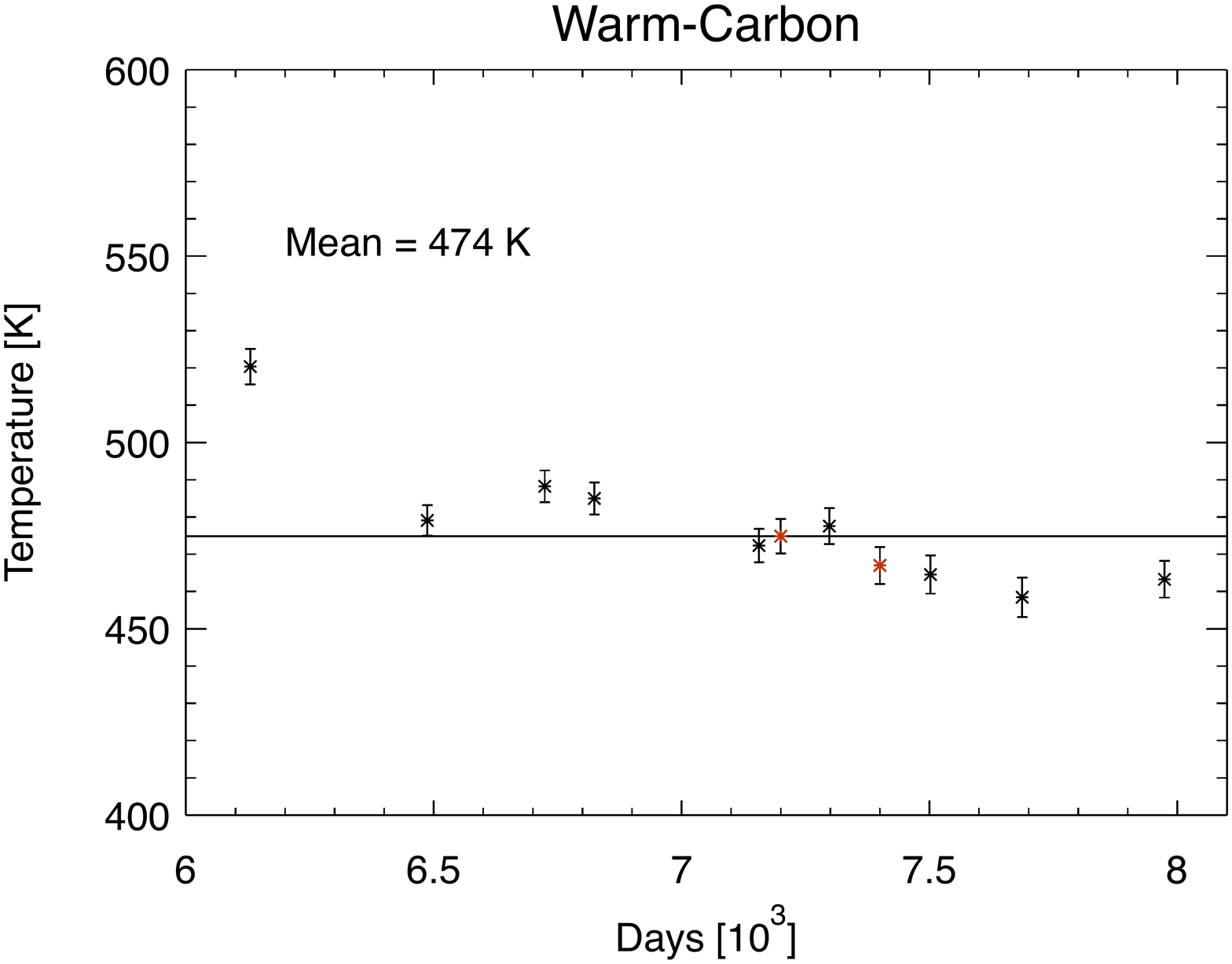}
  \end{center}
  \caption{Time evolution of the amorphous warm carbon mass (left) and temperature (right). Data points for days 7200 and 7400 are marked in red. The error bars are obtained through Monte-Carlo simulation with 20000 draws.} 
  \label{fig:evolution_warmcarbon}
\end{figure*}

\begin{figure*}
\begin{center}
 \includegraphics[width=8.5cm]{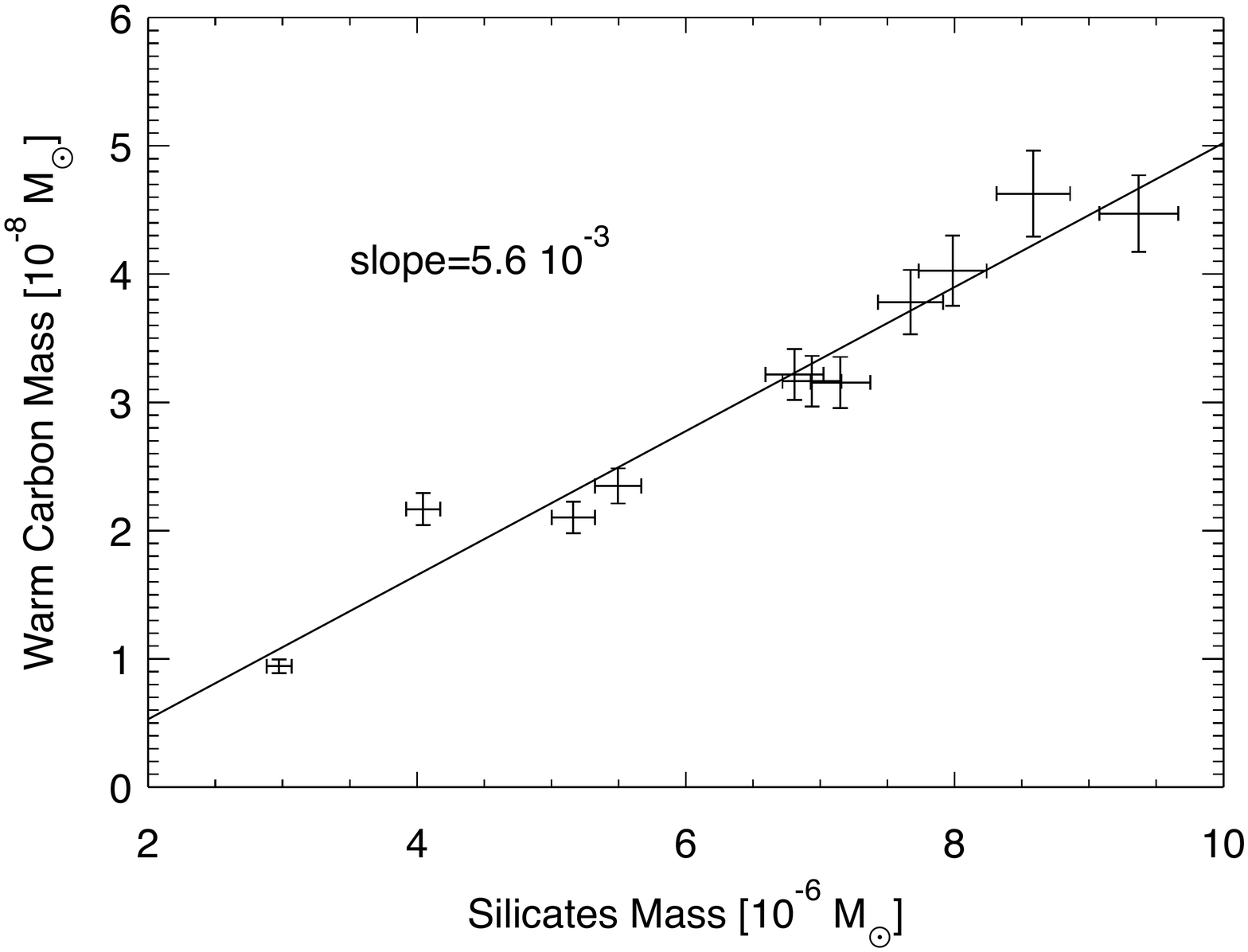}
 \end{center}
  \caption{The linear relationship between the mass of the warm carbon dust and the mass of the silicates dust during the period analyzed. } 
\label{fig:masses_wc_Si}
\end{figure*}

\begin{figure*}
\begin{center}
\includegraphics[width=8.5cm]{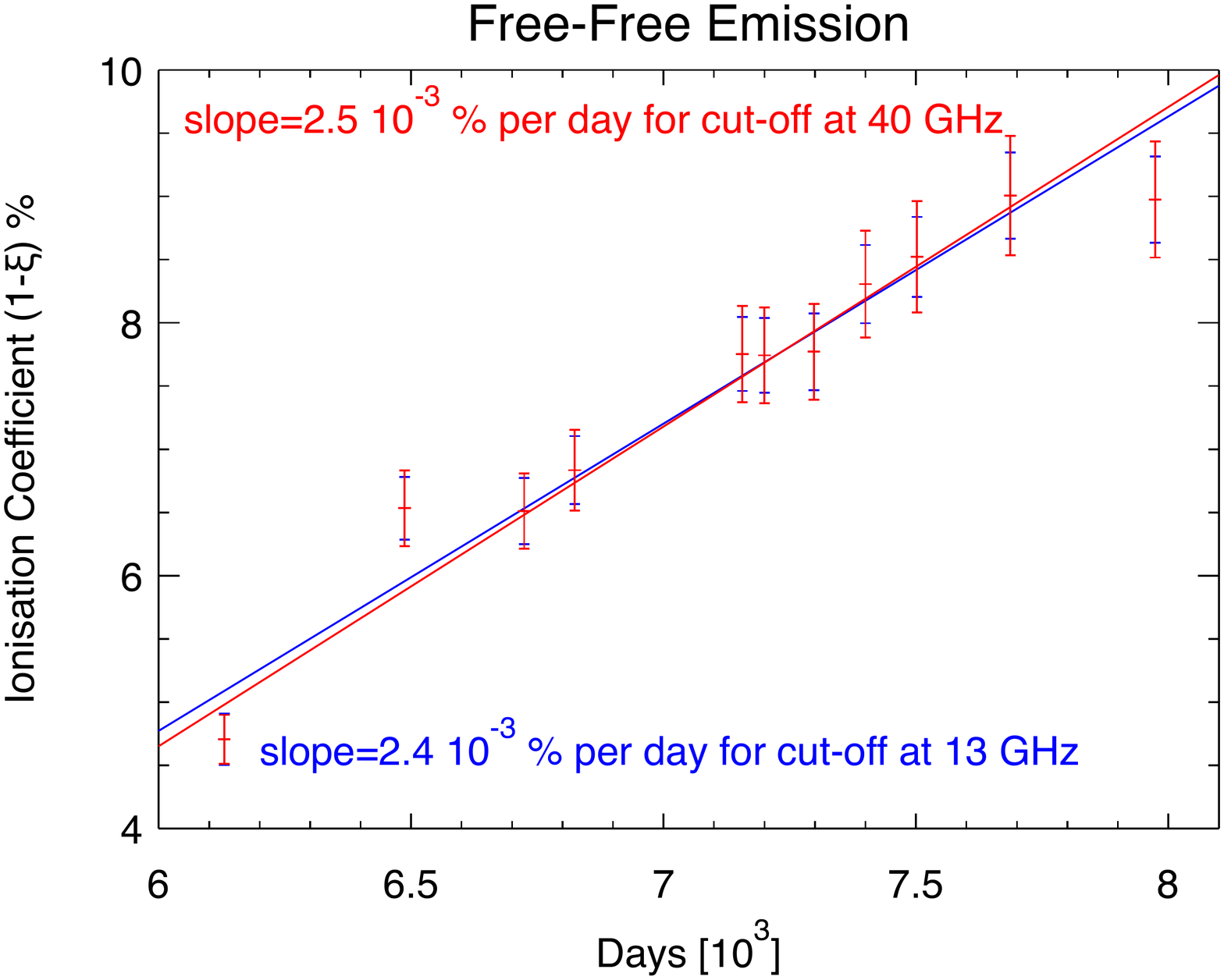}
\includegraphics[width=8.5cm]{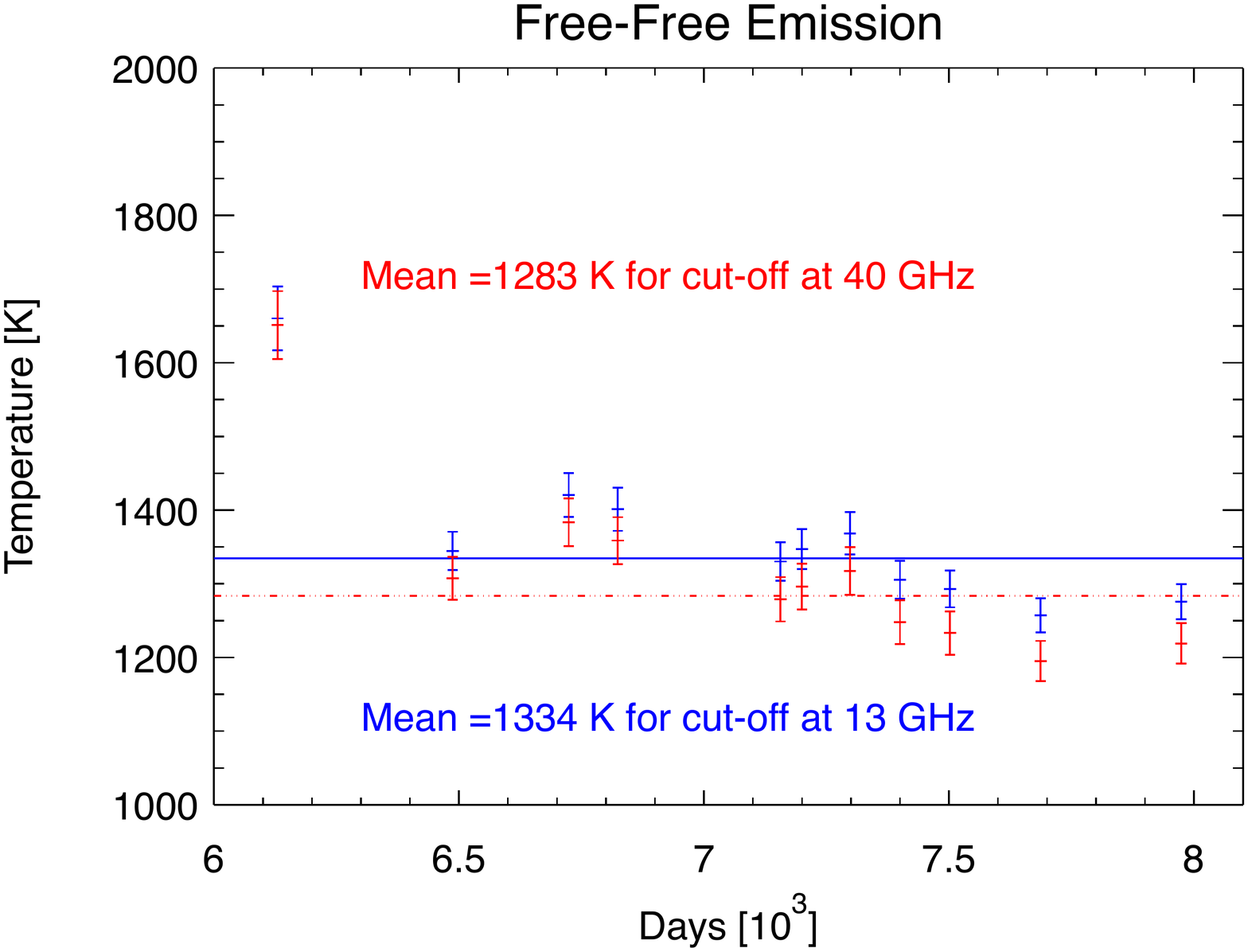}
  \end{center}
  \caption{Evolution with time of the $(1 - \xi)$ ionization coefficient (left) and of the free-free temperature (right) in the two cases self-absorbed. Inside the uncertainties due to our fitting procedure, the ionization coefficients are identical for both cut-off cases, while the temperature is slightly higher in the 13~GHz case. The error bars are obtained through Monte-Carlo simulation with 20000 draws.}
  \label{fig:evolution_ff}
\end{figure*}

\begin{figure*}
\begin{center}
  \includegraphics[width=8.5cm] {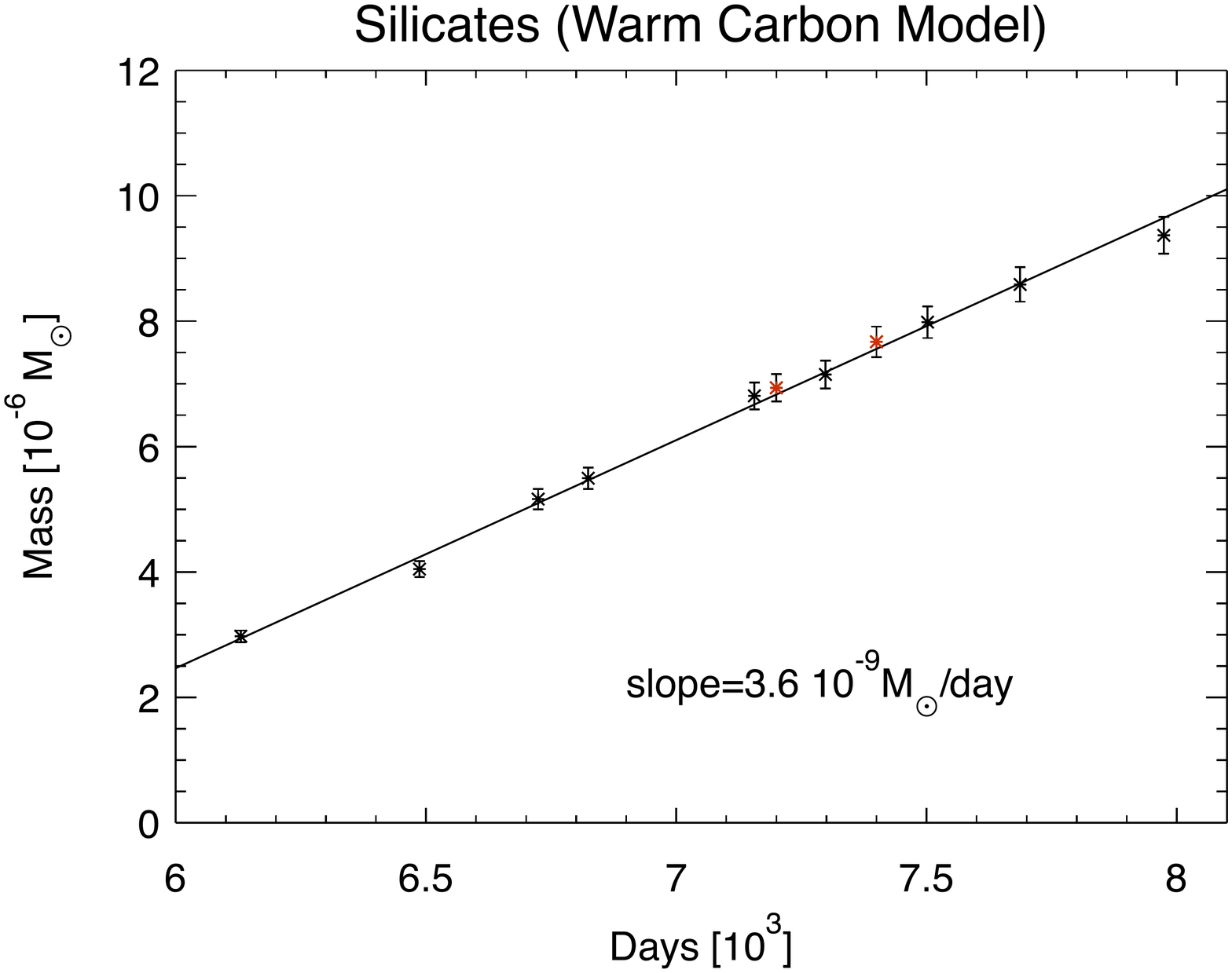}
  \includegraphics[width=8.5cm]{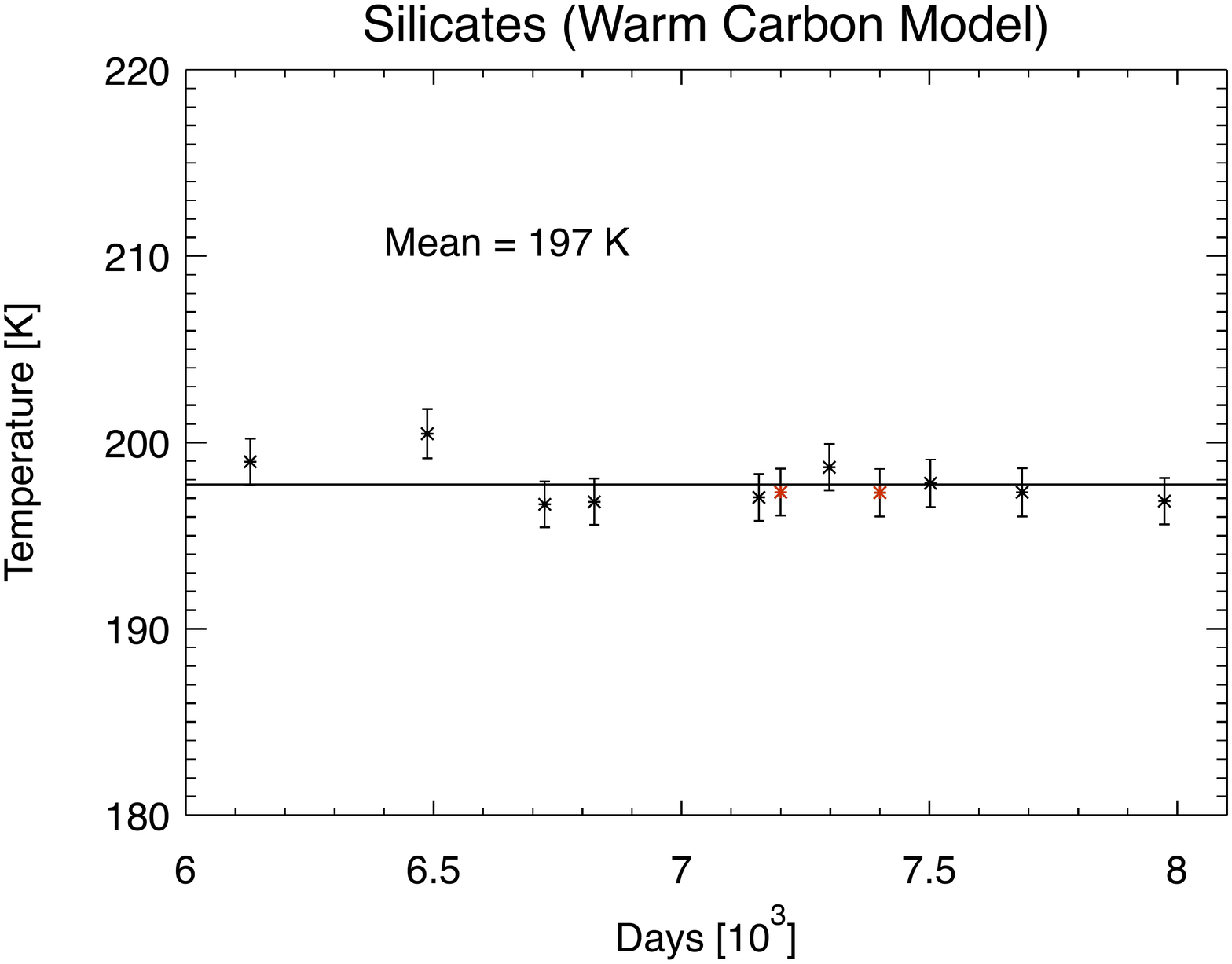}
  \includegraphics[width=8.5cm]
{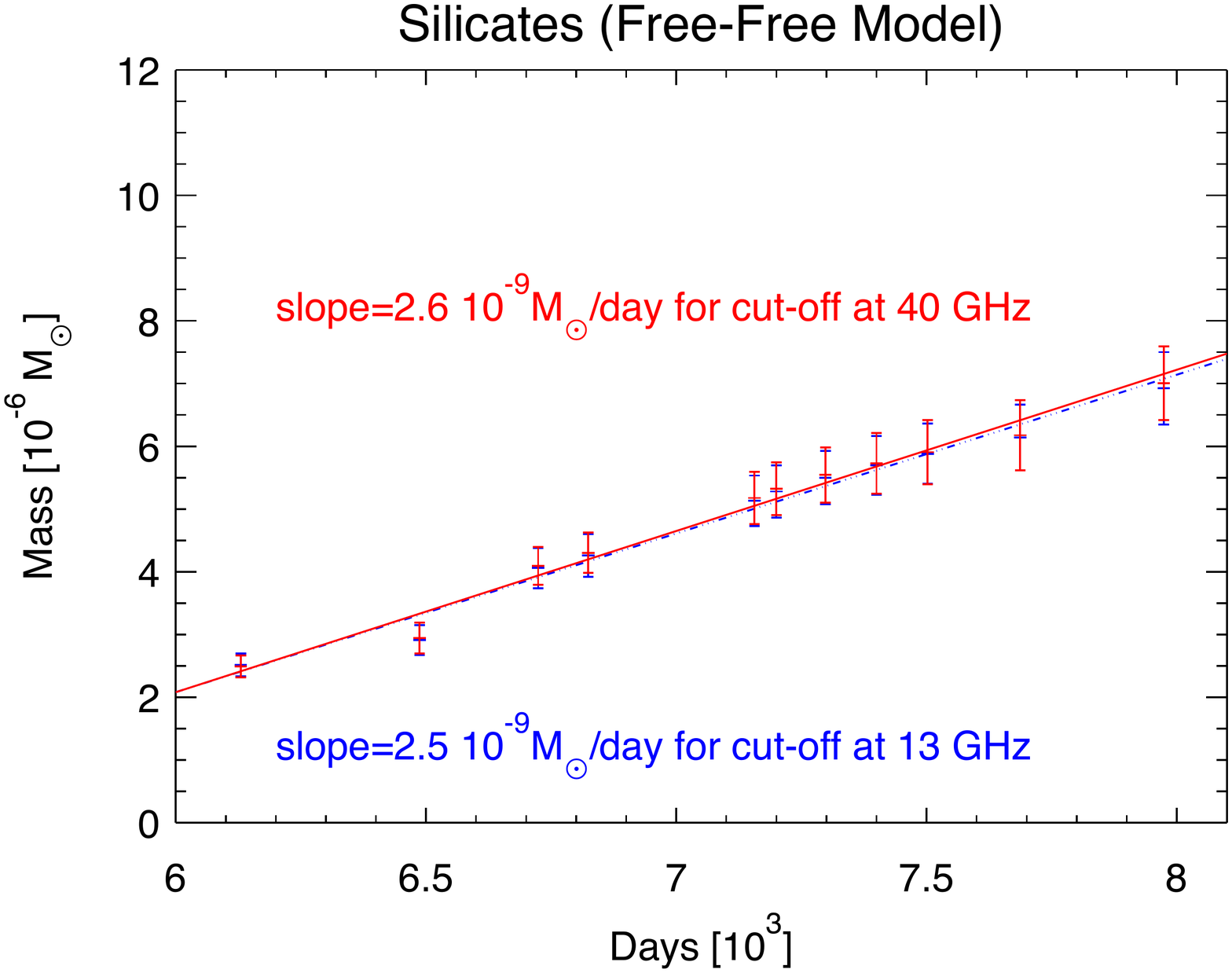}
\includegraphics[width=8.5cm]{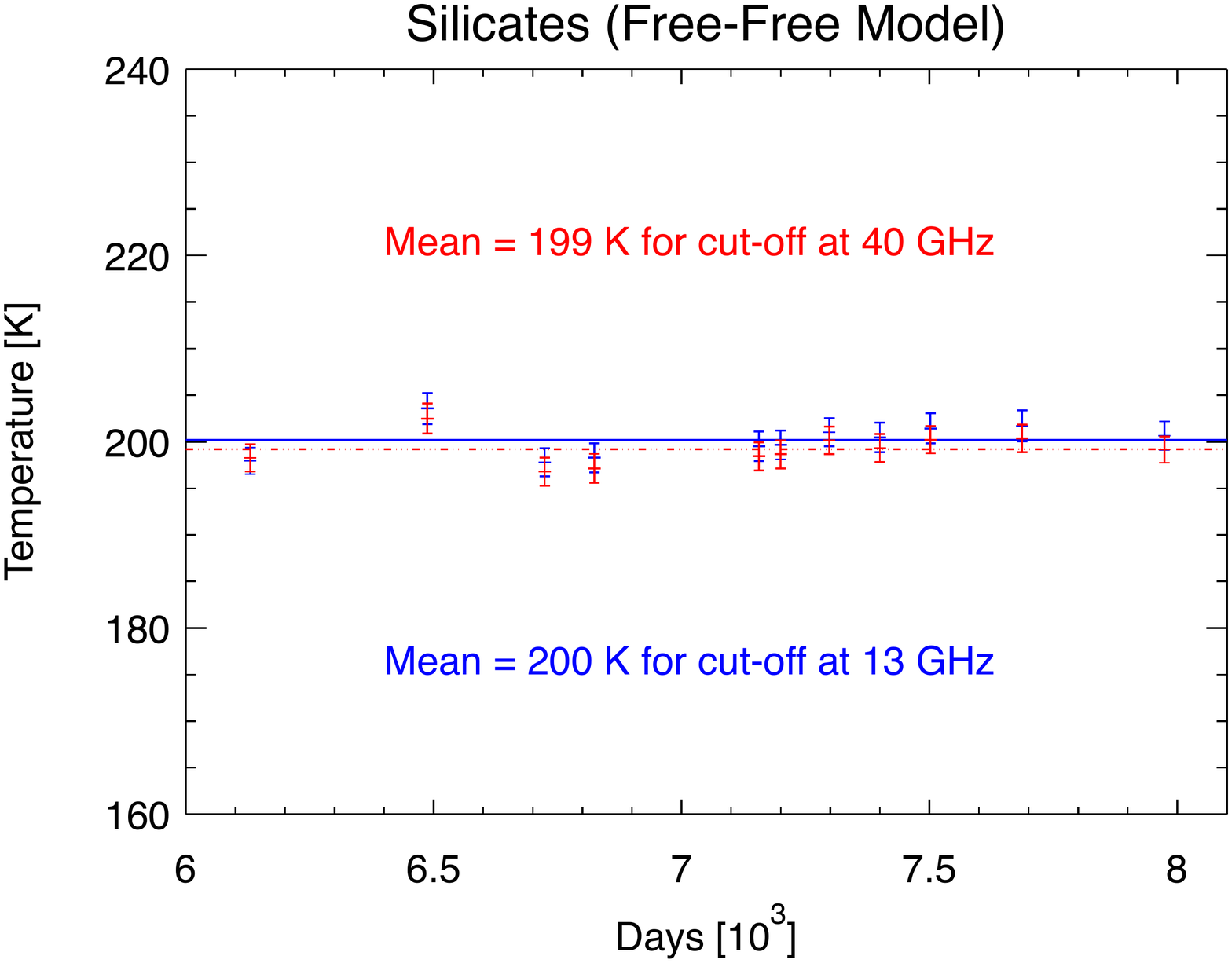}
  \end{center}
  \caption{Time evolution of the silicates dust mass (left) and temperature (right) in the case of a warm carbon dust component (top) and in the case of a free-free radiation (bottom) with a 13~GHz cut-off (blue, plain line) and a 40~GHz cut-off (red, doted line). These parameters are identical for both cut-off frequencies of the free-free emission, inside the uncertainties due to our fitting procedure. The error bars are obtained through Monte-Carlo simulation with 20000 samples.} 
  \label{fig:evolution_silicates}
\end{figure*}

\begin{figure}
\begin{center}
\includegraphics[width=8.5cm]
{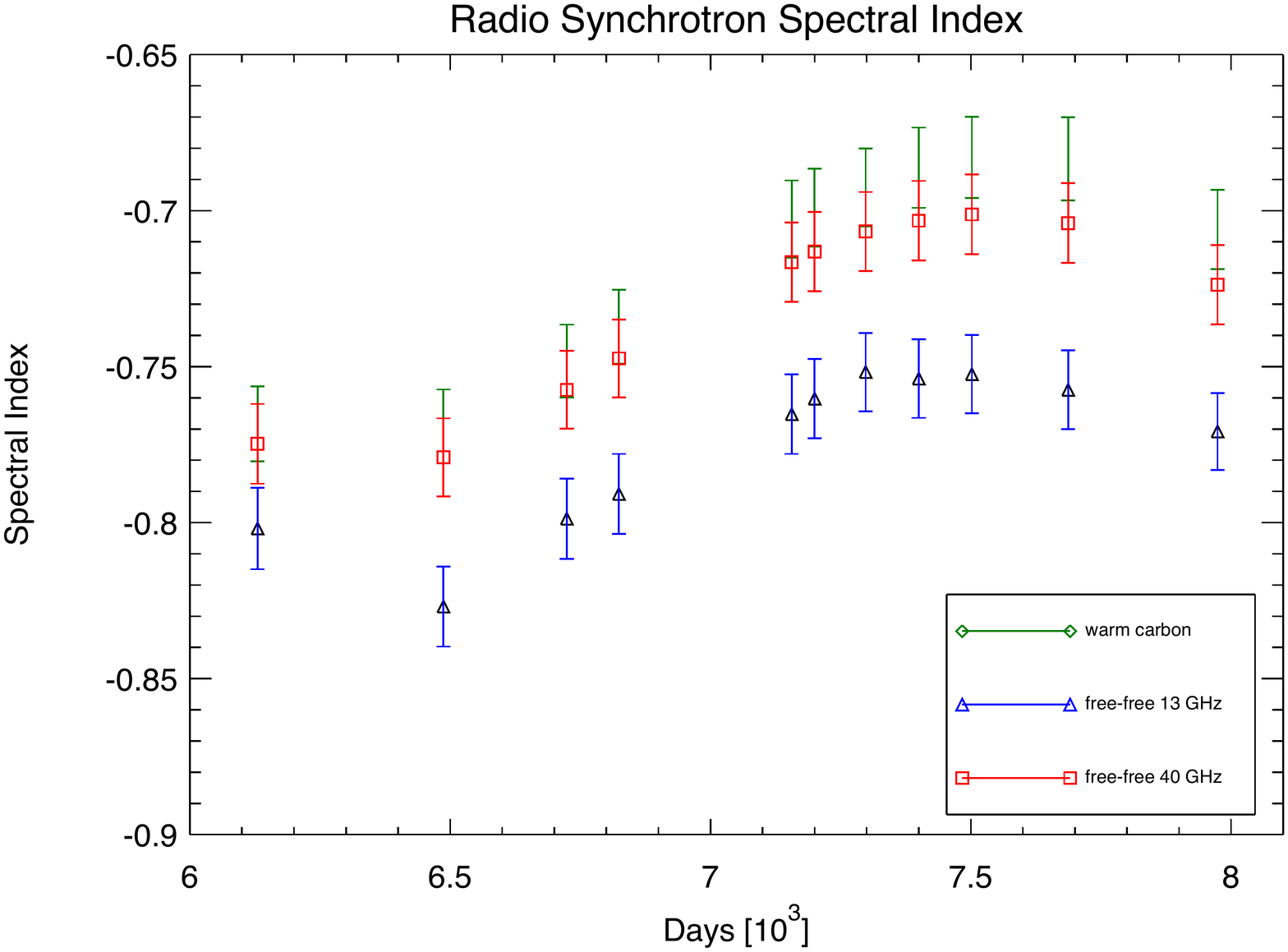}
  \end{center}
  \caption{Time evolution of the radio synchrotron spectral index (Green: with a warm carbon dust component; Red: free-free emission with cut-off at 40~GHz; Blue: free-free emission with cut-off at 13~GHz). Note that the synchrotron radio emission is de-correlated from the warm carbon dust emission and therefore this fit is equivalent to adjusting the synchrotron emission with the cold dust component only.}
  \label{fig:evolution_synchrotron_alpha}
\end{figure}

\begin{figure}
\begin{center}
\includegraphics[width=8.5cm]{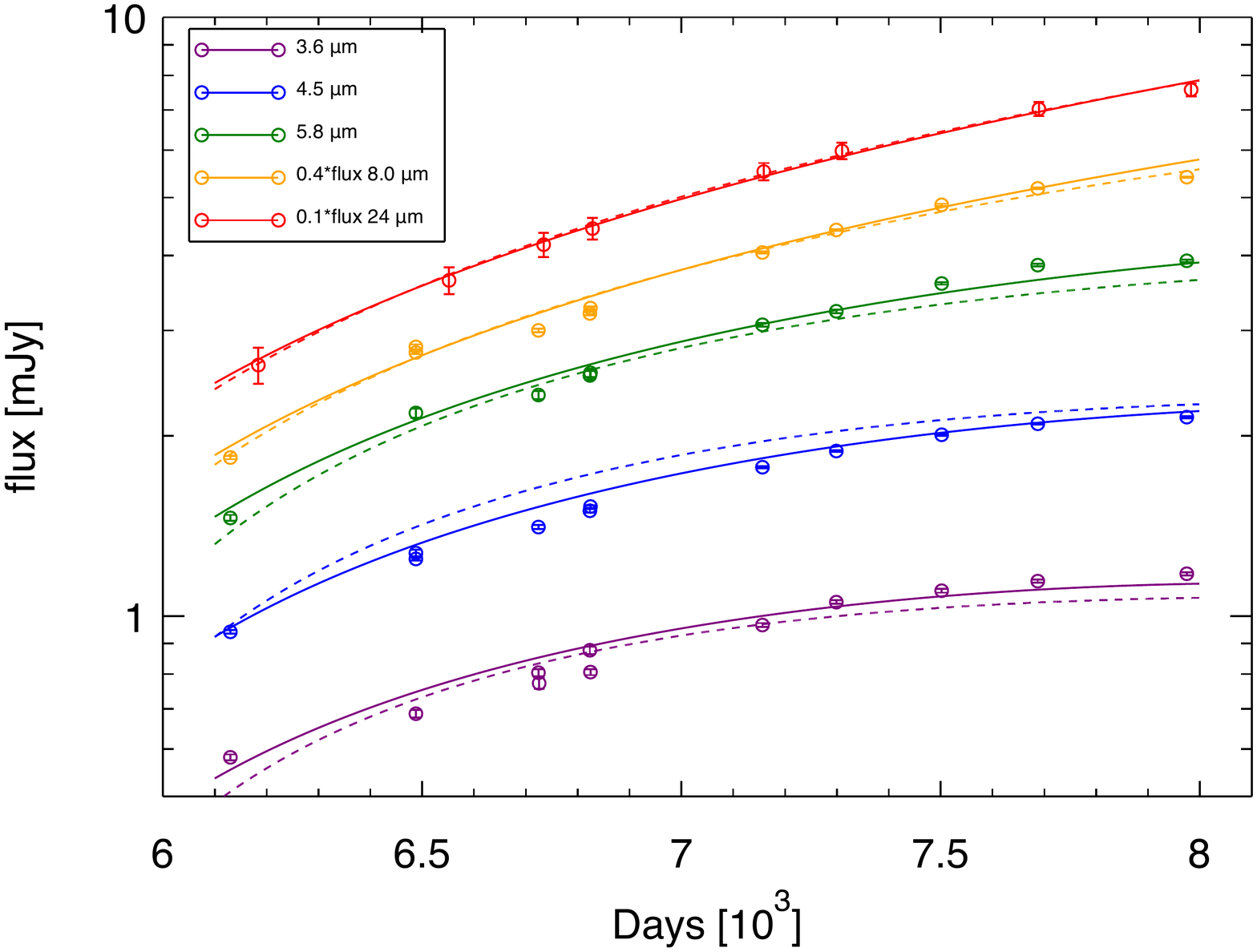}
  \end{center}
  \caption{Computed light curves from our models with amorphous carbon dust (dotted line) and self-absorbed free-free emission (plain line) compared to the {\it Spitzer} data. The free-free curves plotted here correspond to the cut-off at 13~GHz. The superposition of the light curves calculated in the 40~GHz case would not allow a difference to be distinguished. The association between colors and wavelengths of each curve is the same as for Figure~\Ref{fig:lc_IRAC} (Purple, blue, green, orange, red = 3.6, 4.5, 5.8, 8, 24 $\mu$m). } 
  \label{fig:lc_models}
\end{figure}

ATCA monitored SNR~1987A until day 8014 \citep{Zanardo2010}. It is extremely risky and hazardous to try to extrapolate the light curves to about 2000 days later. 
Especially since the light curves in the infrared domain covered by {\it Spitzer} all show after day 8000 first a lowering of the flux increase then a clear decrease  (see Figure~\Ref{fig:lc_IRAC}), which would consequently make any extrapolation of the ATCA data after that date quite suspicious. We note that, most probably for that reason, \cite{Matsuura2019} do not take into account the synchrotron radio data to build their SED, whereas they are of primary importance for us. 

\begin{figure} 
\begin{center}
\includegraphics[width=8.5cm]
{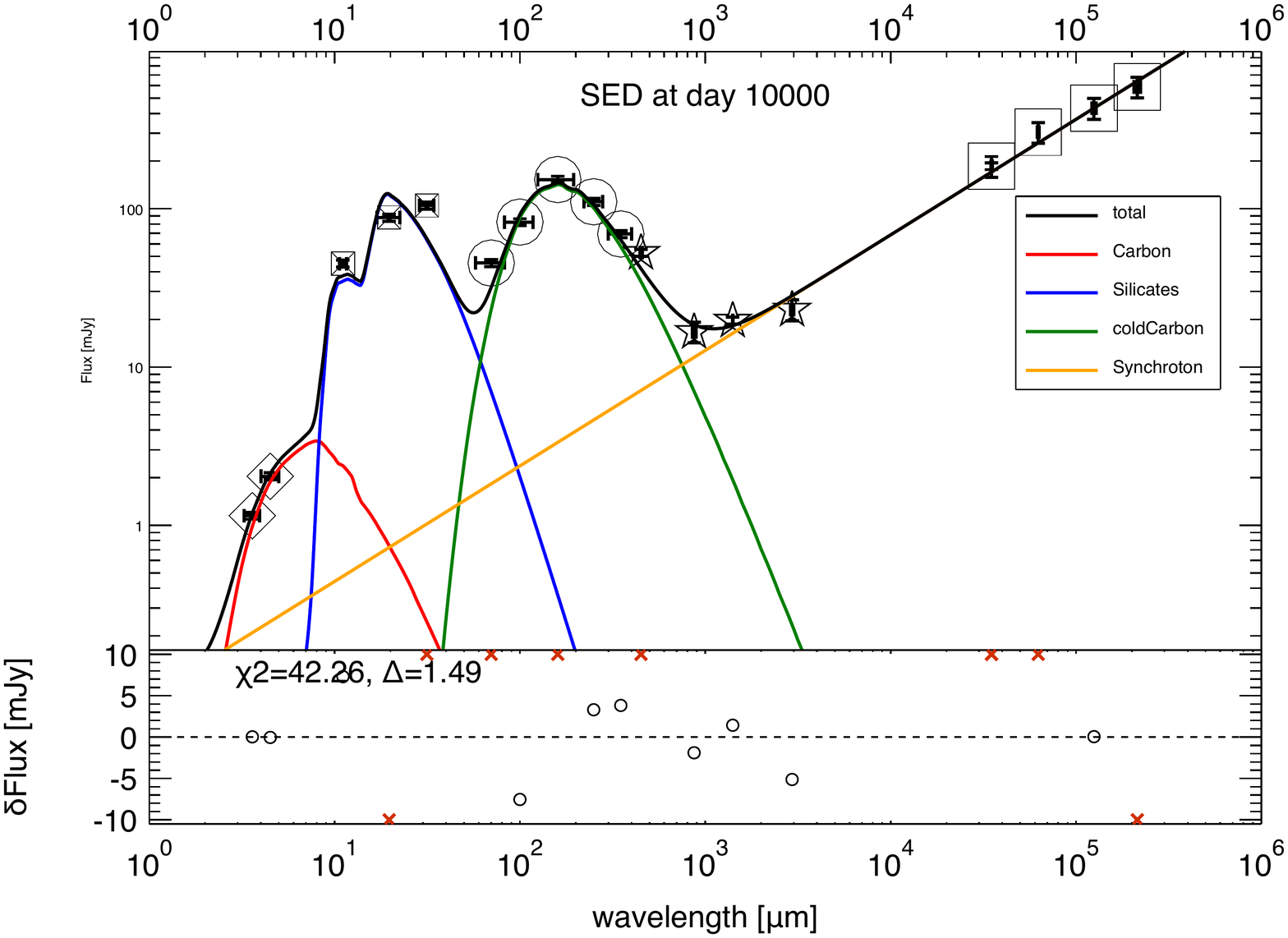} 
\includegraphics[width=8.5cm]
{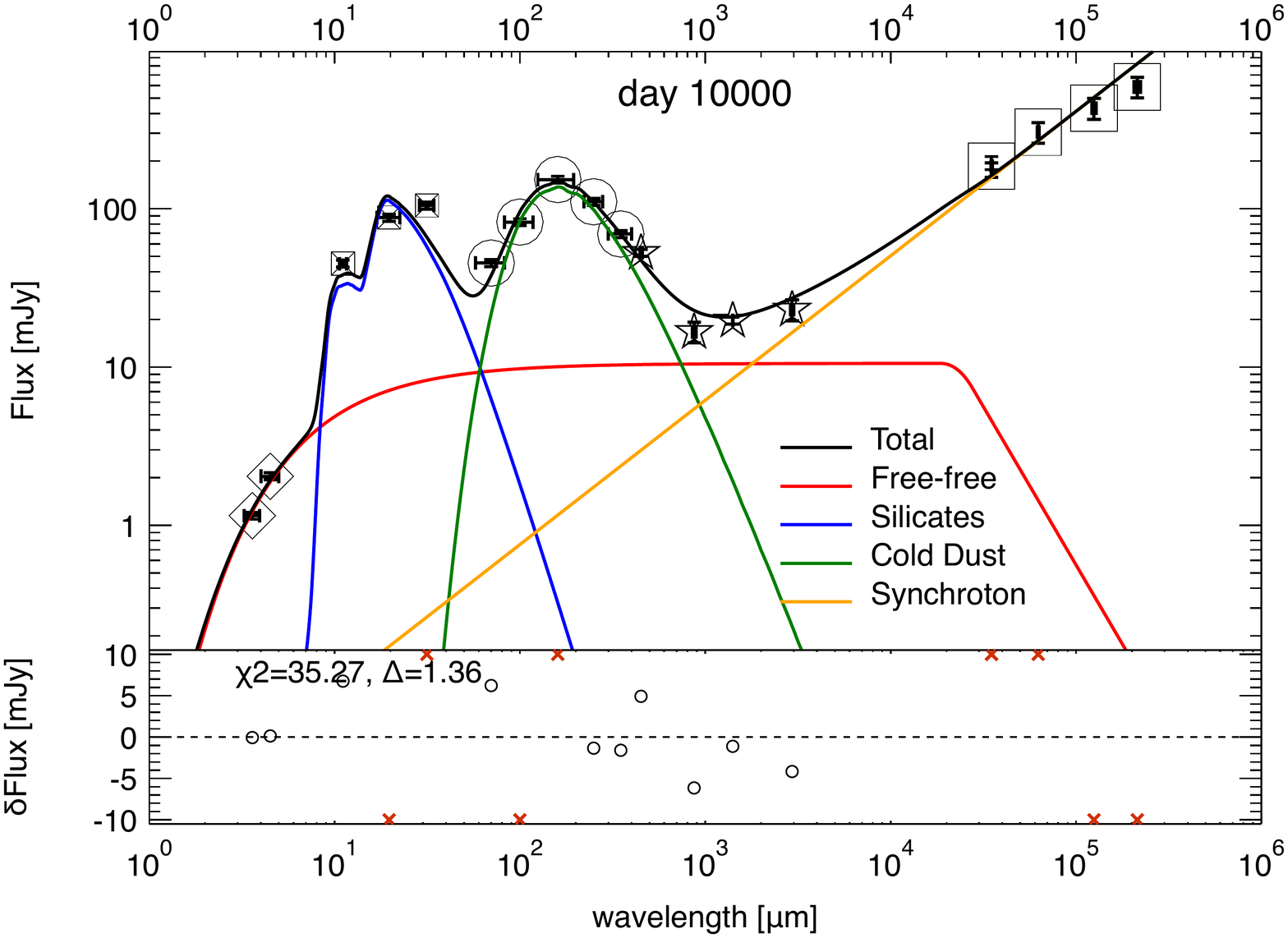}
\end{center}
 \caption{The Spectral Energy Distribution of SNR~1987A at day 10,000 with our resulting fits. The short wavelengths range has been fitted with a warm carbon component (top) and with a self-absorbed free-free radiation with a cut-off frequency of 13~GHz (bottom). The SOFIA 30~$\micron$ data point is slightly out of our adjustment. This could be due to the difference in the dates of the observations (this data point has been obtained at day 10732). \cite{Matsuura2019} use this measurement to better fit the SED at day~10,000 introducing an additional dust component at T = 85~K (their model 2 and 4). Our ``warm carbon'' model is essentially the same as \cite{Matsuura2019}'s ``Model 3'' in Table 2 of these authors. The difference being that we simply accept the 30~\micron\ data point as an outlier.}
 \label{fig:sed10000}
\end{figure}

Last but not least, at day 10,000 there is very little data available in the near- and mid-IR  wavelengths range to constrain, first either a dust component or a free-free emission, second the curve attesting the presence of silicates dust. However, in order to verify our hypothesis that the mass and the temperature of the cold dust in the ejecta do not vary significantly during the period analyzed, and to evaluate the value of the synchrotron spectral index, we nevertheless accepted the challenge of trying to extrapolate the ATCA curves, and to fit the near-IR wavelengths with so few data points (see Table~\ref{tab:data10000} and Figure~\ref{fig:sed10000}), while we are aware of the great limitations of that exercise.

Outside of all these difficulties our fit for the day 10,000 yields $T_{\rm Cold~Dust}$ = 23~K for the warm carbon and the 13~GHz self absorbed free-free models, and  $M_{\rm Cold~Dust}$ = 0.29 and 0.28~M$_\sun$ for the warm carbon dust and free-free emission, respectively (see Table~\ref{tab:parameters_d10000_fit}). Comparing these values with the ones that we have computed for days 7200 and 7400 proves, {\it a posteriori}, that our hypothesis that the temperature of the cold dust in the ejecta did not vary significantly between days 7200 and 10,000, is justified. Regarding the mass of this component, it would have slightly increased in the warm carbon model (0.29~$M_\sun$ instead of 0.25~$M_\sun$), and considerably increased in the free-free model (0.29~$M_\sun$ instead of 0.17)~$M_\sun$. We emphasize however, that these results should be taken with caution.  

As for the synchrotron spectral index, we found $\alpha = -0.91$ in the free-free case and $\alpha = -0.79$ in the warm carbon hypothesis. Although our fit is far from perfect (due to the challenging extrapolation of the ATCA data points), it is very doubtful that it could reach the value $\alpha = -0.5$ at day 11500 as predicted by \cite{Staveley2007} in either model. 

%\section{Tables}

\begin{deluxetable*}{rrrcccl}
\tabletypesize{\small}
\tablecaption{Telescopes used in our study, Dates of Observations and References}
\label{tab:telused}
\tablehead{
\colhead{$\lambda_0$ ($\micron$) } & \colhead{$\lambda_{\rm min}$ ($\micron$)} & 
\colhead{$\lambda_{\rm max}$ ($\micron$)} & 
\colhead{Instr.} &
\colhead{Reference} & 
\colhead{Days after outburst} & 
\colhead{Symbol Used} }
\startdata
3.13 & 2.5 & 4.0 & AKARI & \cite{Seok2008} & 7190 -- 7194 & 
{\begin{tikzpicture}{}
\draw(0,0) circle (0.1); 
\end{tikzpicture}}
\\
3.60 & 3.2 & 3.9 & {\it Spitzer} & \cite{Arendt2020}& 6130 – 10035 & 
\begin{tikzpicture}
 \node[diamond,
 draw = black, 
 scale=0.6] (d) at (0,0) {}; 
\end{tikzpicture}\\
4.50 & 4.0 & 5.0 & {\it Spitzer} & \cite{Arendt2020}& 6130 – 10035 & 
\begin{tikzpicture}
 \node[diamond,
 draw = black, 
 scale=0.6] (d) at (0,0) {}; 
\end{tikzpicture}\\
5.80 & 5.2 & 6.3 & {\it Spitzer} & \cite{Arendt2020} & 6130 – 10035 & 
\begin{tikzpicture}
 \node[diamond,
 draw = black, 
 scale=0.6] (d) at (0,0) {}; 
\end{tikzpicture}\\
6.95 & 5.5 & 8.7 & AKARI & \cite{Seok2008} & 7190 -- 7194 & 
{\begin{tikzpicture}{}
\draw(0,0) circle (0.1); 
\end{tikzpicture}}
\\
8.0 & 6.4 & 9.3 & {\it Spitzer} &\cite{Arendt2020} & 6130 – 10035 & 
\begin{tikzpicture}{}
 \node[diamond,
 draw = black, 
 scale=0.6] (d) at (0,0) {}; 
\end{tikzpicture}\\
10.19 & 8.3 & 15.3 & AKARI & \cite{Seok2008} & 7190 – 7194 & {\begin{tikzpicture}{}
\draw(0,0) circle (0.1); \end{tikzpicture}}
\\
11.1 & 10.625 & 11.575 & SOFIA & \cite{Matsuura2019} & 10731 – 10733 & $\boxtimes$\\ 
15.23 & 12.2 & 21.5 & AKARI & \cite{Seok2008} & 7190 -- 7194 &  
{\begin{tikzpicture}{}
\draw(0,0) circle (0.1); \end{tikzpicture}}
\\
19.7 & 16.95 & 22.45 & SOFIA & \cite{Matsuura2019} & 10731 – 10733 & $\boxtimes$\\
23.4 & 20.3 & 26.5 & AKARI & \cite{Seok2008} & 7190 – 7194 & {\begin{tikzpicture}{}
\draw(0,0) circle (0.1); \end{tikzpicture}}
\\
24.0 & 20.3 & 26.5 & {\it Spitzer} & \cite{Arendt2020} & 6130 – 10035 & 
\begin{tikzpicture}
 \node[diamond,
 draw = black, 
 scale=0.6] (d) at (0,0) {}; 
\end{tikzpicture}\\
31.5 & 28.65 & 34.35 &  SOFIA & \cite{Matsuura2019} & 10731 – 10733 & $\boxtimes$\\
70.0 & 60.0   & 85.0   & {\it Herschel}/PACS & \cite{Matsuura2015} & 9374 & $\bigtriangledown$\\
100 & 85.0 & 130.0 & {\it Herschel}/PACS & \cite{Matsuura2011} & 8467 – 8564, 9090 & $\bigtriangledown$\\
160 & 130.0 & 210.0 & {\it Herschel}/PACS & \cite{Matsuura2011} & 8467 – 8564, 9090 & $\bigtriangledown$\\
250 & 197.0 & 298.0 & {\it Herschel}/SPIRE & \cite{Matsuura2011} & 8467 – 8564, 9090 & $\bigtriangledown$\\
350 & 277.3 & 423.5 & {\it Herschel}/SPIRE & \cite{Matsuura2011} & 8467 – 8564, 9090  & $\bigtriangledown$\\
350 & 300 & 400 & APEX/SABOCA & \cite{Laki2012b} & 8907 – 8969  & $\bigtriangleup$\\
450 & 440 & 453 & ALMA & \cite{Zanardo2014} & 9351 & $\largestar$  \\
870 & 769.0 & 967.0 & APEX/LABOCA & \cite{Laki2011} & 7400 \& 8907 & $\bigtriangleup$\\
870 & 849 & 890 & ALMA & \cite{Zanardo2014} & 9294 & $\largestar$ \\
1406 & 1403 & 1407 & ALMA & \cite{Zanardo2014} & 9289 & $\largestar$ \\
2939 & 2883 & 2998 & ALMA & \cite{Zanardo2014} & 9174 & $\largestar$ \\
3200 & 2855 & 3526 & ATCA & \cite{Laki2011} & 6714 & {\begin{tikzpicture}{}
\draw(0,0) rectangle (0.2,0.2);
\end{tikzpicture}}
\\
35000 & 34600 & 35200 & ATCA & \cite{Zanardo2010} & 918 – 8014 & {\begin{tikzpicture}{}
\draw(0,0) rectangle (0.2,0.2);
\end{tikzpicture}}
\\
62500 & 61700 & 63200 & ATCA & \cite{Zanardo2010} & 918 – 8014 & 
{\begin{tikzpicture}{}
\draw(0,0) rectangle (0.2,0.2);
\end{tikzpicture}}
\\
125000 & 121900 & 128100 & ATCA & \cite{Zanardo2010} & 918 – 8014  & 
{\begin{tikzpicture}{}
\draw(0,0) rectangle (0.2,0.2);
\end{tikzpicture}}
\\
214130 & 205350 & 222910 & ATCA & \cite{Zanardo2010} & 918 – 8014 & 
{\begin{tikzpicture}{}
\draw(0,0) rectangle (0.2,0.2);
\end{tikzpicture}}
\\
\enddata
\end{deluxetable*}

\begin{deluxetable*}{rrrrcc}
\tabletypesize{\small}
\tablecaption{Flux densities at day 7200 }
\label{tab:data7200}
\tablehead{
\colhead{$\lambda_0$} & \colhead{$\lambda_{\rm min}$} & \colhead{$\lambda_{\rm max}$} & \colhead{Flux} & 
\colhead{Instr.} &
\colhead{Reference} \\
\colhead{($\micron$)} & \colhead{($\micron$)} & \colhead{($\micron$)} &
\colhead{(mJy)} & 
\colhead{} &
\colhead{}  }
\startdata
3.13 & 2.5 & 4.0 & 1.5 $\pm$ 0.10& AKARI & \cite{Seok2008} \\
3.60 & 3.2 & 3.9 & 1.00 $\pm$ 0.01 & {\it Spitzer} & Interpolated between 7156 and 7299 from \cite{Arendt2020} \\
4.50 & 4.0 & 5.0 & 1.81 $\pm$ 0.01& {\it Spitzer} & Interpolated between 7156 and 7299 from\cite{Arendt2020} \\
5.80 & 5.2 & 6.3 & 3.12 $\pm$ 0.02& {\it Spitzer} & Interpolated between 7156 and 7299 from \cite{Arendt2020} \\
6.95 & 5.5 & 8.7 & 4.9 $\pm$ 0.50& AKARI & \cite{Seok2008} \\
8.0 & 6.4 & 9.3 & 10.35 $\pm$ 0.04& {\it Spitzer} & Interpolated between 7156 and 7299 from \cite{Arendt2020} \\
10.19 & 8.3 & 15.3 & 32.0 $\pm$ 3.0& AKARI & \cite{Seok2008} \\
15.23 & 12.2 & 21.5 & 43.0 $\pm$ 4.0& AKARI & \cite{Seok2008} \\
23.4 & 20.3 & 26.5 & 65.0 $\pm$ 7.0& AKARI & \cite{Seok2008} \\
24.0 & 20.3 & 26.5 & 56.5 $\pm$ 1.9& {\it Spitzer} & Interpolated between 7159 and 7310 from \cite{Arendt2020} \\
100 & 85.0 & 130.0 & 70.5 $\pm$ 8.5 & {\it Herschel} & \cite[Days 8467,…,8564]{Matsuura2011}  \\
160 & 130.0 & 210.0 & 125.3 $\pm$ 16.1 & {\it Herschel} & \cite[Days 8467,…,8564]{Matsuura2011}  \\
250 & 197.0 & 298.0 & 131.7 $\pm$ 12.1 & {\it Herschel} & \cite[Days 8467,…,8564]{Matsuura2011}  \\
350 & 277.3 & 423.5 & 49.3 $\pm$ 18.0 & {\it Herschel} & \cite[Days 8467,…,8564]{Matsuura2011}  \\
350  & 300  & 400  & 44.0 $\pm$ 7 & APEX/SABOCA & \cite[Days 8907 – 8969]{Laki2012b}  \\
870 & 769.0 & 967.0 & 21. $\pm$ 4.0 & APEX/LABOCA & \cite[Day 7401]{Laki2011} \\
35000 & 34600 & 35200 & 83.6 $\pm$ 4.3 & ATCA &  \cite[Day 7202]{Zanardo2010} \\
62500 & 61700 & 63200 & 126.1  $\pm$ 5.0 & ATCA & \cite[Day 7202]{Zanardo2010} \\
125000 & 121900 & 128100 & 203.6  $\pm$ 6.6 & ATCA & \cite[Day 7202]{Zanardo2010} \\
214130 & 205350 & 222910 & 303.3 $\pm$ 9.5 & ATCA & \cite[Day 7202]{Zanardo2010} \\
\multicolumn{6}{c}{ATCA data have been interpolated after smoothing the light curves from \cite{Zanardo2010} by a polynomial of degree 3}     \\
\enddata  
\end{deluxetable*}

\begin{deluxetable*}{rrrrcc}
\tabletypesize{\footnotesize}
\tablewidth{0pt}
\tablecaption{Flux densities at day 7400}
\label{tab:data7400}
\tablehead{
\colhead{$\lambda_0$} & \colhead{$\lambda_{\rm min}$} & \colhead{$\lambda_{\rm max}$} & \colhead{Flux} & 
\colhead{Instr.} & 
\colhead{Reference} \\
\colhead{($\micron$)} & \colhead{($\micron$)} & \colhead{($\micron$)} &
\colhead{(mJy)} & 
\colhead{} & 
\colhead{} }
\startdata
3.60 & 3.2 & 3.9 & 1.07 $\pm$ 0.01& {\it Spitzer} & Interpolated between 7299 and 7502 from \cite{Arendt2020} \\
4.50 & 4.0 & 5.0 & 1.95 $\pm$ 0.01& {\it Spitzer} & Interpolated between 7299 and 7502 from  \cite{Arendt2020} \\
5.80 & 5.2 & 6.3 & 3.45 $\pm$ 0.02& {\it Spitzer} & Interpolated between 7299 and 7502 from  \cite{Arendt2020} \\
8.0 & 6.4 & 9.3 & 11.51 $\pm$ 0.04& {\it Spitzer} & Interpolated between 7299 and 7502 from   \cite{Arendt2020} \\
24.0 & 20.3 & 26.5 & 62.55 $\pm$ 1.9& {\it Spitzer} & Interpolated between 7310 and 7490 from  \cite{Arendt2020} \\
100 & 85.0 & 130.0 & 70.5 $\pm$ 8.5 & {\it Herschel}/PACS & \cite{Matsuura2011} (Days 8467,…,8564)\\
160 & 130.0 & 210.0 & 125.3 $\pm$ 16.1 & {\it Herschel}/PACS & \cite{Matsuura2011} (Days 8467,…,8564) \\
250 & 197.0 & 298.0 & 131.7 $\pm$ 12.1 & {\it Herschel}/SPIRE & \cite{Matsuura2011} (Days 8467,…,8564) \\
350 & 277.3 & 423.5 & 49.3 $\pm$ 18.0 & {\it Herschel}/SPIRE & \cite{Matsuura2011} (Days 8467,…,8564) \\
350 & 300 & 400 & 44.0  $\pm$ 7 & APEX/SABOCA & \cite{Laki2012b} (Days 8907 – 8969) \\
870 & 800.0 & 952.0 & 21. $\pm$ 4. & APEX/LABOCA & \cite{Laki2011} (Day 7401) \\
35000 & 34600 & 35200 & 91.7  $\pm$ 4.6 & ATCA & Interpolated between 7370 and 7437 from \cite{Zanardo2010} after smoothing\\
62500 & 61700 & 63200 & 137.8 $\pm$ 5.5 & ATCA & Interpolated between 7370 and 7437 from \cite{Zanardo2010} after smoothing \\
125000 & 121900 & 128100 & 219.6  $\pm$ 7.0 & ATCA & Interpolated between 7370 and 7437 from \cite{Zanardo2010} after smoothing \\
214130 & 205350 & 222910 & 327.2  $\pm$ 10.3 & ATCA & Interpolated between 7370 and 7437 from \cite{Zanardo2010} after smoothing\\ 
\multicolumn{6}{c}{ATCA data have been interpolated after smoothing the light curves from \cite{Zanardo2010} by a polynomial of degree 3}  
\enddata
\end{deluxetable*}

\begin{deluxetable*}{ccc} 
\tabletypesize{\small}
\tablewidth{0pt}
\tablecaption{Parameters for the Warm Carbon Model (errors are obtained through Monte-Carlo simulation with 5000 draws).}
\label{tab:parameters_wc}
\tablehead{
\colhead{} & \colhead{Day 7200} & \colhead{Day 7400} }
\startdata
Mass & 3.1~$\pm$~1.6~$10^{-8}$${\rm M}_\sun$ & 3.7~$\pm$~1.6~$10^{-8}$${\rm M}_\sun$\\
Temperature & 475~$\pm$~20~K &  469~$\pm$~29~K  \\
$M_{\rm Silicates}$ & 6.9~$\pm$~0.4~$10^{-6}$${\rm M}_\sun$ & 7.6~$\pm$~0.6~$10^{-6}$${\rm M}_\sun$\\
$T_{\rm Silicates}$ & 197~$\pm$~3~K & 197~$\pm$~4~K\\
$M_{\rm Cold~Dust}$ & 0.23$\pm$~0.01~${\rm M}_\sun$ & 0.24~$\pm$~0.03~${\rm M}_\sun$ \\
$T_{\rm Cold~Dust}$ & 23.4~$\pm$~0.2~K & 23~$\pm$~0.70~K \\
Synchrotron Index & -0.70~$\pm$~0.04 & -0.69~$\pm$~0.05 \\
$\chi^2$ & 1987 & 671 \\
$\Delta$ & 10. & 6.9  
\enddata
\end{deluxetable*}

\begin{deluxetable*}{ccc} 
\tabletypesize{\small}
\tablewidth{0pt}
\tablecaption{Parameters for the free-free model self-absorbed at 13~GHz (2.3 $\times$ 10$^4$\micron). Errors are obtained through Monte-Carlo simulation with 5000 draws.}
\label{tab:parameters_ffcut13}
\tablehead{
\colhead{} & \colhead {Day 7200} &\colhead {Day 7400}}
\startdata
$(1 - \xi)$  & 0.073~$\pm$~0.003 & 0.081~$\pm$~0.003\\
Temperature & 1444~$\pm$~53~K & 1319~$\pm$~48~K  \\
$M_{\rm Silicates}$ & 5.5~$\pm$~0.4~$10^{-6}$${\rm M}_\sun$ & 5.9~$\pm$~0.5~$10^{-6}$${\rm M}_\sun$\\
$T_{\rm Silicates}$ & 199~$\pm$~2.6~K & 199~$\pm$~3~K\\
$M_{\rm Cold~Dust}$ & 0.16~$\pm$~0.01~{\rm M}$_\sun$ & 0.18~$\pm$~0.01$~{\rm M}_\sun$ \\
$T_{\rm Cold~Dust}$ & 24.~$\pm$~0.3~K & 23.~$\pm$~0.3~K \\
Synchrotron $\alpha$ & -0.85~$\pm$~0.02 & -0.75~$\pm$~0.02 \\
$\chi^2$ & 2578 & 602 \\ 
$\Delta$ & 11.4 & 6.3
\enddata
\end{deluxetable*}

\begin{deluxetable*}{ccc} 
\tabletypesize{\small}
\tablewidth{0pt}
\tablecaption{Parameters for the free-free model self-absorbed at 40~GHz (7500~\micron). Errors are obtained through Monte-Carlo simulation with 5000 draws. }
\label{tab:parameters_ffcut40}
\tablehead{
\colhead{} & \colhead {Day 7200} &\colhead {Day 7400}}
\startdata
$(1 - \xi)$  & 0.073~$\pm$~0.003 & 0.083~$\pm$~0.004\\
Temperature & 1416~$\pm$~53~K & 1275~$\pm$~52~K  \\
$M_{\rm Silicates}$ & 5.5~$\pm$~0.4~$10^{-6}$${\rm M}_\sun$ & 5.9~$\pm$~0.4~$10^{-6}$${\rm M}_\sun$\\
$T_{\rm Silicates}$ & 200~$\pm$~3.~K & 199~$\pm$~3.~K\\
$M_{\rm Cold~Dust}$ & 0.15~$\pm$~0.01~${\rm M}_\sun$ & 0.17~$\pm$~0.01~${\rm M}_\sun$ \\
$T_{\rm Cold~Dust}$ & 24.~$\pm$~0.3~K & 23.~$\pm$~0.3~K \\
Synchrotron $\alpha$ & -0.75~$\pm$~0.02 & -0.71~$\pm$~0.02 \\
$\chi^2$ & 2579 & 567 \\ 
$\Delta$ & 11.4 & 6.2
\enddata
\end{deluxetable*}

\begin{deluxetable*}{rrrrcc}[ht]
\tabletypesize{\small}
\tablewidth{0pt}
\tablecaption{Flux densities at day $\sim$10000}
\label{tab:data10000}
\tablehead{
\colhead{$\lambda_0$}&\colhead{$\lambda_{\rm min}$}&\colhead{$\lambda_{\rm max}$}&\colhead{Flux}&\colhead{Instr.}&\colhead{Reference} \\
\colhead{($\micron$)}&\colhead{($\micron$)}&\colhead{($\micron$)}&
\colhead{(mJy)}&\colhead{}&\colhead{} }
\startdata
3.60 & 3.2 & 3.9 & 1.15 $\pm$ 0.01& {\it Spitzer} & Interpolated between 9810 and 10035 from \cite{Arendt2020} \\
4.50 & 4.0 & 5.0 & 2.04 $\pm$ 0.02& {\it Spitzer} & Interpolated between 9810 and 10035 from \cite{Arendt2020} \\
11.1 & 10.625 & 11.575 &  45. $\pm$ 7  &  SOFIA & \cite[Days 10731 -- 10733]{Matsuura2019} \\
19.7 & 16.95  & 22.45  &  88. $\pm$ 9  &  SOFIA & \cite[Days 10731 -- 10733]{Matsuura2019} \\
31.5 & 28.65  & 34.35  & 105. $\pm$ 14 &  SOFIA & \cite[Days 10731 -- 10733]{Matsuura2019} \\
70.0 & 60.0   & 85.0   &  45.4 $\pm$ 3.4 & {\it HERSCHEL}/PACS & \cite[Day 9374]{Matsuura2015} \\
100.0 & 85.0   & 120.0   &  82.4 $\pm$ 4.5 & {\it HERSCHEL}/PACS & \cite[Day 9090]{Matsuura2015} \\
160.0 & 130.0   & 200.0   &  153.0 $\pm$ 9 & {\it HERSCHEL}/PACS & \cite[Day 9090]{Matsuura2015} \\
250.0 & 214.0   & 272.0   &  110.7 $\pm$ 25.2 & {\it HERSCHEL}/SPIRE & \cite[Day 9122]{Matsuura2015} \\
350.0 & 300.0   & 400.0   &  69.3 $\pm$ 22.8 & {\it HERSCHEL}/SPIRE & \cite[Day 9122]{Matsuura2015} \\
450.0 & 440.0 & 453.0 & 52.8 $\pm$ 14.2 & ALMA & \cite[Day 9351]{Zanardo2014} \\
870.0 & 849.0 & 890.0 & 16.7 $\pm$ 10.5 & ALMA &  \cite[Day 9294]{Zanardo2014} \\
1406 & 1403 & 1407 & 19.7 $\pm$ 10.0 & ALMA & \cite[Day 9289]{Zanardo2014} \\
2939 & 2883 & 2998 & 23.1 $\pm$ 13.1 & ALMA & \cite[Day 9174]{Zanardo2014} \\
35000 & 34600. & 35200. &  186.0 $\pm$ 5 & ATCA &  \cite[Day 9915]{Cendes2018} \\
35000 & 34600. & 35200. &  185.8 $\pm$ 10.6 & ATCA  & \cite[Extrapolated from Fig.2]{Zanardo2017} \\
62500 & 61700 & 63200 & 305 $\pm$ 18 & ATCA  & \cite[Extrapolated from Fig.2]{Zanardo2017}  \\
125000 & 121900 & 128100 & 432.3 $\pm$ 20 & ATCA  & \cite[Extrapolated from Fig.2]{Zanardo2017} \\
214130 & 205350 & 222910 & 588.9 $\pm$ 20 & ATCA  & \cite[Extrapolated from Fig.2]{Zanardo2017} 
\enddata
\end{deluxetable*}

\begin{deluxetable*}{ccc} 
\tabletypesize{\small}
\tablewidth{0pt}
\tablecaption{Parameters obtained from our fits at days 10,000 with the warm carbon and the 13~GHz self-absorbed free-free models. Errors are obtained through Monte-Carlo simulation with 5000 draws.}
\label{tab:parameters_d10000_fit}
\tablehead{
\colhead{} & \colhead{Free-Free} &\colhead{Warm Carbon}}
\startdata
$(1 - \xi)$ $|$ ~Mass & 0.06~$\pm$~0.005 & 2.5~$\pm$1.5~$10^{-8}$${\rm M}_\sun$\\
Temperature & 1856~$\pm$~56~K & 494~$\pm$~19~K  \\
$M_{\rm Silicates}$ & 2.6~$\pm$~0.2~$10^{-5}$${\rm M}_\sun$ & 2.9~$\pm$~0.5~$10^{-5}$${\rm M}_\sun$\\
$T_{\rm Silicates}$ & 159~$\pm$~2~K & 158~$\pm$~3~K\\
$M_{\rm Cold~Dust}$ & 0.29~$\pm$~0.01$~{\rm M}_\sun$ & 0.30~$\pm$0.01$~{\rm M}_\sun$ \\
$T_{\rm Cold~Dust}$ & 23~$\pm$~0.2~K & 23$\pm$~0.2~K \\
Synchrotron $\alpha$ & -0.91~$\pm$~0.01 & -0.79~$\pm$~0.01 \\
$\chi^2$ & 35 & 42 \\
$\Delta$ & 1.4 & 1.5
\enddata
\end{deluxetable*}

\end{document}